\documentclass[12pt]{article}
\pdfoutput=1

\usepackage{mathtools,amssymb,amsthm,amsfonts}
\usepackage{booktabs,multicol,multirow,adjustbox,threeparttablex,makecell}
\usepackage{graphicx}
\usepackage[margin=1in]{geometry}
\usepackage{setspace}
\usepackage{xcolor}
\usepackage{hyperref}
\usepackage{nicefrac}
\usepackage{enumitem}
\usepackage{lmodern,microtype}

\hypersetup{
	colorlinks,
    urlcolor={blue!50!black},
    linkcolor={blue!50!black},
    citecolor={blue!50!black},
}
\setlist{
	listparindent=\parindent,
	parsep=0pt,
}

\newcommand{\tgfe}{\text{\normalfont tgfe}}
\newcommand{\ugfe}{\text{\normalfont ugfe}}
\newcommand{\ig}{\text{\normalfont ig}}
\newcommand{\rf}{\text{\normalfont rf}}
\newcommand{\tfs}{\text{\normalfont fs}}
\newcommand{\tss}{\text{\normalfont ss}}

\let\Pr\relax\DeclareMathOperator{\Pr}{Pr}

\DeclareMathOperator{\IC}{IC}
\DeclareMathOperator{\K}{(K)}
\DeclareMathOperator{\Kmax}{(K^{\max})}
\DeclareMathOperator{\Kp}{(K^0)}
\DeclareMathOperator{\G}{(G)}
\DeclareMathOperator{\Gmax}{(G^{\max})}

\DeclareMathOperator{\GK}{(G,K)}
\DeclareMathOperator{\GmaxKmax}{(G^{\max},K^{\max})}
\DeclareMathOperator{\GKp}{(G,K^0)}
\DeclareMathOperator{\GpKp}{(G^0,K^0)}

\DeclareMathOperator*{\cov}{Cov}
\DeclareMathOperator*{\Exp}{\mathbb E}
\DeclareMathOperator*{\tr}{tr}
\DeclareMathOperator{\diag}{diag}
\DeclareMathOperator*{\plim}{plim}

\newtheoremstyle{note}
{0.5em}
{0.5em}
{\itshape}
{}
{\bfseries\upshape}
{. \phantomsection}
{.5em}
{}
\theoremstyle{note}
\newtheorem{theorem}{Theorem}
\newtheorem{assumption}{Assumption}
\newtheorem{remark}{Remark}
\newtheorem{lemma}{Lemma}

\allowdisplaybreaks

\usepackage{datetime}
\newdateformat{monthyear}{\monthname[\THEMONTH] \THEYEAR}

\usepackage[style=authoryear-comp, maxbibnames=99, uniquename = false, hyperref = true]{biblatex}
\DeclareBibliographyCategory{appx}
\addtocategory{appx}{BaiNg2002,BonhommeManresa2015,MerlevedePeligradRio2011}
\DeclareBibliographyCategory{onlyappx}
\addtocategory{onlyappx}{BonhommeManresa2015,MerlevedePeligradRio2011}

\makeatletter
\renewcommand\maketitle{\par
  \begingroup
  	\setcounter{footnote}{0}
    \thispagestyle{plain}
    \renewcommand\thefootnote{\@fnsymbol\c@footnote}%
    \if@twocolumn
      \ifnum \col@number=\@ne
        \@maketitle
      \else
        \twocolumn[\@maketitle]%
      \fi
    \else
      \newpage
      \global\@topnum\z@   
      \@maketitle
    \fi
    \thispagestyle{plain}\@thanks
  \endgroup
  \global\let\@thanks\@empty
}
\makeatother

\usepackage{minitoc}

\title{Latent group structure in linear panel data models \\ with endogenous regressors\thanks{
Okui acknowledges financial support from financial support from the BK21 PLUS program, the School of Social Sciences at Seoul National University, and the Japan Society for the Promotion of Science through KAKENHI Grant Nos. 22H00833, 22K20154, and 23H00804.
Choi acknowledges a student travel grant from the International Association for Applied Econometrics Annual Conference (IAAE2022) and financial support from the BK21 FOUR program and the Department of Economics at Seoul National University.
The authors would like to thank Daniel Czarnowske, Xavier D'Haultfoeuille, Ron Smith, Yohei Yamamoto, and participants in a TEDS seminar, the 2022 International Panel Data Conference (IPDC2022), IAAE2022, Erasmus University of Rotterdam, Tilburg, the 2022 Asian Meeting of Econometric Society in Tokyo (AMEST2022), Connecticut, the Spring Meeting of the Japanese Economic Association, and Keio University for their valuable comments and suggestions. Part of this research was conducted while Okui and Choi were at Seoul National University. All errors are ours.}}
\author{Junho Choi\thanks{Department of Economics, University of Wisconsin-Madison. Email: \href{mailto: junho.choi@wisc.edu}{junho.choi@wisc.edu}} \ and Ryo Okui\thanks{Corresponding author. Graduate School of Economics and Faculty of Economics, University of Tokyo. Email: \href{mailto:okuiryo@e.u-tokyo.ac.jp}{okuiryo@e.u-tokyo.ac.jp}}}
\date{May 2024}

\begin{document}
\maketitle

\begin{abstract}

This paper concerns the estimation of linear panel data models with endogenous regressors and a latent group structure in the coefficients. We consider instrumental variables estimation of the group-specific coefficient vector. We show that direct application of the Kmeans algorithm to the generalized method of moments objective function does not yield unique estimates. We newly develop and theoretically justify two-stage estimation methods that apply the Kmeans algorithm to a regression of the dependent variable on predicted values of the endogenous regressors. The results of Monte Carlo simulations demonstrate that two-stage estimation with the first stage modeled using a latent group structure achieves good classification accuracy, even if the true first-stage regression is fully heterogeneous. We apply our estimation methods to revisiting the relationship between income and democracy.


\end{abstract}

\newpage

\begin{mtchideinmaintoc}

\section{Introduction}

This paper considers the estimation of linear panel data models with endogenous regressors where the coefficients exhibit a grouped pattern of heterogeneity. Instrumental variables (IVs) estimation becomes nontrivial because of the latent group structure. We examine various approaches for estimating such models. We particularly examine when these methods can provide consistent group membership and coefficient estimators.

Endogeneity is an important concern in empirical research in economics and other social sciences. It is also relevant for panel data analysis. Regressors are endogenous when they are correlated with the error term. This is important as least squares methods do not yield consistent estimators in the presence of endogenous regressors. IVs are employed to solve this issue. A valid IV is correlated with the endogenous regressors, but uncorrelated with the error term in the model. IV estimation is a standard tool in econometrics and is discussed in many textbooks.

In addition, heterogeneity across units has recently gained attention in the econometrics literature. In our setting, the coefficients exhibit a latent group structure.
 An advantage of panel data is the ability to investigate heterogeneity across units because a panel data set contains multiple observations per unit. However, allowing complete heterogeneity may not be practical in many applications. One reason is that there may not be many observations per unit. However, even if there are many observations per unit to enable us to estimate a model separately for each unit, it would be difficult to summarize the features of heterogeneity across many units if each unit has its own estimate. A grouped pattern of heterogeneity is a useful modeling device for understanding heterogeneity. We divide the set of units into a small number of groups and assume that units in each group are homogeneous, but two units belonging to two different groups are different. Such a structure facilitates estimation and interpretation.

We thus examine how a model with endogenous regressors, and a grouped pattern of heterogeneity, can be estimated. Despite the importance of endogeneity and the recent popularity of models with a latent group structure, the literature lacks a comprehensive analysis. The present paper fills this gap. In particular, we show that na\"ively combining IV methods with Kmeans \parencite{MacQueen1965}, arguably the best-known clustering method, does not yield a uniquely defined estimator. We then consider alternative approaches, including those newly proposed here, that yield consistent estimators. 

We first demonstrate that it is not possible to determine a unique minimizer of the generalized method of moments (GMM) objective function when both the coefficient vector and the group membership structure are unknown. That is, na\"ive extensions of the Kmeans procedure, or the group fixed effects (GFE) estimator of \textcite{BonhommeManresa2015} for regression models, fail to yield unique estimates, and the estimators lack bases for their statistical and asymptotic analyses. Specifically, we show that for any hypothetical group membership structure, there may exist a coefficient vector that sets the GMM objective function to zero, at least asymptotically. This result is in stark contrast to the case of models without endogeneity where a least squares objective function can be employed to estimate the group memberships and coefficients simultaneously.

We develop two-stage least squares (2SLS) estimation to consistently estimate the group memberships and consider various specifications for the first-stage regression. This procedure appears novel in the literature on latent group structures. 
We then consider other approaches under which the estimation of the group membership structure and the coefficients are separated.

In the 2SLS estimation approach, we estimate a first-stage regression that describes the relationship between the endogenous regressors and IVs and then apply the GFE estimator of \textcite{BonhommeManresa2015} to a second-stage regression of the dependent variable on the predicted values of the endogenous regressors from the first stage. We consider three modeling frameworks for the first-stage regression. The simplest version is a homogeneous regression model. 
The second modeling framework allows the first-stage regression to exhibit a grouped pattern of heterogeneity. The first-stage regression is estimated by GFE, and the predicted values are constructed based on estimated group-specific coefficients. The group structures in the first and second stages are allowed to be different. The third modeling framework is to allow complete heterogeneity across units. The first-stage regression is estimated for each unit. It turns out that this approach coincides with minimizing the sum of the GMM objective functions for each unit with a specific choice of weighting matrices.\footnote{The estimators in \textcite{SuShiPhillips2016} and \textcite{Mehrabani2022} are based on this objective function, but they consider penalized estimations. In the context of generalized estimating equation, \textcite{itosugasawa2023} consider a similar estimator.} We prove the consistency of these 2SLS estimators of both the coefficient vector and the group membership structure.

The other two procedures provide consistent group membership estimators under some additional conditions but do not directly deliver a consistent coefficient estimator. The first procedure, referred to as the ``ignoring endogeneity approach,'' is to apply the GFE estimator to the model for the dependent variable and the endogenous regressors while ignoring the endogeneity. This provides a consistent group membership estimator if the grouped pattern in the linear projection of the dependent variable on the endogenous regressors is identical to that in the true coefficient vector. While this assumption may be debatable, this approach is suggested, for example, in \textcite[][Section S.4]{BonhommeManresa2015Supp}. Once an estimate of the group membership structure is obtained, the coefficient vector is estimated by GMM for each group. The second procedure, referred to as the ``reduced-form approach,'' is to consider the reduced-form equation of the dependent variable on the IVs. This approach requires that the group pattern in the reduced-form equation is identical to that for the coefficients of interest. To our best knowledge, this approach has not hitherto been considered in the literature. The coefficients are estimated by group-by-group GMM estimations after the group membership structure is estimated.

We also consider several extensions. First, we consider models with unit-specific fixed effects. Here, we merely need to apply the estimation procedures to the within or first-differencing transformed models. 
We use dynamic linear panel data models for illustration.
Second, we consider models with time-varying group-specific fixed effects. Because the dimension of the coefficient vectors now grow as the data becomes longer, a different theoretical analysis is required. 
We illustrate the advantages of the 2SLS estimation approach in these extensions.

We compare the performance of our estimators using simulations.
Performance is measured using the average classification accuracy and the average deviation from the true coefficient vector.
Four data-generating processes (DGPs) are considered, differentiated by the first-stage modeling.
Every estimator performs well when predicted to be so by asymptotic theory, confirming the usefulness of the asymptotic analyses.
Among the estimators, the 2SLS estimators, allowing heterogeneity in the first stage, show good performance under all DGPs. Interestingly, even in the DGPs where each unit has different first-stage coefficients, the 2SLS estimators, assuming a first-stage group structure, provide comparable performance to the estimator allowing complete heterogeneity.
The ignoring endogeneity approach shows excellent performance when the group structures of the estimating model and the structural equation coincide. However, when the correlation between the first- and second-stage errors is high and varies across units in addition to the complete heterogeneity in the first stage, the heterogeneity pattern in the correlation contaminates the group structure, and the ignoring endogeneity approach performs poorly. In such cases, the 2SLS estimator provides reliable estimates, and imposing a group structure in the first stage improves the estimation.

Lastly, we apply our estimators to revisit the empirical analysis in \textcite{AcemogluJohnsonRobinsonYared08}, which examines the relationship between income growth and democracy.
The authors use dynamic linear models to estimate the causal effect of income on democracy.
They employ the savings rate and trade-weighted world income as IVs to address the endogeneity of income growth because of reverse causality.
We extend these models and allow the coefficients to differ across groups.
The number of groups is set to two. The IV is trade-weighted world income.
When only the coefficient of income growth is allowed to exhibit a group structure, all the procedures produce compatible estimates, which are positive and mostly statistically significant. Re-running a separate IV regression for each group does not change the results qualitatively. 
Overall, larger groups tend to demonstrate a greater effect of income. However, when we introduce group-specific fixed effects additionally, the procedures yield estimates that differ to some extent---particularly when including the lagged democracy in the model. The coefficient estimates are different across (second-stage) groups when the number of groups in the first stage is small, but they become similar as we increase the number of groups in the first stage. This result underscores the importance of modeling first-stage heterogeneity 
in correctly eliciting the heterogeneity pattern in the structural parameters.

\subsection*{Related literature}

Latent group structure has been a popular topic in the recent econometrics literature. An (admittedly incomplete) list of papers on this topic includes \textcite{Sun2015}, \textcite{HahnMoon2010}, \textcite{LinNg2012}, \textcite{BonhommeManresa2015}, \textcite{SuShiPhillips2016}, \textcite{AndoBai2016}, \textcite{VogtLinton2016}, \textcite{WangPhillipsSu2018}, \textcite{Mehrabani2022}, and \textcite{mugnier2022simple}. They primarily focus on models with exogenous regressors. The present paper considers cases with endogenous regressors and their IV estimation.

While IV estimation for models with a latent group structure has been examined in the existing literature, discussion of these models is typically included only in extensions or appendices of main papers. See, for example, \textcite{BonhommeManresa2015Supp, SuShiPhillips2016, Mehrabani2022}. An application is found in \textcite{Czarnowske2022}, which uses the method in \textcite{SuShiPhillips2016} to estimate heterogeneous production functions. Our aim is to provide a comprehensive account of an application of group heterogeneity to IV estimation. In many economic applications, endogeneity is a valid concern, and IV estimation is required. The present paper expands the scope of the heterogeneity analysis with a latent group structure in the existing economics literature.

We also develop a new estimation approach, namely the 2SLS approach. This approach has several advantages over existing methods.  In particular, it enables us to consider various extensions. For example, it can accommodate time-varying parameters while allowing general heterogeneity patterns in the first stage. We note that controlling for time-varying group-specific intercepts is the motivation of \textcite{BonhommeManresa2015}. Generic clustering approaches such as that by \textcite{YuGuVolgushev2022} cannot be applied here because it is impossible to obtain preliminary estimates from unit-by-unit estimation in the presence of time-varying parameters. It is thus necessary to develop estimation methods tailored to handle time-varying parameters.

The proposed estimator can exploit the heterogeneity in the first stage. \textcite{adadiegushen2023} also consider first-stage equations with group heterogeneity but assume that the group structure is known. \textcite{wiemann2023} applies the Kmeans algorithm to cluster categorical instruments to improve the efficiency of the estimator. For our part, we allow latent group structure in the first stage but require panel data. 

Regarding proof techniques and estimation ideas, our paper relates to the structural break literature. There is a rich body of research on structural break detection using time series data. The proofs of the negative results for the na\"ive approaches follow the argument of \textcite{HallHanBoldea2012}. They document the failure of the GMM procedure to detect structural breaks in time series analysis. We consider different econometric situations, and the proofs are not identical. Nonetheless, their mathematical insights are useful in deriving our results. Furthermore, some of the estimation methods we consider have their origins in the structural break literature. The 2SLS approach to detect structural breaks is examined in \textcite{HallHanBoldea2012} and \textcite{PerronYamamoto2014}. The approach based on biased estimation (by ignoring endogeneity) is considered in \textcite{PerronYamamoto2015, QianSu2014}. The present paper demonstrates that although the results in the structural break literature are at first very different in scope from ours, they are useful in investigating econometric problems for models with group patterns of heterogeneity.

While some research papers consider combining structural breaks with a grouped pattern of heterogeneity, such as \textcite{OkuiWang2022, LumsdaineOkuiWang2022}, they have very different scopes from the present paper. They consider the estimation of break points that affect group-specific coefficient vectors and/or group membership structures. The present paper does not consider structural breaks; we merely utilize their theoretical and mathematical insights for models with endogenous regressors.

\paragraph{Organization of the paper}

Section \ref{sec:model} introduces a panel data model with endogenous regressors and a latent group structure. Section \ref{sec:naive} demonstrates that the na\"ive approaches of minimizing the GMM objective function with respect to the coefficient vector and the group membership structure fail to provide unique estimates. Section \ref{sec:estimators} introduces various estimation procedures that provide consistent group membership estimators. Section \ref{sec:extensions} considers various extensions. The first involves models with individual fixed effects, and the second comprises dynamic panel data models. The third extension concerns models with time-varying group-specific coefficients. The results of the Monte Carlo simulations are discussed in Section \ref{sec:monte}. We illustrate the estimation procedures by an empirical example in Section \ref{sec:empirical}. Section \ref{sec:conclusion} concludes. The mathematical proofs are included in the Appendix.


\section{Model}\label{sec:model}

We consider a linear panel data model with possibly endogenous regressors. The coefficients are heterogeneous across observational units, and the heterogeneity exhibits a grouped pattern. IVs that are uncorrelated with the error term and relevant to the endogenous regressors are available.

Suppose that we observe panel data $(y_{it}, x_{it}', z_{it}')$, $i= 1, \dots, N$ and $t= 1, \dots, T$, where $y_{it}$ is a scalar dependent variable, $x_{it}$ is a vector of regressors, and $z_{it}$ is a vector of IVs. Elements of $x_{it}$ and $z_{it}$ may overlap. The index $i$ refers to an observational unit, and $t$ denotes a time period. Let $d$ and $m$ be the dimension of $x_{it}$ and $z_{it}$, respectively.

We consider the following linear model with group-specific coefficients:
\begin{align}
	y_{it} = x_{it}'\beta_i^0 + u_{it}, \label{eq:model}
\end{align}
where $\beta_i^0$ is the true unit-specific coefficient, which we explain in more detail below, and $u_{it}$ is an error term.
We suspect that $x_{it}$ may be endogenous, i.e., $x_{it}$ may be correlated with $u_{it}$. IVs, $z_{it}$, are exogenous in the sense that they are uncorrelated with the error term $u_{it}$ and are also relevant such that $z_{it}$ and $x_{it}$ are correlated.

The departure from the standard IV regression model is that we allow the coefficients to be heterogeneous across $i$. Specifically, we consider a grouped pattern of heterogeneity. Observational units are divided into $G$ groups. Let $\mathbb{G}\equiv\{1,\dots,G\}$ denote the set of groups. Units in the same group share the same value of the coefficient vector. Two units in different groups have different values of the coefficient. Unit $i$ belongs to $g_i^0 \in \mathbb{G}$. Group memberships $\{g_i^0 \}_{i=1}^N$ are unobservable and need to be inferred from data. We manipulate the notation, and a value of the coefficient for group $g$ is denoted as $\beta_g$. The true coefficient for $i$ is
\begin{align*}
	\beta_i^0 = \beta_{g_i^0}^0.
\end{align*}

A grouped pattern of heterogeneity is a useful modeling device. 
This is because allowing complete heterogeneity may not be attractive in practice because information about the coefficient of a given unit can then be obtained only from that specific unit. By grouping units, we can combine the information from multiple units in a group to obtain more precise coefficient estimates. Grouping methods also enhance the interpretability of the estimated coefficients because the dimension of $\{\beta_g\}_{g\in \mathbb{G}}$ is $dG$ and is not large when $G$ is small.

We are interested in estimating $\beta^0 \equiv \{ \beta_{g}^0 \}_{g\in \mathbb{G}} \in \mathbb{R}^{dG}$ and $\gamma^0 \equiv \{ g_i^0 \}_{i=1}^N \in \mathbb{G}^N$. In contrast to standard IV estimation, the group membership structure $\gamma^0$ is unknown, which makes the estimation nontrivial. In the next section, we argue that na\"ive applications of standard GMM estimation may not yield unique estimates. Various alternative estimation procedures are introduced and examined in Section \ref{sec:estimators}.

\section{Na\"ive approaches do not work} \label{sec:naive}

This section considers the estimation procedures that minimize the GMM objective function with respect to the coefficient vector and the group membership structure. We argue that such na\"ive approaches do not provide unique estimates. It is easy to gauge the problem when the dimension of the regressor vector and that of the IVs are equal. A conceptually similar yet technically more involved argument demonstrates that this approach fails to yield a unique minimizer of the objective function asymptotically, even when the dimension of the IVs is larger than that of the regressors.

\subsection{Na\"ive procedures}

Consider estimating $\beta^0$ and $\gamma^0$ by minimizing the GMM objective function. Let $\beta \in \mathbb{R}^{dG}$ and $\gamma = \{g_i \}_{i=1}^N \in \mathbb{G}^N$ denote a value of the coefficient parameter and a group membership assignment structure, respectively. A na\"ive GMM estimator is
\begin{align}
	(\hat \beta , \hat \gamma )	= \arg \min_{\beta, \gamma} \biggl(  \sum_{i=1}^N \sum_{t=1}^T z_{it} (y_{it} - x_{it}'\beta_{g_i} ) \biggr)' \hat W  \biggl( \sum_{i=1}^N \sum_{t=1}^T z_{it} (y_{it} - x_{it}'\beta_{g_i} ) \biggr), \label{eq:naive-gmm-obj1}
\end{align}
where $\hat W$ is an $m \times m$ positive definite matrix possibly depending on the data. Consider also estimating $\beta_g$ for each $g$ by minimizing the group-specific GMM objective functions:
\begin{align}
	(\hat \beta , \hat \gamma )	= \arg \min_{\beta, \gamma} \sum_{g\in \mathbb{G}} \biggl(  \sum_{g_i =g} \sum_{t=1}^T z_{it} (y_{it} - x_{it}'\beta_g ) \biggr)' \hat W_g  \biggl( \sum_{g_i=g} \sum_{t=1}^T z_{it} (y_{it} - x_{it}'\beta_g ) \biggr), \label{eq:naive-gmm-obj2}
\end{align}
where $\hat W_g$ is an $m \times m$ positive definite matrix which depends on $g$ and possibly on the data. When $\gamma^0$ is known, solving these minimization problems under $\gamma = \gamma^0$ provides a standard GMM estimator of $\beta^0$. Thus, its consistency and asymptotic normality can be shown under conditions established in the literature. The problem is that $\gamma$ is unknown and needs to be estimated together with $\beta$.

We demonstrate the failure of these estimators: that is, they are not consistent and may not even be uniquely determined.
A common motivation behind these approaches is an expectation (which turns out to be wrong) that
\begin{align}
(\hat{\beta},\hat{\gamma})\rightarrow_p &\arg\min_{\beta,\gamma} \plim_{N,T\rightarrow\infty} \biggl(  \sum_{i=1}^N \sum_{t=1}^T z_{it} (y_{it} - x_{it}'\beta_{g_i} ) \biggr)' \hat W  \biggl( \sum_{i=1}^N \sum_{t=1}^T z_{it} (y_{it} - x_{it}'\beta_{g_i} ) \biggr) \label{eq:gmm-id1}
\intertext{and}
(\beta^0,\gamma^0)=	&\arg\min_{\beta,\gamma} \plim_{N,T\rightarrow\infty}\biggl(\sum_{i=1}^N \sum_{t=1}^T z_{it} (y_{it} - x_{it}'\beta_{g_i} ) \biggr)' \hat W  \biggl( \sum_{i=1}^N \sum_{t=1}^T z_{it} (y_{it} - x_{it}'\beta_{g_i} ) \biggr) \label{eq:gmm-id2}
\end{align}
would hold. This is a typical argument for the asymptotic analysis of an extremum estimator: the minimizer of a sample objective function converges to the minimizer of the limit of the objective function, and the latter is the true value of the parameter. We now show that one of these two conditions might not be true. When the dimension of the regressors and that of the IVs are equal, $(\hat{\beta},\hat{\gamma})$ cannot be uniquely defined (Theorem \ref{th-undef}), which renders \eqref{eq:gmm-id1} unrealizable. When the dimension of the IVs is larger than that of the regressors, the probability limits of the objective functions may have zeros at parameter values other than $(\beta^0,\gamma^0)$ (Theorem \ref{th-unid-obj1} and \ref{th-unid-obj2}), which violates \eqref{eq:gmm-id2} because the true parameter value is not the unique minimizer.

\subsection{Just-identified cases}

We first consider the case of $m=d$. We refer to this as a ``just-identified'' case because, under a known group membership structure, solving \eqref{eq:naive-gmm-obj2} is the standard IV/GMM estimation under just-identification. However, when the group membership structure is unknown and needs to be estimated jointly with the coefficient vector, the na\"ive approaches fail and do not provide a unique solution to the minimization problem.

Here, we show that for any group membership structure $\gamma$, a value of $\beta$ exists that sets the objective functions to zero.
Take any $\gamma = \{g_i\}_{i=1}^N$ and consider
\begin{align}
	\hat \beta_g (\gamma) \equiv \biggl( \sum_{g_i=g} \sum_{t=1}^T z_{it} x_{it}' \biggr)^{-1} \sum_{g_i=g} \sum_{t=1}^T z_{it} y_{it}, \label{eq:beta-just-id}
\end{align}
for each $g\in \mathbb{G}$. This is the IV estimator based on the observations assigned to group $g$ under $\gamma$.
It is easy to see that
\begin{align*}
	\sum_{g_i=g} \sum_{t=1}^T z_{it} (y_{it} - x_{it}' \hat \beta_{g} (\gamma))  &=0
\intertext{for all $g$, which implies}
	\sum_{i=1}^N \sum_{t=1}^T z_{it} (y_{it} - x_{it}' \hat \beta_{g_i} (\gamma))  &=0.
\end{align*}

Both of the objective functions in \eqref{eq:naive-gmm-obj1} and \eqref{eq:naive-gmm-obj2} are zero under $\{ \hat \beta_g (\gamma) \}_{g\in \mathbb{G}}$ for any $\gamma$.
The value of $(\beta, \gamma)$ that minimizes the objective functions is not unique, and
the estimators cannot be determined uniquely. Consequently, the estimators \eqref{eq:naive-gmm-obj1} and \eqref{eq:naive-gmm-obj2} are not consistent. The discussion so far is summarized in the following theorem.
\begin{theorem} \label{th-undef}
	Suppose that $m=d$. Let $\hat \beta_g (\gamma)$ satisfy
	\begin{align*}
		 \sum_{g_i=g} \sum_{t=1}^T z_{it} x_{it}' \hat \beta_g (\gamma) = \sum_{g_i=g} \sum_{t=1}^T z_{it} y_{it}.
	\end{align*}
	Then, for any $\gamma \in \mathbb{G}^N$,
	\begin{align*}
		\biggl(  \sum_{i=1}^N \sum_{t=1}^T z_{it} (y_{it} - x_{it}' \hat \beta_{g_i} (\gamma ) \biggr)'
		&\hat W  \biggl( \sum_{i=1}^N \sum_{t=1}^T z_{it} (y_{it} - x_{it}' \hat\beta_{g_i} (\gamma ) \biggr) =0
		\intertext{and}
		\sum_{g\in \mathbb{G}} \biggl(  \sum_{g_i =g} \sum_{t=1}^T z_{it} (y_{it} - x_{it}'\hat \beta_g (\gamma ) )\biggr)'
		&\hat W_g  \biggl( \sum_{g_i=g} \sum_{t=1}^T z_{it} (y_{it} - x_{it}'\hat \beta_g (\gamma ) ) \biggr) =0
	\end{align*}
	hold. Note that if $  \sum_{g_i=g} \sum_{t=1}^T z_{it} x_{it}'$ is invertible, $\hat \beta_g (\gamma)$ is given in \eqref{eq:beta-just-id}.
\end{theorem}

\subsection{Over-identified cases}

This section considers the cases when $m>d$. We refer to these as ``over-identified'' cases because if the group membership structure were known in advance, solving \eqref{eq:naive-gmm-obj2} reduces to standard IV/GMM estimation under over-identification. Note that in the case of over-identification, no parameter value sets the objective functions to zero in finite samples in general.
We argue that the objective functions for the na\"ive approaches do not have a unique minimizer asymptotically.

For simplicity, we consider cases with $G=2$ only. Considering cases with $G>2$ makes the analysis more complicated but adds nothing conceptually. Set $\mathbb{G} = \{1, 2\}$.

We use double asymptotics: both $N$ and $T$ tend to infinity at the same time, that is, $N, T\rightarrow \infty$. This represents the following situation. Let $T\equiv T(N)$ be an increasing function of $N$ such that $T (N) \to \infty$ as $N \to \infty$. We understand that $\plim_{N, T\to \infty} $ gives the limit of a sequence indexed by $N$ and $T$ under $N\to \infty$ with any choice of $T (\cdot)$. If there are conditions on the relative magnitude of $N$ and $T$, then we consider a class of $T(\cdot)$ that is compatible with those conditions. 

We introduce several notations that are needed to represent the limit of the objective functions.
For $g\in\mathbb G$, let 
\begin{align*}
	\lambda_g^0 \equiv \lim_{N \to \infty} \frac{1}{N} \sum_{i=1}^N \mathbf{1} \{g_i^0 =g\}
\end{align*}
where $\mathbf{1}\{\cdot\}$ is the indicator function that returns one if the argument is true and zero otherwise.
Thus, $\lambda_1^0$ and $\lambda_2^0$ are the (asymptotic) fractions of units belonging to groups 1 and 2, respectively.
Note that $\lambda_1^0 + \lambda_2^0 =1$.
We also define the asymptotic fractions of units that are assigned to their true groups under a group membership structure $\gamma$.
Here, we manipulate the notation and let $\gamma \equiv \{ g_i \}_{i=1}^{\infty}\in\mathbb G^\mathbb N$, where $\mathbb{N}$ is the set of natural numbers. Note that, in finite samples, the first $N$ elements of $\gamma$ are relevant.
For $g\in\mathbb G$, let
\begin{align*}
	\lambda_{gg} (\gamma) \equiv \lim_{N \to \infty} \frac{1}{N} \sum_{i=1}^N\mathbf{1} \{g_i^0 =g\} \mathbf{1} \{g_i =g\}
\end{align*}
The (asymptotic) fraction of units that belong to group 1 and are assigned to group 1 under $\gamma$ is $\lambda_{11} (\gamma)$. Similarly, $\lambda_{22} (\gamma)$ is the fraction of units that belong to group 2 and are assigned to group 2. We note that $\lambda_{11} (\gamma^0) = \lambda_1^0$ and $\lambda_{22} (\gamma^0) = \lambda_2^0$.

Next, we define the matrices that represent the relationship between the IVs and the regressors for each group.
For $g\in\mathbb G$, let
\begin{align*}
	M_g \equiv \plim_{N, T\to \infty} \frac{1}{\sum_{i=1}^n \mathbf{1} \{g_i^0 =g\}} \sum_{i=1}^n \frac{1}{T} \sum_{t=1}^T \mathbf{1} \{g_i^0=g\} z_{it} x_{it}'
\end{align*}
Roughly speaking, $M_1$ and $M_2$ may be interpreted as the expected values of $z_{it}x_{it}'$ for groups 1 and 2, respectively.
In addition, for each $(g,\tilde g)\in\mathbb G^2$, let
\begin{align*}
	M_{g\tilde g}(\gamma) \equiv& \plim_{N, T\to \infty} \frac{1}{\sum_{i=1}^n \mathbf {1}\{g_i^0 =g\}\mathbf{1}\{g_i=\tilde g\}} \sum_{i=1}^n \frac{1}{T} \sum_{t=1}^T \mathbf{1} \{g_i^0=g\}\mathbf{1}\{g_i=\tilde g\}z_{it} x_{it}'.
\end{align*}
Similarly, we can understand $M_{g\tilde g}(\gamma)$ as the conditional expected values of $z_{it}x_{it}'$ for the units in group $g$ that are assigned to group $\tilde g$ under $\gamma$.

Lastly, we define the vectors that measure the exogeneity of the IVs for each assigned group. Specifically, for each $g\in\mathbb G$, let
\begin{align*}
	L_g(\gamma)\equiv \plim_{N,T\rightarrow\infty}\frac{1}{\sum_{i=1}^N\mathbf{1}\{g_i=g\}}\sum_{i=1}^N\frac{1}{T}\sum_{t=1}^T\mathbf{1}\{g_i=g\}z_{it}u_{it}.
\end{align*}

We now make the following sets of assumptions:
\begin{assumption}\label{assumption:basic}
\begin{enumerate}[label=\normalfont(\alph*)]
	\item The limits $\lambda_1^0$ and $\lambda_2^0$ are well-defined.
	\item The limits $M_1$ and $M_2$ are well-defined and finite.
\end{enumerate}
\end{assumption}
\begin{assumption}\label{assumption:basic-sample}
\begin{enumerate}[label=\normalfont(\alph*),ref=(\alph*)]
	\item \label{assumption:basic-sample:lambda} The limits $\lambda_{11} (\gamma)$ and $\lambda_{22}(\gamma)$ are well-defined.
	\item \label{assumption:basic-sample:M} For each $(g,\tilde g)\in\mathbb G^2$, the limit $M_{g\tilde g}(\gamma)$ is well-defined and finite. Also, for each $g\in\mathbb G$, $M_{g1}(\gamma)=M_{g2}(\gamma)$.
	\item For each $g\in\mathbb G$, the limit $L_g(\gamma)$ is well-defined and finite. 
\end{enumerate}
\end{assumption}
In Assumption \ref{assumption:basic-sample}, we restrict the set of possible group assignments $\gamma$. That $\lambda_{11} (\gamma)$ and $\lambda_{22}(\gamma)$ are well-defined is a weak condition. The condition $M_{g1}(\gamma)=M_{g2}(\gamma)$ for both $g=1$ and $g=2$ indicates that the group assignment is not correlated with the strength of the IVs within each group. The condition is satisfied when the first-stage relationship between $x_{it}$ and $z_{it}$ is homogeneous. We note that the first-stage relationship is allowed to be heterogeneous, but then we consider group assignments that are not correlated with the first-stage relationship. Importantly, there remain infinitely many group assignments even after Assumption \ref{assumption:basic-sample} is imposed.

The following theorem demonstrates that for any group membership assignment satisfying certain conditions, there exists a value of the coefficient vector that sets the limit of the objective function in \eqref{eq:naive-gmm-obj1} zero.
\begin{theorem}\label{th-unid-obj1}
	Suppose that Assumption \ref{assumption:basic} holds. Assume that $\plim_{N,T\to \infty} \hat W = W$, where $W$ is a symmetric positive definite matrix, and that $\plim_{N,T \to \infty}  (NT)^{-1}\sum_{i=1}^N\sum_{t=1}^T z_{it} u_{it}=0$. Then, if $(M_2-M_1)(\beta_2^0 - \beta_1^0)=0$,
	\begin{align}
		\plim_{N,T\to \infty} \biggl( \frac{1}{NT} \sum_{i=1}^N \sum_{t=1}^T z_{it} (y_{it} - x_{it}'\beta_{g_i} ) \biggr)' \hat W  \biggl( \frac{1}{NT} \sum_{i=1}^N \sum_{t=1}^T z_{it} (y_{it} - x_{it}'\beta_{g_i} ) \biggr) \label{eq:unid-obj1}
	\end{align}
	is zero at
	\begin{align*}
		(\beta_1(\gamma),\beta_2 (\gamma)) &\equiv
		\biggl(\frac{\lambda_{11} (\gamma ) \beta_1^0 + (\lambda_2^0 - \lambda_{22} (\gamma ) ) \beta_2^0}{\lambda_{11} (\gamma ) + \lambda_2^0 - \lambda_{22} (\gamma)},\frac{\lambda_{22} (\gamma ) \beta_2^0 + (\lambda_1^0 - \lambda_{11} (\gamma ) ) \beta_1^0}{\lambda_{22} (\gamma ) + \lambda_1^0 - \lambda_{11} (\gamma)}\biggr)
	\end{align*}
	for any $\gamma\in\mathbb G^{\mathbb N}$ satisfying Assumptions \ref{assumption:basic-sample}\ref{assumption:basic-sample:lambda}--\ref{assumption:basic-sample:M}. 
\end{theorem}

A similar result holds for the objective function in \eqref{eq:naive-gmm-obj2}, as shown in the following theorem.
\begin{theorem}\label{th-unid-obj2}
	Suppose that Assumption \ref{assumption:basic} holds. Assume that for each $g \in \mathbb{G}$, $\plim_{N,T\to \infty} \hat W_g = W_g$, where $W_g$ is a symmetric positive definite matrix. Then, if $(M_2-M_1)(\beta_2^0 - \beta_1^0)=0$,
	\begin{align*}
		\plim_{N,T\to \infty} \sum_{g\in \mathbb{G}} \biggl( \frac{1}{NT} \sum_{g_i =g} \sum_{t=1}^T z_{it} (y_{it} - x_{it}'\beta_g ) \biggr)' \hat W_g  \biggl( \frac{1}{NT} \sum_{g_i=g} \sum_{t=1}^T z_{it} (y_{it} - x_{it}'\beta_g ) \biggr)
	\end{align*}
	is zero at
	\begin{align*}
		(\beta_1 (\gamma), \beta_2 (\gamma)) \equiv \biggl(\frac{\lambda_{11} (\gamma ) \beta_1^0 + (\lambda_2^0 - \lambda_{22} (\gamma) ) \beta_2^0}{\lambda_{11} (\gamma ) + \lambda_2^0 - \lambda_{22} (\gamma)},
		\frac{\lambda_{22} (\gamma ) \beta_2^0 + (\lambda_1^0 - \lambda_{11} (\gamma) ) \beta_1^0}{\lambda_{22} (\gamma ) + \lambda_1^0 - \lambda_{11} (\gamma)}\biggr)
	\end{align*}
	for any $\gamma\in\mathbb G^{\mathbb N}$ that satisfies Assumption \ref{assumption:basic-sample} and $L_1(\gamma)=L_2(\gamma)=0$.
\end{theorem}

These theorems indicate that there are infinitely many values of the parameters that can set the limits of the objective functions to zero. Note that the true value also satisfies this property, so the moment condition is correct. However, the moment condition does not provide sufficient information to identify the parameters. Accordingly, the estimators are not expected to be consistent.

The corresponding proofs are included in the Appendix. These proofs broadly follow the argument in \textcite{HallHanBoldea2012}. Their article discusses a failure of break detection with GMM in time series contexts. While the situations are different, we can utilize their mathematical insights in the present context.

We impose exogeneity conditions. In Theorem \ref{th-unid-obj1}, the IVs and the error term are assumed to be uncorrelated at least asymptotically. Theorem \ref{th-unid-obj2} imposes a different condition that $L_1(\gamma)=L_2(\gamma) =0$. These conditions are satisfied, for example, when the exogeneity holds for each unit: $\plim_{T \to \infty}  T^{-1}\sum_{t=1}^T z_{it} u_{it}=0$ uniformly over $i$.

The key condition is $(M_2-M_1)(\beta_2^0 - \beta_1^0)=0$, namely that $\beta_2^0 - \beta_1^0$ is in the null space of $M_2-M_1$. In practice, it is difficult, if not impossible, to know whether $(M_2-M_1)(\beta_2^0 - \beta_1^0)=0$ in advance because the group structure is latent and the estimation of $M_2$ and $M_1$ requires knowledge of the true group membership structure. As $\beta_2^0$ and $\beta_1^0$ are two of the parameters that we want to estimate, we should always take into account the possibility that this failure may occur in any application.

Indeed, it is not difficult to imagine situations in which $(M_2-M_1)(\beta_2^0 - \beta_1^0)=0$ holds.\footnote{Obviously, $(M_2-M_1)(\beta_2^0 - \beta_1^0)=0$ occurs when $\beta_2^0=\beta_1^0$. However, in such a case, there is no group heterogeneity in the coefficients. It is natural that we cannot uniquely identify a group membership structure.}
For example, if $M_2=M_1$, then $(M_2-M_1)(\beta_2^0 - \beta_1^0)=0$ holds regardless of the value of $\beta^0$. This case implies that the first stage does not have a group structure, i.e., when we can postulate the first stage $x_{it}' = z_{it}'\Pi + v_{it}$ with a common coefficient matrix $\Pi$ across $i$, where $v_{it}$ is an error term that is uncorrelated with $z_{it}$.
Even when $M_2 \neq M_1$, $(M_2-M_1)(\beta_2^0 - \beta_1^0)=0$ may occur.
For example, consider a case in which there are two regressors and three IVs and the DGP satisfies the following:
\begin{align*}
	M_2 = \begin{pmatrix}
		1 & 0 \\
		1 & 0 \\
		0 & 1
	\end{pmatrix}, \quad
	M_1 = \begin{pmatrix}
		0 & 1 \\
		0 & 1 \\
		1 & 0 
	\end{pmatrix}, \quad
	\beta_2 = \begin{pmatrix}
		1  \\
		1  	
	\end{pmatrix}, \text{ and }
	\beta_1 = \begin{pmatrix}
		0  \\
		0  
	\end{pmatrix}.
\end{align*}
In Group 2, both regressors have coefficients equal to one, while in Group 1, they do not affect the dependent variable. In Group 2, the first and second IVs are related to the first regressor but not the second, and the third IV is related to the second regressor, but not the first. The relationship is alternated in Group 1. In this case, $(M_2-M_1)(\beta_2^0 - \beta_1^0)=0$ holds. 

\begin{remark}[Computational problem]
	Even if we were perfectly certain that $(M_2-M_1) (\beta_2^0 - \beta_1^0) \neq 0$, both estimation approaches discussed above encounter a computational problem. Indeed, it is not clear whether an efficient way to solve the minimization problems exists. For example, consider the following Kmeans type algorithm:
	\begin{enumerate}
		\item Set $\gamma^{(0)}$ and $s=1$. 
		\item Update the coefficient: For all $g\in\mathbb{G}$,
		\begin{align*}
			\hat \beta_g^{(s)}	= \arg \min_{\beta_g}\biggl(  \sum_{g_i^{(s-1)} =g} \sum_{t=1}^T z_{it} (y_{it} - x_{it}'\beta_g ) \biggr)' \hat W_g \biggl( \sum_{g_i^{(s-1)}=g} \sum_{t=1}^T z_{it} (y_{it} - x_{it}'\beta_g ) \biggr).
		\end{align*}
		\item Update the group membership estimate:
		\begin{align*}
			\gamma^{(s)} =  \arg \min_{\gamma} \sum_{g\in \mathbb{G}} \biggl(  \sum_{g_i =g} \sum_{t=1}^T z_{it} (y_{it} - x_{it}'\hat \beta_g^{(s)} ) \biggr)' \hat W_g \biggl( \sum_{g_i=g} \sum_{t=1}^T z_{it} (y_{it} - x_{it}'\hat \beta_g^{(s)} ) \biggr).
		\end{align*}
		\item $s=s+1$.
		\item Repeat 2--4 until convergence.
	\end{enumerate}
	This procedure has a problem at Step 3 because it is not clear how to solve the minimization problem quickly. The only exception is the case of just-identification, where, as observed above, the algorithm would converge in the first step. But the minimizer then depends on the initial group membership assignment and fails to be a consistent estimator.
\end{remark}
\begin{remark}
	\textcite[][footnote 3]{SuShiPhillips2016} report an observation related to our findings. They note that they fail to establish an asymptotic theory for the fully pooled criterion, which corresponds to the objective function in \eqref{eq:naive-gmm-obj1} in our case. However, they do not define it as an identification failure, nor do they provide any conditions under which it occurs. Another related observation is given by \textcite[][page 5]{Mehrabani2022} who states that ``...because group membership of individual units is unknown, we cannot apply the usual GMM objective function here,'' but they do not explain the source of the problem.
\end{remark}

\section{Alternative estimation strategies}\label{sec:estimators}
In this section, we propose alternative approaches that produce consistent estimators. They are two-step procedures. We first fit the endogenous regressors using IVs. Then, we apply the Kmeans (i.e., GFE) estimator to the structural model \eqref{eq:model} whose endogenous variables are substituted with the corresponding fitted values from the first-stage regression.

We consider three different approaches to model the first-stage regression:
(i) The first is to assume that the first-stage coefficient is homogeneous across units:
 \begin{align*}
	x_{it} = \Pi^{0\prime} z_{it} + v_{it},
 \end{align*}
where $\Pi^0$ is a $m\times d$ coefficient matrix, and $v_{it}$ is a first-stage error term.
(ii) The second is to assume that the first-stage coefficients exhibit a grouped pattern of heterogeneity $\kappa^0\equiv \{ k_i^0 \}_{i=1}^N\in\mathbb K^N=\{1,\dots,K\}^N$:
\begin{align}
	x_{it} = \Pi_{k_i^0}^{0\prime}z_{it} + v_{it}. \label{eq-gfs:model}
\end{align}
We allow $\gamma^0$ and $\kappa^0$ (also $G$ and $K$) to differ. $\Pi_{k}^{0}$ is a $m\times d$ coefficient matrix for group $k$. $K=1$ corresponds to the homogeneous first stage mentioned above.
(iii) Lastly, we consider the model that allows full heterogeneity across units, that is,
\begin{align}
	x_{it} = \Pi_i^{0\prime}z_{it} + v_{it},	 \label{eq-ufs:model}
\end{align}
where $\Pi_i^{0}$ is the $d\times m$ unit-specific first-stage coefficient parameter matrix.
The relative merit of this modeling is robustness to misspecification, though it may cause some efficiency loss because of the large number of parameters in the first stage.

We develop estimation procedures tailored to each of the first-stage models we postulate. In the following subsections, we elaborate on the estimation procedures and show that they yield consistent estimators. The proofs for the theorems are in the Appendix.
They broadly follow the argument given in \textcite{BonhommeManresa2015,BonhommeManresa2015Supp}.

\subsection{Homogeneous first stage}

We first consider the model in which the first-stage coefficient is homogeneous. In this case, the two-stage estimation runs the first-stage regression of $x_{it}$ and $z_{it}$ and then applies the GFE in the second stage using the fitted values from the first stage:
\begin{enumerate}[label=(\arabic*)]
	\item Regress $x_{it}$ on $z_{it}$ and obtain
	\begin{align*}
		\hat\Pi \equiv \biggl(\frac{1}{NT}\sum_{i=1}^N\sum_{t=1}^Tz_{it}z_{it}'\biggr)^{-1}\frac{1}{NT}\sum_{i=1}^N\sum_{t=1}^Tz_{it}x_{it}'.
	\end{align*}
	Construct $\hat x_{it} \equiv \hat \Pi' z_{it}$.
	\item Apply the GFE estimation to the structural model \eqref{eq:model} whose endogenous variables $x_{it}$ are substituted with the corresponding fitted values $\hat x_{it}$:
	\begin{align*}
		\min_{\beta\in \mathcal B^G, \gamma\in\mathbb G^N}\ \frac{1}{NT}	\sum_{i=1}^N\sum_{t=1}^T(y_{it}-\hat x_{it}'\beta_{g_i})^2.
	\end{align*}
	The minimizer $(\hat\beta,\hat\gamma)\equiv( \{ \hat\beta_g \}_{g=1}^G, \{\hat g_i \}_{i=1}^N)$ of this problem is the final estimate.
\end{enumerate}
Because this approach is a special case of that based on \eqref{eq-gfs:model}, the asymptotic properties are examined in more general settings, and we do not mention them specifically for this approach.
Abusing terminology, we specifically refer to this procedure as ``2SLS'' estimation.

However, this approach may encounter a weak instrument problem \parencite[see e.g., ][]{AndrewsStockSun2019} in the presence of heterogeneity, even when IVs are relevant for each unit. The first-stage coefficient is a weighted average of heterogeneous coefficients and may not be sufficiently away from zero even when the coefficient for each unit is large. For example, if half of the units have positive first-stage coefficients and the other half has negative coefficients, then the first-stage coefficient in the homogeneous regression may be close to zero. We thus consider approaches that explicitly account for possible heterogeneity in the first stage. 

\subsection{First stage with group structure}

Next, we consider the first-stage model \eqref{eq-gfs:model} where the coefficients exhibit a group structure.
The two-step estimation procedure proceeds in the following way, which, as a whole, we call the ``two-stage GFE'' (TGFE) estimation:
\begin{enumerate}[label=(\arabic*)]
	\item Apply the GFE estimation to first-stage model \eqref{eq-gfs:model}:
	\begin{align*}
		\min_{\Pi\in\mathbf \Pi^K,\kappa\in\mathbb K^N}\ \frac{1}{NT}\sum_{i=1}^N\sum_{t=1}^T\|x_{it}-\Pi_{k_i}'z_{it}\|^2,
	\end{align*}
	where $\mathbf \Pi\subseteq \mathbb R^{md}$ denotes the common parameter space of the group-specific first-stage coefficient matrices $\Pi_k$'s.
	Let $(\hat\Pi^\tgfe,\hat\kappa^\tgfe)\equiv (\{\hat\Pi_k^\tgfe\}_{k=1}^K,\{\hat k_i^\tgfe\}_{i=1}^N)$ be the minimizer of this problem.
	Construct the fitted values $\hat x_{it}^\tgfe\equiv \hat \Pi_{\hat k_i^\tgfe}^{\tgfe\prime}z_{it}$.
	\item Apply the GFE estimation to the structural model \eqref{eq:model} whose endogenous variables $x_{it}$ are substituted with the fitted values $\hat x_{it}^\tgfe$:
	\begin{align*}
		\min_{\beta\in \mathcal B^G, \gamma\in\mathbb G^N}\ \frac{1}{NT}	\sum_{i=1}^N\sum_{t=1}^T(y_{it}-\hat x_{it}^{\tgfe\prime}\beta_{g_i})^2.
	\end{align*}
	The minimizer $(\hat\beta^\tgfe,\hat\gamma^\tgfe)\equiv (\{\hat\beta_g^\tgfe \}_{g=1}^G, \{\hat g_i^\tgfe \}_{i=1}^N)$ of this problem is the final estimate.
\end{enumerate}
Separate GFE estimations are conducted in the first and second steps. By doing so, the procedure allows the group heterogeneity in the first-stage regression to be different from that in the coefficients of interest.

We examine the asymptotic properties of the TGFE estimator. We introduce a set of assumptions.
The superscript ``fs'' that appears in our notation indicates that it is associated with the first stage. Similarly, the superscript ``ss'' denotes that it is related to the second stage. 
We make assumptions similar to those in \textcite{BonhommeManresa2015,BonhommeManresa2015Supp}.
Below, $\|\cdot\|$ denotes the Frobenius norm, and $\|\cdot\|_2$ the spectral norm.
\begin{assumption} \label{ass-gfs:beta-cnst} %
For some constant $M>0$,
\begin{enumerate}[label=(\alph*),ref=(\alph*)]
	\item \label{ass-gfs:cpt-param} Let $\mathbf \Pi$ and $\mathcal B$ be compact sets in $\mathbb R^{m\times d}$ and $\mathbb R^d$ respectively. For all $k\in\mathbb K$, $\Pi_k^0\in\mathbf \Pi$, and for all $g\in\mathbb G$, $\beta_g^0\in\mathcal B$.
	\item \label{ass-gfs:wkdep-zv} $\Exp[(NT)^{-1}\sum_{i=1}^N\|\sum_{t=1}^Tz_{it}v_{it}'\|^2]\leq M$.
	\item \label{ass-gfs:wkdep-zu} $\Exp[(NT)^{-1}\sum_{i=1}^N\|\sum_{t=1}^Tz_{it}u_{it}'\|^2]\leq M$.
	\item \label{ass-gfs:finmt-z} $\Exp[(NT)^{-1}\sum_{i=1}^N\|\sum_{t=1}^Tz_{it}z_{it}'\|_2] \leq M$.
	\item \label{ass-gfs:relev-ss} There exists $\underline \rho^\tss$, which depends on data, such that $\min_{\gamma\in\mathbb G^N}\max_{\tilde g\in\mathbb G}\rho^\tss(\gamma,g,\tilde g)\geq \underline \rho^\tss $, for all $g\in\mathbb G$, where $\rho^\tss(\gamma,g,\tilde g)$ is the smallest eigenvalue of
	\begin{align*}
		M^\tss(\gamma,g,\tilde g)
	 	\equiv &\frac{1}{N}\sum_{i=1}^N\mathbf 1\{g_i^0=g\}\mathbf 1 \{g_i=\tilde g\}\frac{1}{T}\sum_{t=1}^T(\Pi_{k_i^0}^{0\prime}z_{it})(\Pi_{k_i^0}^{0\prime}z_{it})'
	\end{align*}
	and $\underline \rho^\tss\rightarrow \rho^{\tss*}>0$ as $N$ and $T$ tend to infinity.
	\item \label{ass-gfs:grpsep-ss} There exists $c^\tss>0$ such that 
	\begin{align*}
		\plim_{N,T\rightarrow \infty}\frac{1}{N}\sum_{i=1}^N\min_{g\neq g}D_{i,g,\tilde g}^\tss >c^\tss,
	\end{align*}
	where $D_{i,g,\tilde g}^\tss \equiv T^{-1}\sum_{t=1}^T((z_{it}'\Pi_{k_i^0}^0)(\beta_g^0-\beta_{\tilde g}^0))^2$.
\end{enumerate}	
\end{assumption}
The compactness of parameter spaces in Assumption \ref{ass-gfs:beta-cnst}\ref{ass-gfs:cpt-param} is standard in econometrics.
Assumptions \ref{ass-gfs:beta-cnst}\ref{ass-gfs:wkdep-zv}--\ref{ass-gfs:wkdep-zu} impose weak dependence of $z_{it}v_{it}'$ and $z_{it}u_{it}$, respectively.
Note also that Assumption \ref{ass-gfs:beta-cnst}\ref{ass-gfs:wkdep-zu} corresponds to the exogeneity of IVs.
To see why, suppose that $(z_{it}u_{it})_{t=1}^T$ is i.i.d. across $i$ and $T^{-1}\sum_{t=1}^T\Exp[z_{it}u_{it}]=a_T$ for some constant $a_T\in\mathbb R^m$.
Then, because $M/T\geq (NT^2)^{-1}\sum_{i=1}^N\sum_{t=1}^T\sum_{s=1}^T \mathbb E[z_{it}z_{is}'u_{it}u_{is}] = \|a_T\|^2$, $a_T$ should shrink to zero as $T$ tends to infinity.
Assumption \ref{ass-gfs:beta-cnst}\ref{ass-gfs:finmt-z} requires that the second-order moment of $z_{it}$ does not explode as $N$ and $T$ tend to infinity.
Assumption \ref{ass-gfs:beta-cnst}\ref{ass-gfs:relev-ss} excludes multicollinearity among the population fitted values $\Pi_{k_i^0}^{0\prime}z_{it}$.
Assumption \ref{ass-gfs:beta-cnst}\ref{ass-gfs:grpsep-ss} is called the ``group separation condition'' in the literature.
It requires that $\beta_g^0$ and $\beta_{\tilde g}^0$ have sufficiently different values.
Note that the IVs need to be relevant for Assumption \ref{ass-gfs:beta-cnst}\ref{ass-gfs:grpsep-ss} to hold. It is violated if $\Pi_k^0=0$.

Under these assumptions, we establish the consistency of $\hat\beta^\tgfe$. We show that its Hausdorff distance from the true parameter $\beta^0$, which is defined as:
\begin{align*}
	d_H(\beta,\beta^0)
	\equiv \max \bigg\{\max_{g\in\mathbb G}\min_{\tilde g\in\mathbb G}\|\beta_g - \beta_{\tilde g}^0\|,\
	\max_{\tilde g\in\mathbb G}\min_{g\in\mathbb G} \|\beta_g - \beta_{\tilde g}^0\|\bigg\},
\end{align*}
converges to zero in probability as $N$ and $T$ diverge.
\begin{theorem}\label{thm-gfs:beta-cnst}
Suppose that Assumption \ref{ass-gfs:beta-cnst} holds for the two-stage model comprising equations \eqref{eq:model} and \eqref{eq-gfs:model}. Then, as $N$ and $T$ tend to infinity,
\begin{align*}
	d_H(\hat\beta^\tgfe,\beta^0)\rightarrow_p 0.
\end{align*}
\end{theorem}

Next, we proceed to the consistency of the estimated group memberships. The notion of consistency here refers to the fact that the probability of exact classification converges to one as $N$ and $T$ tend to infinity. Recall that the first and second stages exhibit different grouped patterns of heterogeneity. Correct classification in the first stage leads to correct classification in the second. We thus first state conditions for the consistency of the group membership estimator in the first stage.
We impose the following set of assumptions. They are used to show the consistency of $\hat\Pi^\tgfe$, which is a prerequisite of the consistency of $\hat \kappa$.
\begin{assumption} \label{ass-gfs:pi-cnst} 
\begin{enumerate}[label=(\alph*),ref=(\alph*)]
	\item \label{ass-gfs:relev-fs} There exists $\underline \rho^\tfs$, which depends on data, such that for all $k\in\mathbb K$, $\min_{\gamma\in\mathbb G^N}\max_{\tilde k\in\mathbb K}\rho^\tfs(\kappa,k,\tilde k)\geq \underline \rho^\tfs$, where $\rho^\tfs(\gamma,k,\tilde k)$ is the smallest eigenvalue of
	\begin{align*}
		M^\tfs(\gamma,g,\tilde g)\equiv \frac{1}{N}\sum_{i=1}^N\mathbf 1\{k_i^0=k\}\mathbf 1 \{k_i=\tilde k\}\frac{1}{T}\sum_{t=1}^Tz_{it}z_{it}'
	\end{align*}
	and $\underline \rho^\tfs\rightarrow \rho^{\tfs*}>0$ as $N$ and $T$ tend to infinity.
	\item \label{ass-gfs:grpsep-fs} There exists $c^\tfs>0$ such that 
	\begin{align*}
		\plim_{N,T\rightarrow \infty}\frac{1}{N}\sum_{i=1}^N\min_{k\neq \tilde k}D_{i,k,\tilde k}^\tfs >c^\tfs 	
	\end{align*}
	where $D_{i,k,\tilde k}^\tfs \equiv T^{-1}\sum_{t=1}^T(z_{it}'(\Pi_{k}^0-\Pi_{\tilde k}^0))^2$.
\end{enumerate}	
\end{assumption}
Assumption \ref{ass-gfs:pi-cnst}\ref{ass-gfs:relev-fs} excludes the multicollinearity between IVs and requires that each group have sufficiently many members in the first stage.
Assumption \ref{ass-gfs:pi-cnst}\ref{ass-gfs:grpsep-fs} is a first-stage version of the group separation condition.
In addition, impose the following set of assumptions for the consistency of $\hat\kappa^\tgfe$.
\begin{assumption} \label{ass-gfs:kappa-cnst}
\begin{enumerate}[label=(\alph*),ref=(\alph*)]
	\item $\lim_{T\rightarrow\infty}\min_{i\in\{1,\dots,N\}}\Exp[\min_{k\neq \tilde k}D_{i,k,\tilde k}^\tfs]>c^\tfs$.
	\item \label{ass-gfs:meantail-z} There exists a constant $M_z>0$ such that as $N$ and $T$ tend to infinity, for all $\delta>0$,
	\begin{align*}
		\max_{i\in\{1,\dots,N\}}\Pr\Biggl(\biggl\|\frac{1}{T}\sum_{t=1}^Tz_{it}z_{it}'\biggr\|_2>M_z\Biggr)	 &= o(T^{-\delta})
	\end{align*}
	\item \label{ass-gfs:meantail-zv} For every positive constant $c$, as $N$ and $T$ tend to infinity, for all $\delta>0$,
	\begin{align*}
		\max_{i\in\{1,\dots,N\}} \Pr\Biggl(\Biggl\|\frac{1}{T}\sum_{t=1}^T z_{it}v_{it}'\Biggr\|>c\Biggr) &= o(T^{-\delta})
	\end{align*}
	\item \label{ass-gfs:mix-fs} There exist positive constants $a^\tfs$ and $d^\tfs$, and a sequence $\alpha^\tfs[t]\leq \exp(-a^\tfs t^{d^\tfs})$ such that, for all $i=1,\dots,N$ and $(k,\tilde k)\in\mathbb K^2$ such that $k\neq \tilde k$,
		$\{(\Pi_k^0 - \Pi_{\tilde k}^0)'z_{it}\}_t$ is a strongly mixing process with mixing coefficients $\alpha^\tfs[t]$. 
	\item \label{ass-gfs:tail-fs} There exist positive constants $b_z^\tfs$ and $d_z^\tfs$ such that, for all $c>0$, $\Pr(\|(\Pi_k^0 - \Pi_{\tilde k}^0)'z_{it}\|>c)\leq \exp(1-(c/b_z^\tfs)^{d_z^\tfs})$.
\end{enumerate}	
\end{assumption}
Assumptions \ref{ass-gfs:kappa-cnst}\ref{ass-gfs:meantail-z}--\ref{ass-gfs:meantail-zv} are satisfied, for example, if $(z_{it},v_{it})$ is Gaussian and independent over time. They indicate that the tails of the distributions of the averages of $||z_{it}||^2$ and $z_{it} v_{it}'$ vanish faster than any polynomial rate. For example, if they vanish at an exponential rate, the assumptions are satisfied.
Assumption \ref{ass-gfs:kappa-cnst}\ref{ass-gfs:mix-fs} restricts the serial dependence of $(\Pi_k^0-\Pi_{\tilde k}^0)'z_{it}$.
Assumption \ref{ass-gfs:kappa-cnst}\ref{ass-gfs:tail-fs} is related to its tail property, and requires that the probability mass on the tail decays at an exponential rate. We note that Assumptions \ref{ass-gfs:kappa-cnst}\ref{ass-gfs:meantail-z}--\ref{ass-gfs:meantail-zv} can be satisfied under conditions on the serial dependence and the tail of the distribution of $z_{it}$ and $z_{it} v_{it}'$ similar to Assumptions \ref{ass-gfs:kappa-cnst}\ref{ass-gfs:mix-fs} and Assumption \ref{ass-gfs:kappa-cnst}\ref{ass-gfs:tail-fs}.
With these sets of assumptions, $\hat\kappa^\tgfe$ is shown to be consistent. The details are in the Appendix.

We now discuss the consistency of $\hat\gamma^\tgfe$.
The following set of assumptions is imposed.
\begin{assumption} \label{ass-gfs:gamma-cnst}
\begin{enumerate}[label=(\alph*),ref=(\alph*)]
	\item $\lim_{T\rightarrow\infty}\min_{i\in\{1,\dots,N\}}\Exp[\min_{g\neq \tilde g}D_{i,g,\tilde g}^\tss]>c^\tss$.
	\item \label{ass-gfs:meantail-zu} For every positive constant $c$, as $N$ and $T$ tend to infinity, for all $\delta>0$,
	\begin{align*}
		\max_{i\in\{1,\dots,N\}} \Pr\Biggl(\Biggl\|\frac{1}{T}\sum_{t=1}^T z_{it}u_{it}\Biggr\|>c\Biggr) = o(T^{-\delta})
	\end{align*}
	\item \label{ass-gfs:mix-ss} There exist positive constants $a^\tss$ and $d^\tss$, and a sequence $\alpha^\tss[t]\leq \exp(-a^\tss t^{d^\tss})$ such that for all $i=1,\dots,N$ and $(g,\tilde g)\in\mathbb G^2$ such that $g\neq \tilde g$, $\{ z_{it}'\Pi_{k_i^0}^0(\beta_g^0-\beta_{\tilde g}^0)\}_t$ is a strong mixing process with mixing coefficients $\alpha^\tss[t]$. 
	\item \label{ass-gfs:tail-ss} There exist positive constants $b_x^\tss$ and $d_x^\tss$ such that, for all $c>0$, $\Pr(|z_{it}'\Pi_{k_i^0}^0(\beta_g^0-\beta_{\tilde g}^0)|>c)\leq \exp(1-(c/b_x^\tss)^{d_x^\tss})$.
\end{enumerate}	
\end{assumption}
Discussions analogous to those for Assumption \ref{ass-gfs:kappa-cnst} apply here.
The following theorem states that the group membership estimators for each are uniformly consistent.
\begin{theorem}\label{thm-gfs:gamma-cnst}
Suppose that Assumptions \ref{ass-gfs:beta-cnst}--\ref{ass-gfs:gamma-cnst} hold for the two-stage model comprising equations \eqref{eq:model} and \eqref{eq-gfs:model}. Then, as $N$ and $T$ tend to infinity, for all $\delta>0$,
\begin{align*}
	\Pr\biggl(\max_{i\in\{1,\dots,N\}}|\hat g_i^\tgfe - g_i^0|>0\biggr)=o(1) + o(NT^{-\delta}).
\end{align*}
\end{theorem}
Note that the required condition on the relative size of $N$ and $T$ is $NT^{-\delta} \to 0$, which is weak because $\delta$ can be as large as a researcher wants. It is violated if, for example, $T=\log N$, but is satisfied when $T$ is of geometric order $N$.

\subsection{Unit-specific first stage}

Finally, we consider the first-stage model \eqref{eq-ufs:model} that allows full heterogeneity across units.
We describe the following two-step estimation procedure, which we hereinafter call the ``unit-specific first-stage GFE'' (UGFE) estimation:
\begin{enumerate}[label=(\arabic*)]
	\item Apply least squares estimation to the first-stage model \eqref{eq-ufs:model}:
	\begin{align*}
		\min_{(\Pi_i)_{i=1}^N\in\mathbf \Pi^N}\frac{1}{NT}\sum_{i=1}^N\sum_{t=1}^T\|x_{it}-\Pi_i'z_{it}\|^2,
	\end{align*}
	where $\mathbf \Pi$ denotes the common parameter space of the unit-specific first-stage coefficient matrices $\Pi_i$'s.
	Let $\{\hat\Pi_i^\ugfe\}_{i=1}^N$ be the minimizer of this problem.
	Construct the fitted values $\hat x_{it}^\ugfe \equiv \hat\Pi_i^{\ugfe\prime}z_{it}$.
	Note that because $ \hat\Pi_i^\ugfe = \arg \min_{\Pi_i\in\mathbf \Pi}T^{-1}\sum_{t=1}^T\|x_{it}-\Pi_i'z_{it}\|^2$, 
	a closed-form solution is available (when $\mathbf \Pi$ is sufficiently large):
	\begin{align*}
		\hat\Pi_i^\ugfe
		= \Biggl(\frac{1}{T}\sum_{t=1}^Tz_{it}z_{it}'\Biggr)^{-1} \frac{1}{T}\sum_{t=1}^Tz_{it}x_{it}'.
	\end{align*}
	\item Apply GFE estimation to the structural model \eqref{eq:model} whose endogenous variables $x_{it}$ are substituted with the corresponding fitted values $\hat x_{it}^\ugfe$: 
	\begin{align*}
		\min_{\beta\in\mathcal B^G,\gamma\in\mathbb G^N}\frac{1}{NT}\sum_{i=1}^N\sum_{t=1}^T(y_{it}-\hat x_{it}^{\ugfe\prime}\beta_{g_i})^2
	\end{align*}
	The minimizer $(\hat\beta^\ugfe,\hat\gamma^\ugfe) \equiv (\{ \hat\beta_g^\ugfe \}_{g=1}^G,\{\hat g_i^\ugfe \}_{i=1}^N)$ of this problem is our final estimate.
\end{enumerate}

The derivation of the asymptotic properties of the UGFE estimator is like that of the TGFE estimator.
Indeed, the following two assumptions are very similar to Assumptions \ref{ass-gfs:beta-cnst} and \ref{ass-gfs:kappa-cnst}--\ref{ass-gfs:gamma-cnst}, respectively, except that $\Pi_k^0$ and $\Pi_{k_i^0}^0$ are now replaced by $\Pi_i^0$.

We first demonstrate the consistency of $\hat\beta^\ugfe$ under the following set of assumptions.
\begin{assumption}\label{ass-ufs:beta-cnst}
\begin{enumerate}[label=(\alph*),ref=(\alph*)]
	\item \label{ass-ufs:cpt-param} Let $\mathbf \Pi$ be a compact set in $\mathbb R^{m\times d}$. For all $i\in\{1,\dots,N\}$, $\Pi_i^0\in\mathbf \Pi$.
	\item \label{ass-ufs:relev-ss} There exists $\underline \rho^\tss$, which depends on data, such that $\min_{\gamma\in\mathbb G^N}\max_{\tilde g\in\mathbb G}\rho^\tss(\gamma,g,\tilde g)\geq \underline \rho^\tss $, for all $g\in\mathbb G$, where $\rho^\tss(\gamma,g,\tilde g)$ is the smallest eigenvalue of
	\begin{align*}
		M^\tss(\gamma,g,\tilde g)
	 	\equiv &\frac{1}{N}\sum_{i=1}^N\mathbf 1\{g_i^0=g\}\mathbf 1 \{g_i=\tilde g\}\frac{1}{T}\sum_{t=1}^T(\Pi_i^{0\prime}z_{it})(\Pi_i^{0\prime}z_{it})'
	\end{align*}
	and $\underline \rho^\tss\rightarrow \rho^{\tss*}>0$ as $N$ and $T$ tend to infinity.
	\item \label{ass-ufs:grpsep-ss} There exists $c^\tss>0$ such that 
	\begin{align*}
		\plim_{N,T\rightarrow \infty}\frac{1}{N}\sum_{i=1}^N\min_{g\neq \tilde g}D_{i,g,\tilde g}^\tss >c^\tss
	\end{align*}
	where $D_{i,g,\tilde g}^\tss \equiv T^{-1}\sum_{t=1}^T(z_{it}'\Pi_i^0(\beta_g^0-\beta_{\tilde g}^0))^2$.
\end{enumerate}	
\end{assumption}
\begin{theorem}\label{thm-ufs:beta-cnst}
Suppose that Assumptions \ref{ass-gfs:beta-cnst}\ref{ass-gfs:cpt-param}--\ref{ass-gfs:finmt-z} and \ref{ass-ufs:beta-cnst} hold for the two-stage model comprising equations \eqref{eq:model} and \eqref{eq-ufs:model}. Then, as $N$ and $T$ tend to infinity,
\begin{align*}
	d_H(\hat\beta^\ugfe,\beta^0)\rightarrow_p 0.
\end{align*}
\end{theorem}

Next comes the consistency of $\hat\gamma^\ugfe$ under the following additional set of assumptions.
\begin{assumption}\label{ass-ufs:gamma-cnst}
\begin{enumerate}[label=(\alph*),ref=\ref{ass:ugfe-2}(\alph*)]
	\item $\lim_{T\rightarrow\infty}\min_{i\in\{1,\dots,N\}}\Exp[\min_{g\neq \tilde g}D_{i,g,\tilde g}^\tss]>c^\tss$.
	\item \label{ass-ufs:mix-ss} There exist positive constants $a^\tss$ and $d^\tss$, and a sequence $\alpha^\tss[t]\leq \exp(-a^\tss t^{d^\tss})$ such that for all $i=1,\dots,N$ and $(g,\tilde g)\in\mathbb G^2$ such that $g\neq \tilde g$, $\{z_{it}'\Pi_i^0(\beta_g^0-\beta_{\tilde g}^0\}_t$ is a strong mixing process with mixing coefficients $\alpha^\tss[t]$. 
	\item \label{ass-ufs:tail-ss} There exist positive constants $b_x^\tss$ and $d_x^\tss$ such that, for all $c>0$, $\Pr(|z_{it}'\Pi_i^0(\beta_g^0-\beta_{\tilde g}^0)|>c)\leq \exp(1-(c/b_x^\tss)^{d_x^\tss})$.
\end{enumerate}	
\end{assumption}
\begin{theorem}\label{thm-ufs:gamma-cnst}
Suppose that Assumptions \ref{ass-gfs:beta-cnst}\ref{ass-gfs:cpt-param}--\ref{ass-gfs:finmt-z}, \ref{ass-gfs:kappa-cnst}\ref{ass-gfs:meantail-z}--\ref{ass-gfs:meantail-zv}, \ref{ass-gfs:gamma-cnst}\ref{ass-gfs:meantail-zu}, and \ref{ass-ufs:beta-cnst}--\ref{ass-ufs:gamma-cnst} hold for the two-stage model comprising equations \eqref{eq:model} and \eqref{eq-ufs:model}. Then, as $N$ and $T$ tend to infinity, for all $\delta>0$,
\begin{align*}
	\Pr\biggl(\max_{i\in\{1,\dots,N\}}|\hat g_i^\ugfe - g_i^0|>0\biggr)=o(1) + o(NT^{-\delta}).
\end{align*}
\end{theorem}

\begin{remark}
	The estimator can be rewritten as
	\begin{align*}
		(\hat\beta^\ugfe,\hat\gamma^\ugfe) = \arg \min_{\beta\in\mathcal B^G, \gamma\in\mathbb G^N} \sum_{i=1}^N\biggl(\sum_{t=1}^T z_{it}(y_{it} - x_{it}'\beta_{g_i})\biggr)\biggl(\sum_{t=1}^T z_{it}z_{it}'\biggr)^{-1}\biggl(\sum_{t=1}^T z_{it}(y_{it} - x_{it}'\beta_{g_i})\biggr),
	\end{align*}
	which implies that we can run the UGFE estimation in a single stage. Note that this objective function is the unpenalized version of the PGMM objective function in \textcite{SuShiPhillips2016} and \textcite{Mehrabani2022} with the weighting matrix for unit $i$ set to $(\sum_{t=1}^T z_{it}z_{it}')^{-1}$. \textcite{itosugasawa2023} consider a similar objective function for generalized estimating equation models.
\end{remark}

\subsection{Choice of number of groups} \label{sec:select-num-grps}
The number of latent groups may not be known a priori, and methods to choose it are required.
We propose data-driven procedures that can select the correct number of groups in the limit.
Our selection procedures follow \textcite{PesaranSmith1994}. We use the residuals from the second-stage estimation. Note that the second-stage residuals are different from the residuals of the structural equation.

We start with the TGFE estimation. Both the first and second stages require choosing the number of groups, and the two stages require different procedures.
The information criteria for each stage are 
\begin{align*}
	\IC^{\tgfe,\tfs}(K) 
	\equiv& \frac{1}{NT}\sum_{i=1}^N\sum_{t=1}^T\|x_{it}-\hat\Pi_{\hat k_i^{\tgfe\K}}^{\tgfe\K\prime}z_{it}\|^2 + \frac{c^{\tgfe,\tfs}(K,N,T)}{NT} \text{ and} \\
	\IC^{\tgfe,\tss}(G,K) \equiv& \frac{1}{NT}\sum_{i=1}^N\sum_{t=1}^T(y_{it}-(z_{it}'\hat\Pi_{\hat k_i^{\tgfe\K}}^{\tgfe\K})\hat\beta_{\hat g_i^{\tgfe\GK}}^{\tgfe\GK})^2
	+ \frac{c^{\tgfe,\tss}(G,N,T)}{NT},
\end{align*}
where the superscripts $\K$, $\G$, and $\GK$ denote that the given estimate is obtained assuming that the numbers of groups in the first and second stages are $K$ and $G$ respectively. $c^{\tgfe,\tfs}$ and $c^{\tgfe,\tss}$ are functions having the stated arguments.
Following \textcite{Mallows1973}, the literature suggests using $c^{\tgfe,\tfs}(K,N,T)=\hat{\sigma^{\tgfe,\tfs}}^2(mdK+N)\log(NT)$ and $c^{\tgfe,\tss}(K,N,T)=\hat{\sigma^{\tgfe,\tss}}^2(dG+N)\log(NT)$, where $\hat{\sigma^{\tgfe,\tfs}}^2\equiv (NT)^{-1}\sum_{i=1}^N\sum_{t=1}^T\|x_{it}-\hat\Pi_{\hat k_i^{\tgfe\Kmax}}^{\tgfe\Kmax\prime}z_{it}\|^2$ and $\smash{\hat{\sigma^{\tgfe,\tss}}^2\equiv (NT)^{-1}\sum_{i=1}^N\sum_{t=1}^T(y_{it}-(z_{it}'\hat\Pi_{\hat k_i^{\tgfe\Kmax}}^{\tgfe\Kmax})\hat\beta_{\hat g_i^{\tgfe\GmaxKmax}}^{\tgfe\GmaxKmax})^2}$.
The information criterion for the first stage is standard. The first term is the sum of squared residuals, and the second is a penalty term. The information criterion for the second stage relies on the residuals from the second-stage GFE estimation, which are different from the residuals from the structural equation: $y_{it}-x_{it}\hat\beta_{\hat g_i^{\tgfe\GmaxKmax}}^{\tgfe\GmaxKmax}$. Let
\begin{align*}
	\hat K^\tgfe \equiv \arg \min_{K \in \{1,\dots,K^{\max}\}}\IC^{\tgfe,\tfs}(K) \text{ and } 
	\hat G^\tgfe(K) \equiv \arg \min_{G\in\{1,\dots,G^{\max}\}}\IC^{\tgfe,\tss}(G,K)
\end{align*}
be the corresponding minimizers. The estimated number of groups is then $\hat G^\tgfe \equiv \hat G^\tgfe(\hat K^\tgfe)$.

Next, we consider the UGFE estimation. Define:
\begin{align*}
	\IC^{\ugfe}(G) \equiv \frac{1}{NT}\sum_{i=1}^N\sum_{t=1}^T(y_{it}-(z_{it}'\hat\Pi_i^\ugfe)\hat\beta_{\hat g_i^{\ugfe\G}}^{\ugfe\G})^2
	+ \frac{c^{\ugfe,\tss}(G,N,T)}{NT}.
\end{align*}
The estimated number of groups is the minimizer $\hat G^\ugfe\equiv \arg \min_{G\in\{1,\dots,G^{\max}\}}\IC^{\ugfe}(G)$ of this criterion function.
As before, we can employ $c^{\ugfe,\tss}(G,N,T)=\hat{\sigma^{\ugfe,\tss}}^2(dG+N)\log(NT)$, where $\hat{\sigma^{\ugfe,\tss}}^2\equiv (NT)^{-1}\sum_{i=1}^N\sum_{t=1}^T(y_{it}-(z_{it}'\hat\Pi_i^\ugfe)\hat\beta_{\hat g_i^{\ugfe\Gmax}}^{\ugfe\Gmax})^2$.

\subsection{Other possible procedures}

We introduce two additional estimation strategies. They are simple and easy to implement. However, they only yield consistent estimators under possibly strong additional assumptions.

\paragraph{Ignoring endogeneity.} The first strategy is to apply GFE estimation to the structural model \eqref{eq:model} ignoring the presence of endogeneity between $x_{it}$ and $u_{it}$. This approach directly solves:
\begin{align*}
	\min_{\beta\in\mathcal B^G,\gamma\in\mathbb G^N}\frac{1}{NT}\sum_{i=1}^N\sum_{t=1}^T(y_{it} - x_{it}'\beta_{g_i})^2.
\end{align*}
Let $(\hat\beta^\ig,\hat\gamma^\ig)$ be the minimizer of this problem.

Even if $\hat\beta^\ig$ suffers from endogeneity bias, $\hat\gamma^\ig$ is still consistent if the coefficient of the linear projection of $y_{it}$ on $x_{it}$ for each $i$---$\plim_{T\to \infty} (\sum_{t=1}^T x_{it} x_{it}')^{-1} \sum_{t=1}^T x_{it} y_{it}$---is also grouped by $\gamma^0$.
This condition is satisfied, for example, when $(x_{it},u_{it})$ are i.i.d. across $i$ and $t$.
Then, the coefficient of the linear projection is $\beta_{g_i^0}^0 + (\mathbb E[x_{it}x_{it}'])^{-1} \mathbb E[x_{it}u_{it}]$, which exhibits the same grouped pattern as $\beta_{g_i^0}^0$.
Once the condition is assumed, we can rely on the existing results in the literature for exogenous cases such as those in \textcite{BonhommeManresa2015}.
That is, we can invoke their assumptions to show the consistency of $\hat\gamma^\ig$.

However, justifying the required condition in any given empirical situation would be difficult. The relevant assumptions should be made concerning the model of linear projection, not to the original model. It is unclear whether an economic theory can support the conditions on linear projection. There are cases in which the relevant condition is violated.
For example, if the degree of endogeneity, namely $\mathbb E[x_{it}u_{it}]$, varies with a distinct grouped pattern from $\beta_{g_i^0}^0$, $\hat\gamma^\ig$ may not be consistent.

This approach is considered in \textcite[Section S4]{BonhommeManresa2015Supp} in the framework of dynamic linear models.
Furthermore, a conceptually similar strategy has been employed for break detection in the time series literature. \textcite{PerronYamamoto2015} and \textcite{QianSu2014}, among others, consider the detection of break points in models with endogenous regressors.
The authors first estimate the break point while ignoring endogeneity and then, given that break point, estimate regime-specific model parameters.

\paragraph{Reduced form.} The second strategy examines the reduced form, namely, the relationship between the dependent variable and the IVs. This solves:
\begin{align*}
	\min_{\xi\in\Xi^G,\lambda\in\mathbb H^N}\frac{1}{NT}\sum_{i=1}^N\sum_{t=1}^T(y_{it} - z_{it}'\xi_{l_i})^2,
\end{align*}
where $\xi\equiv \{\xi_g \}_{g=1}^G\subseteq \Xi^G$ and $\lambda\equiv \{l_i\}_{i=1}^N\in \mathbb G^N$.
Let $(\hat\xi^\rf,\hat\lambda^\rf)$ be the minimizer of this problem. Coefficient estimates are obtained by IV estimation for each group given by $\hat\lambda^\rf$.

If the reduced form and the structural model share the same group structure, $\hat\lambda^\rf$ is consistent for $\gamma^0$.
To examine the situations in which consistency holds, consider the first-stage model \eqref{eq-gfs:model}. The reduced form is:
\begin{align*}
	y_{it} = z_{it}'(\Pi_{k_i^0}^0\beta_{g_i^0}^0) + (v_{it}'\beta_{g_i^0}^0 + u_{it}).
\end{align*}
When $\Pi_{k_i^0}^0\beta_{g_i^0}^0$ exhibit the same grouped pattern as $\beta_{g_i^0}^0$, the consistency of $\hat\lambda^\rf$ is guaranteed by \textcite[Section S4.2]{BonhommeManresa2015Supp}.
This condition is satisfied if, for example, the first-stage model has no group structure, that is, $\Pi_k^0=\Pi^0$.
However, when the first-stage model and the structural model exhibit different group structures, the reduced form may have a different group structure, and the condition is violated.

\section{Extensions}\label{sec:extensions}
This section discusses two extensions of our baseline model.
We begin with models with unit-specific fixed effects.
In this case, we simply need to consider the within or first-differencing transformed models and apply the estimation procedures discussed above to transformed models. 
As an important example, we discuss dynamic linear panel data models.
Then, we consider models with time-varying group-specific fixed effects. This extension is important because previous results concerning IV estimation did not handle this case, yet it is practically relevant. 
We establish a new theory to accommodate the situation that the dimension of the coefficient vector increases as $T$ increases. 
As in the previous section, we separate our theoretical results based on the model we assume for the first-stage regression.

\subsection{Unit-specific fixed effects}\label{sec:extensions:ufe}
The first extension adds unit-specific fixed effects to our baseline model. 
Specifically, we consider the model
\begin{align}
	y_{it} = x_{it}'\beta_{g_i^0}^0 + \eta_i + u_{it},
\end{align}
where $\{\eta_i\}_{i=1}^N$ are unit-specific fixed effects and are allowed to be correlated with the IVs, $z_{it}$.
The estimation starts with eliminating the unit-specific fixed effects using the within or first-differencing transformation, each of which transforms the model into
\begin{align*}
	\ddot y_{it} &= \ddot x_{it}' \beta_{g_i^0}^0 + \ddot u_{it}
\intertext{or}
	\Delta y_{it} &= \Delta x_{it}' \beta_{g_i^0}^0 + \Delta u_{it},
\end{align*}
where, for a random variable or vector $w_{it}$, $\ddot w_{it} \equiv w_{it} - \sum_{t=1}^T w_{it}/T$ and $\Delta w_{it} \equiv w_{it} - w_{it-1}$.
The coefficient vectors $\{\beta_g^0\}_{g=1}^G$ and the group membership structure $\{g_i^0\}_{i=1}^N$ are then estimated by applying a method discussed in Section \ref{sec:estimators} to the transformed model.

Note that we need to impose the relevant assumptions on the transformed variables instead of the original ones in level. 
For example, in order to use the TGFE estimator, Assumptions \ref{ass-gfs:beta-cnst}--\ref{ass-gfs:gamma-cnst} should be applied for $(\ddot x_{it},\ddot u_{it})$ or $(\Delta x_{it},\Delta u_{it})$, not for $(x_{it},u_{it})$.
In particular, the exogeneity of IVs, $z_{it}$, needs to hold with respect to transformed error terms ($\ddot u_{it}$ or $\Delta u_{it}$).
This requirement implies that the choice of the transformation depends on the IVs, $z_{it}$, or alternatively, that the choice of instruments depends on a transformation. 
The consistency of the TGFE estimator is then secured by Theorems \ref{thm-gfs:beta-cnst}--\ref{thm-gfs:gamma-cnst}.

\paragraph{Dynamic linear panel data models.}
As an illustration, we examine a panel autoregressive model with lag order 1 (panel AR(1) model):
\begin{align}
	y_{it} = \beta_{g_i^0}^0 y_{i,t-1} + \eta_i + u_{it}, \label{eq-par1}
\end{align}
where $\{\beta_g^0\}_{g=1}^G$ are the group-specific AR(1) coefficients.
The arguments that follow can be readily generalized for models with more lags.

Applying our procedures to this model encounters an additional challenge. The number of IVs differs across time periods. The first-differencing transformation transforms model \eqref{eq-par1} into
\begin{align*}
	\Delta y_{it} = \beta_{g_i^0}^0\Delta y_{it-1} + \Delta u_{it}.
\end{align*}
Because $\Delta y_{it-1}$ and $\Delta u_{it}$ are systematically correlated, the literature \parencites{AndersonCheng1981,AndersonCheng1982,ArellanoBond91,HoltzEakinNeweyRosen88} suggests using various combination of lags $y_{it-2},\dots,y_{i0}, \Delta y_{it-2},\dots,\Delta y_{i1}$ as IVs for $\Delta y_{it-1}$.
No issue arises if one uses the same number of lags for each period. 
The estimation methods in Section \ref{sec:estimators} will suffice.
However, if we intend to vary the number for efficiency gain, the direct application of our methods becomes infeasible, since they are based on the fixed dimension of the IVs.
Nevertheless, a simple solution is available.

The solution we propose is to combine IVs linearly.
Let $z_{it}$ be an $m_t$ dimensional IV vector for $\Delta y_{it-1}$.
We suggest using linearly transformed $\Gamma_t'z_{it}$, where $\Gamma_t$ is an $m_t \times d$ matrix that possibly depends on data, as new IVs.
Because their dimension is fixed at $d$, our methods are then executable.
A candidate for $\Gamma_t$ is
\begin{align*}
	\biggl(\frac{1}{N}\sum_{i=1}^N z_{it} z_{it}'\biggr)^{-1}\frac{1}{N}\sum_{i=1}^N z_{it}\Delta y_{it-1},
\end{align*}
which is obtained by running the cross-sectional regression of $\Delta y_{it-1}$ on $z_{it}$ at period $t$.\footnote{\textcite{Wooldridge05,Wooldridge10} shows that, in the absence of a latent group structure, that is, when $G=1$, the estimation using instruments $(\sum_{i=1}^N z_{it}\Delta y_{it-1})'(\sum_{i=1}^N z_{it} z_{it}')^{-1}z_{it}$ is equivalent to GMM estimation using all the available moment restrictions, that is, $\mathbb E(z_{i1}\Delta u_{i1})=0, \dots,\mathbb E(z_{iT}\Delta u_{iT})=0$.}

\subsection{Time-varying group-specific fixed effects}
The second extension adds time-varying group-specific effects to our baseline model:
\begin{align}
	y_{it} = x_{it}'\beta_{g_i^0}^0 + \alpha_{g_i^0 t}^0 +  u_{it}. \label{eq-gfe:model}
\end{align}
where $\{\alpha_{gt}^0\}_{t=1}^T$ represents the time profile shared by the units in group $g$.
The main motivation of including $\alpha_{g_i^0 t}^0$ is to avoid omitted variable bias caused by time-varying unobserved confounders. \textcite{BonhommeManresa2015} and \textcite{mugnier2022simple} consider models with time-varying group-specific effects. Furthermore, we allow the coefficients on $x_{it}$ to exhibit a grouped pattern of heterogeneity.

The first-stage models \eqref{eq-gfs:model} and \eqref{eq-ufs:model} are adjusted to
\begin{align}
	x_{it}=& \Pi_{k_i^0}^{0\prime}z_{it} + \mu_{k_i^0 t}^0 +v_{it} \label{eq-gfe-gfs:model}
	\intertext{and}
	x_{it}=& \Pi_i^{0\prime}z_{it} + \mu_{k_i^0 t}^0 + v_{it} \label{eq-gfe-ufs:model}
\end{align}
respectively. Note that, in model \eqref{eq-gfe-ufs:model}, time-varying effects are kept group-specific. 
If they were unit-specific, then the first-stage coefficients $\{\Pi_i^0\}_{i=1}^N$ would not be identified.

\subsubsection{First stage with group structure}
We start by considering the first-stage model \eqref{eq-gfe-gfs:model}.
We define the following two-step procedure, which we also call the ``two-stage GFE'' (TGFE) estimation:
\begin{enumerate}[label=(\arabic*)]
	\item Apply the GFE estimation to the first-stage model \eqref{eq-gfe-gfs:model}:
	\begin{align*}
		\min_{\Pi\in\mathbf \Pi^K,\mu\in\mathcal M^{KT},\kappa\in\mathbb K^N}\ \frac{1}{NT}\sum_{i=1}^N\sum_{t=1}^T\|x_{it}-\Pi_{k_i}'z_{it}-\mu_{k_i t}\|^2,
	\end{align*}
	where $\mu\equiv (\mu_k)_{k=1}^K$, $\mu_k\equiv (\mu_{kt})_{t=1}^T$, and $\mathcal M$ denotes the common parameter space of $\mu_{kt}$'s.
	Let $(\hat\Pi^\tgfe,\hat\mu^\tgfe,\hat\kappa^\tgfe)\equiv (\{\hat\Pi_k^\tgfe\}_{k=1}^K,\{\hat\mu_k^\tgfe\}_{k=1}^K,\{\hat k_i^\tgfe\}_{i=1}^N)$ be the minimizer of this problem.
	Construct the fitted values $\hat x_{it}^\tgfe\equiv \hat \Pi_{\hat k_i^\tgfe}^{\tgfe\prime}z_{it} + \hat\mu_{\hat k_i^\tgfe}^\tgfe$.
	\item Apply the GFE estimation to the structural model \eqref{eq-gfe-gfs:model} whose endogenous variables $x_{it}$ are substituted with the fitted values $\hat x_{it}^\tgfe$:
	\begin{align*}
		\min_{\beta\in \mathcal B^G,\alpha\in\mathcal A^{GT},\gamma\in\mathbb G^N}\ \frac{1}{NT}	\sum_{i=1}^N\sum_{t=1}^T(y_{it}-\hat x_{it}^{\tgfe\prime}\beta_{g_i}-\alpha_{g_i t})^2,
	\end{align*}
	where $\alpha\equiv \{(\alpha_{gt})_{t=1}^T\}_{g=1}^G$ and $\mathcal A$ denotes the common parameter space of $\alpha_{gt}$'s.
	The minimizer $(\hat\beta^\tgfe,\hat\alpha^\tgfe,\hat\gamma^\tgfe)\equiv (\{\hat\beta_g^\tgfe \}_{g=1}^G, \{(\hat\alpha_{gt})_{t=1}^T\}_{g=1}^G,\{\hat g_i^\tgfe \}_{i=1}^N)$ of this problem becomes the final estimate.
\end{enumerate}

Next, we study its asymptotic properties.
We introduce the following additional set of assumptions to Assumption \ref{ass-gfs:beta-cnst}.
\begin{assumption} \label{ass-gfe-gfs:beta-cnst}
For some constant $M>0$,
\begin{enumerate}[label=(\alph*),ref=(\alph*)]
	\item \label{ass-gfe-gfs:cpt-param} Let $\mathcal M$ and  $\mathcal A$ be compact sets in $\mathbb R^d$ and $\mathbb R$, respectively. For all $k\in\mathbb K$, $\mu_{kt}^0\in\mathcal M$, and for all $g\in\mathbb G$, $\alpha_{gt}^0\in\mathcal A$, 
	\item \label{ass-gfe-gfs:wkcdep-u} $(NT)^{-1}\sum_{i=1}^N\sum_{j=1}^N\sum_{t=1}^T|\Exp[u_{it}u_{jt}]|\leq M$.
	\item \label{ass-gfe-gfs:wkdep-u} $|(N^2T)^{-1}\sum_{i=1}^N\sum_{j=1}^N\sum_{t=1}^T\sum_{s=1}^T\cov[u_{it}u_{jt},u_{is}u_{js}]|\leq M$
	\item \label{ass-gfe-gfs:wkcdep-v} $(NT)^{-1}\sum_{i=1}^N\sum_{j=1}^N\sum_{t=1}^T|\Exp[v_{it}'v_{jt}]|\leq M$.
	\item \label{ass-gfe-gfs:wkdep-v} $|(N^2T)^{-1}\sum_{i=1}^N\sum_{j=1}^N\sum_{t=1}^T\sum_{s=1}^T\cov[v_{it}'v_{jt},v_{is}'v_{js}]|\leq M$
	\item \label{ass-gfe-gfs:relev-ss} There exists $\underline \rho^\tss$, which depends on data, such that $\min_{\gamma\in\mathbb G^N}\max_{\tilde g\in\mathbb G}\rho^\tss(\gamma,g,\tilde g)\geq \underline \rho^\tss $, for all $g\in\mathbb G$, where $\rho^\tss(\gamma,g,\tilde g)$ is the smallest eigenvalue of
	\begin{align*}
		&M^\tss(\gamma,g,\tilde g)\\
	 	\equiv& \frac{1}{N}\sum_{i=1}^N\mathbf 1\{g_i^0=g\}\mathbf 1 \{g_i=\tilde g\}\\
	 	 \times &
		\begin{pmatrix}
			\frac{1}{T}\sum_{t=1}^T(\Pi_{k_i^0}^{0\prime}z_{it} + \mu_{k_i^0t}^0)(\Pi_{k_i^0}^{0\prime}z_{it} + \mu_{k_i^0t}^0)' & \frac{1}{\sqrt{T}}(\Pi_{k_i^0}^{0\prime}z_{i1} + \mu_{k_i^01}^0) & \cdots & \frac{1}{\sqrt{T}}(\Pi_{k_i^0}^{0\prime}z_{iT} + \mu_{k_i^0T}^0)\\
			\frac{1}{\sqrt{T}}(\Pi_{k_i^0}^{0\prime}z_{i1} + \mu_{k_i^01}^0)' & 1 & \cdots & 0 \\
			\vdots & 0 & \ddots & 0 \\
			\frac{1}{\sqrt{T}}(\Pi_{k_i^0}^{0\prime}z_{iT} + \mu_{k_i^0T}^0)' & 0 & \cdots & 1 \\
		\end{pmatrix}
	\end{align*}
	and $\underline \rho^\tss\rightarrow \rho^{\tss*}>0$ as $N$ and $T$ tend to infinity.
	\item \label{ass-gfe-gfs:grpsep-ss} There exists $c^\tss>0$ such that
	\begin{align*}
		\plim_{N,T\rightarrow \infty}\frac{1}{N}\sum_{i=1}^N\min_{g\neq \tilde g}D_{i,g,\tilde g}^\tss >c^\tss	
	\end{align*}
	where $D_{i,g,\tilde g}^\tss \equiv T^{-1}\sum_{t=1}^T((z_{it}'\Pi_{k_i^0}^0+\mu_{k_i^0 t}^0)(\beta_g^0-\beta_{\tilde g}^0) + (\alpha_{g t}^0 - \alpha_{\tilde g t}^0))^2$.
\end{enumerate}	
\end{assumption}
In Assumption \ref{ass-gfe-gfs:beta-cnst}\ref{ass-gfe-gfs:cpt-param}, we require the support of the group-specific time fixed effects to be compact.
Assumptions \ref{ass-gfe-gfs:beta-cnst}\ref{ass-gfe-gfs:wkcdep-u} and \ref{ass-gfe-gfs:beta-cnst}\ref{ass-gfe-gfs:wkcdep-v} restrict the degree of cross-sectional dependence of the first- and second-stage errors, 
and Assumptions \ref{ass-gfe-gfs:beta-cnst}\ref{ass-gfe-gfs:wkdep-u} and \ref{ass-gfe-gfs:beta-cnst}\ref{ass-gfe-gfs:wkdep-v} bear on their weak dependence along the time dimension.

With the aid of these assumptions, we establish the consistency of $(\hat\beta^\tgfe,\hat\alpha^\tgfe)$. Define:
\begin{align*}
	d_H(\alpha,\alpha^0) \equiv \max\biggl\{\max_{g\in\mathbb G}\min_{\tilde g\in\mathbb G}\frac{1}{T}\sum_{t=1}^T\|\alpha_{gt}-\alpha_{\tilde gt}^0\|^2,\max_{\tilde g\in\mathbb G}\min_{g\in\mathbb G}\frac{1}{T}\sum_{t=1}^T\|\alpha_{gt}-\alpha_{\tilde gt}^0\|^2\biggr\}.
\end{align*}

\begin{theorem} \label{thm-gfe-gfs:beta-cnst}
Suppose that Assumptions \ref{ass-gfs:beta-cnst}\ref{ass-gfs:cpt-param}--\ref{ass-gfs:finmt-z} and \ref{ass-gfe-gfs:beta-cnst} hold for the two-stage model comprising equations \eqref{eq-gfe:model} and \eqref{eq-gfe-gfs:model}. Then, as $N$ and $T$ tend to infinity,
\begin{align*}
	d_H(\hat\beta^\tgfe,\beta^0) \rightarrow_p 0
	\text{ and }
	d_H(\hat\alpha^\tgfe,\alpha^0) \rightarrow_p 0.
\end{align*}
\end{theorem}

Our next result concerns the uniform consistency of $\hat\gamma^\tgfe$, for which the following assumptions are employed.
\begin{assumption}\label{ass-gfe-gfs:pi-cnst} 
\begin{enumerate}[label=(\alph*),ref=(\alph*)]
	\item \label{ass-gfe-gfs:relev-fs} There exists $\underline \rho^\tfs$, which depends on data, such that for all $k\in\mathbb K$, $\min_{\gamma\in\mathbb G^N}\max_{\tilde k\in\mathbb K}\rho^\tfs(\kappa,k,\tilde k)\geq \underline \rho^\tfs$, where $\rho^\tfs(\gamma,k,\tilde k)$ is the smallest eigenvalue of
	\begin{align*}
		M^\tfs(\kappa,k,\tilde k)
	 	\equiv &\frac{1}{N}\sum_{i=1}^N\mathbf 1\{k_i^0=k\}\mathbf 1 \{k_i=\tilde k\}
		\begin{pmatrix}
			\frac{1}{T}\sum_{t=1}^Tz_{it}z_{it}' & \frac{1}{\sqrt{T}}z_{i1} & \cdots & \frac{1}{\sqrt{T}}z_{iT}\\
			\frac{1}{\sqrt{T}}z_{i1}' & 1 & \cdots & 0 \\
			\vdots & 0 & \ddots & 0 \\
			\frac{1}{\sqrt{T}}z_{iT}' & 0 & \cdots & 1 \\
		\end{pmatrix}
	\end{align*}
	and $\underline \rho^\tfs\rightarrow \rho^{\tfs*}>0$ as $N$ and $T$ tend to infinity.
	\item \label{ass-gfe-gfs:grpsep-fs} There exists $c^\tfs>0$ such that 
	\begin{align*}
		\plim_{N,T\rightarrow \infty}\frac{1}{N}\sum_{i=1}^N\min_{k\neq \tilde k}D_{i,k,\tilde k}^\tfs >c^\tfs	
	\end{align*}
	where $D_{i,k,\tilde k}^\tfs \equiv T^{-1}\sum_{t=1}^T\|(\Pi_{k}^0-\Pi_{\tilde k}^0)'z_{it} + (\mu_{k t}^0 - \mu_{\tilde k t}^0)\|^2$.
\end{enumerate}	
\end{assumption}

\begin{assumption}\label{ass-gfe-gfs:kappa-cnst} 
\begin{enumerate}[label=(\alph*),ref=(\alph*)]
	\item \label{ass-gfe-gfs:grpsep-fs-uni} $\lim_{T\rightarrow\infty}\min_{i\in\{1,\dots,N\}}\Exp[\min_{k\neq \tilde k}D_{i,k,\tilde k}^\tfs]>c^\tfs$. 
	\item \label{ass-gfe-gfs:mix-fs} There exist positive constants $a^\tfs$ and $d^\tfs$, and a sequence $\alpha^\tfs[t]\leq \exp(-a^\tfs t^{d^\tfs})$ such that, for all $i\in\{1,\dots,N\}$ and  $(k,\tilde k)\in\mathbb K^2$ such that $k\neq \tilde k$, $\{v_{it}\}_t$, $\{\mu_{k_i^0t}^0v_{it}'\}_t$, $\{v_{it}'(\mu_{k t}^0 - \mu_{\tilde  kt}^0)\}_t$, and $\{(\Pi_k^0 - \Pi_{\tilde k}^0)'z_{it} + (\mu_{k t}^0 - \mu_{\tilde k t}^0)\}_t$ are strongly mixing processes with mixing coefficients $\alpha^\tfs[t]$. Moreover, $\Exp[\mu_{k_i^0t}^0v_{it}']=0$ and $\Exp[v_{it}'(\mu_{k t}^0 - \mu_{\tilde  kt}^0)]=0$. 
	\item \label{ass-gfe-gfs:tail-fs} There exist positive constants $b_v^\tfs$ and  $b_z^\tfs$, and $d_v^\tfs$ and $d_{z}^\tfs$ such that, for all $c>0$,
		$\Pr(|v_{it}|>c)\leq \exp(1-(c/b_v^\tfs)^{d_v^\tfs})$
		and $\Pr(\|(\Pi_k^0 - \Pi_{\tilde k}^0)'z_{it} + (\mu_{k t}^0 - \mu_{\tilde k t}^0)\|>c)\leq \exp(1-(c/b_z^\tfs)^{d_{z}^\tfs})$.
\end{enumerate}	
\end{assumption}

\begin{assumption}\label{ass-gfe-gfs:gamma-cnst} 
\begin{enumerate}[label=(\alph*),ref=(\alph*)]
	\item $\lim_{T\rightarrow\infty}\min_{i\in\{1,\dots,N\}}\Exp[\min_{k\neq \tilde k}D_{i,k,\tilde k}^\tss]>c^\tss$.
	\item \label{ass-gfe-gfs:mix-ss} There exist positive constants $a^\tss$ and $d^\tss$, and a sequence $\alpha^\tss[t]\leq \exp(-a^\tss t^{d^\tss})$ such that, for all $i\in\{1,\dots,N\}$ and $(g,\tilde g)\in\mathbb G^2$ such that $g\neq \tilde g$, 
		$\{u_{it}\}_t$, $\{\mu_{k_i^0t}^0u_{it}\}_t$,
		$\{v_{it}(\alpha_{g t}^0 - \alpha_{\tilde g t}^0)\}_t$,
		$\{u_{it}(\alpha_{g t}^0 - \alpha_{\tilde g t}^0)\}_t$, and
		$\{(z_{it}'\Pi_{k_i^0 t}^0+\mu_{k_i^0 t}^0)(\beta_g^0-\beta_{\tilde g}^0) + (\alpha_{g t}^0 - \alpha_{\tilde g t}^0)\}_t$ are strongly mixing with mixing coefficients $\alpha^\tss[t]$. 
		Moreover, $\Exp[\mu_{k_i^0t}^0u_{it}]=0$, $\Exp[v_{it}(\alpha_{g t}^0 - \alpha_{\tilde g t}^0)]=0$, and $\Exp[u_{it}(\alpha_{g t}^0 - \alpha_{\tilde g t}^0)]=0$.
	\item \label{ass-gfe-gfs:tail-ss} There exist positive constants $b_{u}^\tss$ and $b_x^\tss$, and $d_{u}^\tss$ and  $d_x^\tss$ such that, for all $c>0$,
		$\Pr(|u_{it}|>c)\leq \exp(1-(c/b_u^\tss)^{d_u^\tss})$ and
		$\Pr(|(z_{it}'\Pi_{k_i^0}^0 + \mu_{k_i^0 t}^0)(\beta_g^0-\beta_{\tilde g}^0) + (\alpha_{g t}^0 - \alpha_{\tilde g t}^0)|>c)\leq \exp(1-(c/b_x^\tss)^{d_x^\tss})$.
\end{enumerate}	
\end{assumption}

We obtain the same rate of uniform convergence as Theorem \ref{thm-gfs:gamma-cnst}.
\begin{theorem}\label{thm-gfe-gfs:gamma-cnst}
Suppose that Assumptions \ref{ass-gfs:beta-cnst}\ref{ass-gfs:cpt-param}--\ref{ass-gfs:finmt-z}, \ref{ass-gfs:kappa-cnst}\ref{ass-gfs:meantail-z}--\ref{ass-gfs:meantail-zv}, \ref{ass-gfs:gamma-cnst}\ref{ass-gfs:meantail-zu}, and \ref{ass-gfe-gfs:beta-cnst}--\ref{ass-gfe-gfs:gamma-cnst} hold for the two-stage model comprising equations \eqref{eq-gfe:model} and \eqref{eq-gfe-gfs:model}. Then, as $N$ and $T$ tend to infinity, for all $\delta>0$,
\begin{align*}
	\Pr\biggl(\max_{i\in\{1,\dots,N\}}|\hat g_i^\tgfe - g_i^0|>0\biggr)=o(1) + o(NT^{-\delta}).
\end{align*}
\end{theorem}

\subsubsection{Unit-specific first stage}
Now we consider the first-stage model \eqref{eq-gfe-ufs:model}, where each unit can have different first-stage linear projection coefficients.
The estimation procedure proceeds in two steps, which is called the ``unit-specific first-stage GFE'' (UGFE) estimation as before:
\begin{enumerate}[label=(\arabic*)]
	\item Apply the least squares estimation to the first-stage model \eqref{eq-gfe-ufs:model}:
	\begin{align*}
		\min_{(\Pi_i)_{i=1}^N\in\mathbf \Pi^N, \mu\in\mathcal M^{KT},\kappa\in\mathbb K^N}\frac{1}{NT}\sum_{i=1}^N\sum_{t=1}^T\|x_{it}-\Pi_i'z_{it}-\mu_{k_i t}\|^2.
	\end{align*}
	Let $(\hat\Pi^\ugfe,\hat\mu^\ugfe,\hat\kappa^\ugfe)\equiv (\{\hat\Pi_i^\ugfe\}_{i=1}^N,\{\hat\mu_k^\ugfe\}_{k=1}^K,\{\hat k_i^\ugfe\}_{i=1}^N)$ denote the minimizer of this problem.
	Construct the fitted values $\hat x_{it}^\ugfe \equiv \hat\Pi_i^{\ugfe\prime}z_{it} + \hat\mu_{\hat k_i^\ugfe}^\ugfe$.
	\item Apply the GFE estimation to the structural model \eqref{eq-gfe:model} whose endogenous variables $x_{it}$ are substituted with the corresponding fitted values $\hat x_{it}^\ugfe$: 
	\begin{align*}
		\min_{\beta\in\mathcal B^G,\alpha\in\mathcal A^{GT},\gamma\in\mathbb G^N}\frac{1}{NT}\sum_{i=1}^N\sum_{t=1}^T(y_{it}-\hat x_{it}^{\ugfe\prime}\beta_{g_i}-\alpha_{g_i t})^2.
	\end{align*}
	The minimizer $(\hat\beta^\ugfe,\hat\alpha^\ugfe,\hat\gamma^\ugfe) \equiv (\{\hat\beta_g^\ugfe \}_{g=1}^G,\{(\hat\alpha_{gt}^\ugfe)_{t=1}^T\}_{g=1}^G, \{\hat g_i^\ugfe\}_{i=1}^N)$ of this problem is the final estimate.
\end{enumerate}

The following assumptions are needed for the consistency of $(\hat\beta^\ugfe,\hat\alpha^\ugfe)$.
\begin{assumption} \label{ass-gfe-ufs:beta-cnst}
\begin{enumerate}[label=(\alph*),ref=(\alph*)]
	\item \label{ass-gfe-ufs:relev-ss} There exists $\underline \rho^\tss$, which depends on data, such that for all $g\in\mathbb G$, $\min_{\gamma\in\mathbb G^N}\max_{\tilde g\in\mathbb G}\rho^\tss(\gamma,g,\tilde g)\geq \underline \rho^\tss$, where $\rho^\tss(\gamma,g,\tilde g)$ is the smallest eigenvalue of
	\begin{align*}
		&M^\tss(\gamma,g,\tilde g)\\
	 	\equiv &\frac{1}{N}\sum_{i=1}^N\mathbf 1\{g_i^0=g\}\mathbf 1 \{g_i=\tilde g\}\\
	 	 \times &
		\begin{pmatrix}
			\frac{1}{T}\sum_{t=1}^T(\Pi_i^{0\prime}z_{it} + \mu_{k_i^0t}^0)(\Pi_i^{0\prime}z_{it} + \mu_{k_i^0t}^0)' & \frac{1}{\sqrt{T}}(\Pi_i^{0\prime}z_{i1} + \mu_{k_i^01}^0) & \cdots & \frac{1}{\sqrt{T}}(\Pi_i^{0\prime}z_{iT} + \mu_{k_i^0T}^0)\\
			\frac{1}{\sqrt{T}}(\Pi_i^{0\prime}z_{i1} + \mu_{k_i^01}^0)' & 1 & \cdots & 0 \\
			\vdots & 0 & \ddots & 0 \\
			\frac{1}{\sqrt{T}}(\Pi_i^{0\prime}z_{iT} + \mu_{k_i^0T}^0)' & 0 & \cdots & 1 \\
		\end{pmatrix}
	\end{align*}
	and $\underline \rho^\tss\rightarrow \rho^{\tss*}>0$ as $N$ and $T$ tend to infinity.
	\item \label{ass-gfe-ufs:grpsep-ss} There exists $c^\tss>0$ such that
	\begin{align*}
		\plim_{N,T\rightarrow \infty}\frac{1}{N}\sum_{i=1}^ND_{i,g,\tilde g}^\tss >c^\tss,
	\end{align*}
	where $D_{i,g,\tilde g}^\tss \equiv T^{-1}\sum_{t=1}^T((z_{it}'\Pi_i^0+\mu_{k_i^0 t}^0)(\beta_g^0-\beta_{\tilde g}^0) + (\alpha_{g t}^0 - \alpha_{\tilde g t}^0))^2$.
\end{enumerate}
\end{assumption}
Our result then follows:
\begin{theorem} \label{thm-gfe-ufs:beta-cnst} Suppose that Assumptions \ref{ass-gfs:beta-cnst}\ref{ass-gfs:cpt-param}--\ref{ass-gfs:finmt-z}, \ref{ass-ufs:beta-cnst}\ref{ass-ufs:cpt-param}, \ref{ass-gfe-gfs:beta-cnst}\ref{ass-gfe-gfs:cpt-param}--\ref{ass-gfe-gfs:wkdep-v}, and \ref{ass-gfe-ufs:beta-cnst} hold for the two-stage model comprising equations \eqref{eq-gfe:model} and \eqref{eq-gfe-ufs:model}. Then, as $N$ and $T$ tend to infinity, 
\begin{align*}
	d_H(\hat\beta^\ugfe,\beta^0)\rightarrow_p 0
	\text{ and }
	d_H(\hat\alpha^\ugfe,\alpha^0)	\rightarrow_p 0.
\end{align*}	
\end{theorem}

The next three assumptions are needed for the uniform consistency of $\hat\gamma^\ugfe$.
\begin{assumption} \label{ass-gfe-ufs:pi-cnst}
\begin{enumerate}[label=(\alph*),ref=(\alph*)]
	\item \label{ass-gfe-ufs:relev-fs} There exists $\underline \rho^\tfs$, which depends on data, such that for all $k\in\mathbb K$, $\min_{\kappa\in\mathbb K^N}\max_{\tilde k\in\mathbb K}\rho(\gamma,k,\tilde k) \geq \underline \rho^\tfs $, where $\rho(\gamma,k,\tilde k)$ is the smallest eigenvalue of
		\begin{align*}
			M_i^\tss(\kappa,k,\tilde k)
		 	\equiv & \frac{1}{N}\sum_{i=1}^N\mathbf 1\{k_i^0=k\}\mathbf 1\{k_i=\tilde k\}I_T
		\end{align*}
		and $\underline \rho^\tfs\rightarrow \rho^{\tfs*}>0$ as $N$ and $T$ tend to infinity.
	\item \label{ass-gfe-ufs:grpsep-fs} There exists $c^\tfs>0$ such that
	\begin{align*}
		\lim_{T\rightarrow\infty} \Exp[\min_{k\neq \tilde k}D_{k,\tilde k}^\tfs] >c^\tfs,
	\end{align*}
	where $D_{k,\tilde k}^\tfs \equiv T^{-1}\sum_{t=1}^T\|\mu_{k t}^0 - \mu_{\tilde k t}^0\|^2$.
\end{enumerate}
\end{assumption}

\begin{assumption}\label{ass-gfe-ufs:kappa-cnst} 
\begin{enumerate}[label=(\alph*),ref=(\alph*)]
	\item \label{ass-gfe-ufs:mix-fs} There exist positive constants $a^\tfs$ and $d^\tfs$, and a sequence $\alpha^\tfs[t]\leq \exp(-a^\tfs t^{d^\tfs})$ such that, for all $i\in\{1,\dots,N\}$ and $(k,\tilde k)\in\mathbb K^2$ such that $k\neq \tilde k$ and
		$\{v_{it} \}_t$, $\{v_{it}\mu_{k_i^0t}^{0\prime}\}_t$, $\{v_{it}'(\mu_{k t}^0 - \mu_{\tilde  kt}^0)\}_t$ are strongly mixing processes with mixing coefficients $\alpha^\tfs[t]$. Moreover, $\Exp[v_{it}\mu_{k_i^0t}^{0\prime}]=0$ and $\Exp[v_{it}'(\mu_{k t}^0 - \mu_{\tilde  kt}^0)]=0$.
	\item \label{ass-gfe-ufs:tail-fs} There exist positive constants $b_v^\tfs$ and $d_v^\tfs$ such that, for all $c>0$,
		$\Pr(\|v_{it}\|>c)\leq \exp(1-(c/b_v^\tfs)^{d_v^\tfs})$.
\end{enumerate}	
\end{assumption}

\begin{assumption}\label{ass-gfe-ufs:gamma-cnst} 
\begin{enumerate}[label=(\alph*),ref=(\alph*)]
	\item $\lim_{T\rightarrow\infty}\min_{i\in\{1,\dots,N\}}\Exp[\min_{g\neq \tilde g}D_{i,g,\tilde g}^\tss]>c^\tss$.
	\item \label{ass-gfe-ufs:mix-ss} There exist positive constants $a^\tss$ and $d^\tss$, and a sequence $\alpha^\tss[t]\leq \exp(-a^\tss t^{d^\tss})$ such that, for all $i\in\{1,\dots,N\}$ and $(g,\tilde g)\in\mathbb G^2$ such that $g\neq \tilde g$, 
		$\{u_{it}\}_t$,
		$\{u_{it}\mu_{k_i^0t}^0\}_t$,
		$\{v_{it}(\alpha_{g t}^0 - \alpha_{\tilde g t}^0)\}_t$,
		$\{u_{it}(\alpha_{g t}^0 - \alpha_{\tilde g t}^0)\}_t$, and
		$\{(z_{it}'\Pi_i^0+\mu_{k_i^0 t}^0)(\beta_g^0-\beta_{\tilde g}^0) + (\alpha_{g t}^0 - \alpha_{\tilde g t}^0)\}_t$ are strongly mixing processes with mixing coefficients $\alpha^\tss[t]$. 
		Moreover, $\Exp[u_{it}\mu_{k_i^0t}^0]=0$, $\Exp[v_{it}(\alpha_{g t}^0 - \alpha_{\tilde g t}^0)]=0$, and $\Exp[u_{it}(\alpha_{g t}^0 - \alpha_{\tilde g t}^0)]=0$.
	\item \label{ass-gfe-ufs:tail-ss} There exist positive constants $b_{u}^\tss$ and $b_x^\tss$, and $d_{u}^\tss$ and  $d_x^\tss$ such that, for all $c>0$,
		$\Pr(|u_{it}|>c)\leq \exp(1-(c/b_u^\tss)^{d_u^\tss})$ and
		$\Pr(|(z_{it}'\Pi_i^0 + \mu_{k_i^0 t}^0)(\beta_g^0-\beta_{\tilde g}^0) + (\alpha_{g t}^0 - \alpha_{\tilde g t}^0)|>c)\leq \exp(1-(c/b_x^\tss)^{d_x^\tss})$.
\end{enumerate}	
\end{assumption}

We arrive at the same rate of uniform convergence as Theorem \ref{thm-ufs:gamma-cnst}.
\begin{theorem} \label{thm-gfe-ufs:gamma-cnst} Suppose that Assumptions \ref{ass-gfs:beta-cnst}\ref{ass-gfs:cpt-param}--\ref{ass-gfs:finmt-z}, \ref{ass-gfs:kappa-cnst}\ref{ass-gfs:meantail-z}--\ref{ass-gfs:meantail-zv}, \ref{ass-gfs:gamma-cnst}\ref{ass-gfs:meantail-zu}, \ref{ass-ufs:beta-cnst}\ref{ass-ufs:cpt-param}, \ref{ass-gfe-gfs:beta-cnst}\ref{ass-gfe-gfs:cpt-param}--\ref{ass-gfe-gfs:wkdep-v}, \ref{ass-gfe-ufs:pi-cnst}--\ref{ass-gfe-ufs:gamma-cnst} hold for the two-stage model comprising equations \eqref{eq-gfe:model} and \eqref{eq-gfe-ufs:model}. Then, as $N$ and $T$ tend to infinity,
\begin{align*}
	\Pr\biggl(\max_{i\in\{1,\dots,N\}}|\hat g_i^\ugfe - g_i^0|>0\biggr) = o(1) + o(NT^{-\delta}).
\end{align*}
\end{theorem}

\section{Monte Carlo simulations}\label{sec:monte}

We evaluate the finite sample performance of our estimators and confirm their consistency using simulations under various specifications.
Let $d=m=1$ and $\mathbb G = \{1,2\}$.
The structural and first-stage equations are then streamlined into $y_{it} = \beta_{g_i^0}^0 x_{it} + u_{it}$ and $x_{it} = \Pi_i^0 z_{it} + v_{it}$, where $\beta_{g_i^0}^0$ and $\Pi_i^0$ are now scalars.

\subsection{Data-generating process}

We consider four DGPs that are differentiated by their first-stage modeling.
We start by specifying the distribution of the underlying variables $(z_{it},v_{it},u_{it})$.
The error terms $(v_{it},u_{it})$ are generated from a jointly normal distribution whose mean is 0, respective variances are $\sigma^2$ and 1, and correlation is $\rho_i$. Note that the correlation can be unit-specific.
An IV, $z_{it}$, is drawn independently of $(v_{it},u_{it})$ and follows a normal distribution with mean 0 and variance $\sigma^2$.
$(z_{it},v_{it},u_{it})$ are drawn independently across $i$ and $t$.
Summing up, the underlying variables are generated by:
\begin{align*}
	\begin{pmatrix}
		z_{it} \\ v_{it} \\ u_{it} 
	\end{pmatrix}
	 \sim_{i.i.d.} \mathcal N\left(\begin{pmatrix}
		0 \\ 0 \\ 0	
	\end{pmatrix}
	, \begin{pmatrix}
		\sigma^2 & 0 & 0 \\
		0 & \sigma^2 & \sigma \cdot \rho_i \\
		0 & \sigma \cdot \rho_i & 1 
	\end{pmatrix}\right),
\end{align*}
for each $t= 1,\dots, T$, and are independently drawn across $i$, where the distribution of $\rho_i$ varies across the DGPs.

Next, we specify the structural equation.
Every DGP shares the same grouped pattern of the coefficient parameter $\beta_i^0$, which is defined as follows:
\begin{align*}
	\beta_i^0 = \begin{cases}
 		\  1 &\text{if $i=1,\dots, N/2$}\\
 		\ -1 &\text{if $i=N/2+1,\dots,N$}
	\end{cases}.
\end{align*}
Namely, $g_i^0=1$ for $i=1,\dots,N/2$ and $g_i^0=2$ for $i=N/2+1,\dots,N$.

We list the first-stage specifications of each DGP:
\begin{description}
	\item[DGP 1:] $\Pi_i^0=1$ for all $i$. $\rho_i=\rho$ for all $i$.
	\item[DGP 2:] $\Pi_i^0=1$ for odd $i$ and $\Pi_i^0=-1$ for even $i$. $\rho_i=\rho$ for all $i$.
	\item[DGP 3:] $\Pi_i^0\sim_{i.i.d}\text{Unif}_{[0.5,1.5]}$ for $i=1,\dots, N/2$ and $\Pi_i^0\sim_{i.i.d}\text{Unif}_{[-1.5,-0.5]}$ for $i=N/2+ 1,\dots, N$. $\rho_i=\rho$ for all $i$.
	\item[DGP 4:] $\Pi_i^0=1$ for all $i$. $\rho_i=\rho$ for $i=1,\dots, N/2$ and $\rho_i=-\rho$ for $i=N/2+ 1,\dots, N$,
\end{description}
where $\rho=-0.5$. We set $\sigma \in\{0.5,0.75\}$.

DGP 1 is our baseline. There, the first stage is homogeneous across units.
In DGP 2, the first stage exhibits a group structure that is orthogonal to that of the structural equation.
In DGP 3, each unit has a different first stage.
We note that while the IV is relevant for each $i$, the cross-sectional correlation between the IV and the regressor is zero in DGPs 2--3.
In DGP 4, the correlation between the two error terms shows a different grouped pattern from that of the structural equation. In this specification, the linear projection of $y_{it}$ on $x_{it}$ exhibits a different group structure from that of the structural equation.

\subsection{Comparison of performance}
We investigate the finite sample performance of our procedures under these DGPs.
We use the Rand index (RI) \parencite{Rand1971} to measure their classification accuracy.
The RI is a commonly used measure for evaluating the similarity between two classifications.
The formal definition is as follows. We take two group membership structures: $\gamma_1=\{g_i^1\}_{i=1}^N$ and $\gamma_2=\{g_i^2\}_{i=1}^N$. Let $N_{11}$ be the number of pairs $(i,j)$ such that $g_i^1=g_j^1$ and $g_i^2=g_j^2$, $N_{10}$ the number of pairs $(i,j)$ such that $g_i^1=g_j^1$ and $g_i^2\neq g_j^2$, $N_{01}$ the number of pairs $(i,j)$ such that  $g_i^1\neq g_j^1$ and $g_i^2= g_j^2$, and $N_{00}$ the number of pairs $(i,j)$ such that  $g_i^1\neq g_j^1$ and $g_i^2\neq g_j^2$. 
The RI between these two classifications is then calculated by $(N_{11}+N_{00})/(N_{11}+N_{10}+N_{01}+N_{00})$.
The average (over Monte Carlo replications) RI between the true group structure $\gamma^0$ and those estimated from each procedure measures its classification performance.

\begin{table}[hp]
\begin{center}
\resizebox{\textwidth}{!}{
\begin{threeparttable}
\caption{Average Rand index}	
\label{tab:ri_table}
\begin{tabular}{lllcccccccccc}
  \toprule 
 \multicolumn{3}{c}{} & \multicolumn{5}{c}{$\sigma=0.5$} & \multicolumn{5}{c}{$\sigma=0.75$} \\
 \cmidrule(l){4-8} \cmidrule(l){9-13}
  & $N$ & $T$ & IG & TSLS & TGFE$_2$ & TGFE$_{N/4}$ & UGFE & IG & TSLS & TGFE$_2$ & TGFE$_{N/4}$ & UGFE \\
 \midrule 
 \midrule\multirow{6}{*}{\textbf{DGP 1}} & 50 & 5 & 0.856 & 0.714 & 0.717 & 0.718 & 0.717 & 0.953 & 0.799 & 0.805 & 0.810 & 0.814 \\ 
   & 50 & 10 & 0.956 & 0.819 & 0.824 & 0.824 & 0.825 & 0.994 & 0.902 & 0.902 & 0.907 & 0.907 \\ 
   & 50 & 20 & 0.998 & 0.936 & 0.935 & 0.934 & 0.934 & 1.000 & 0.977 & 0.978 & 0.977 & 0.977 \\ 
   & 100 & 5 & 0.861 & 0.713 & 0.718 & 0.722 & 0.723 & 0.943 & 0.789 & 0.795 & 0.801 & 0.802 \\ 
   & 100 & 10 & 0.959 & 0.823 & 0.828 & 0.830 & 0.830 & 0.995 & 0.903 & 0.904 & 0.907 & 0.907 \\ 
   & 100 & 20 & 0.997 & 0.930 & 0.930 & 0.929 & 0.929 & 1.000 & 0.981 & 0.980 & 0.981 & 0.981 \\ 
  \midrule\multirow{6}{*}{\textbf{DGP 2}} & 50 & 5 & 0.840 & 0.496 & 0.701 & 0.709 & 0.710 & 0.948 & 0.493 & 0.795 & 0.800 & 0.799 \\ 
   & 50 & 10 & 0.959 & 0.493 & 0.840 & 0.847 & 0.845 & 0.996 & 0.491 & 0.911 & 0.911 & 0.910 \\ 
   & 50 & 20 & 0.996 & 0.491 & 0.945 & 0.944 & 0.944 & 1.000 & 0.490 & 0.981 & 0.981 & 0.982 \\ 
   & 100 & 5 & 0.853 & 0.498 & 0.708 & 0.720 & 0.721 & 0.951 & 0.497 & 0.805 & 0.811 & 0.811 \\ 
   & 100 & 10 & 0.962 & 0.496 & 0.833 & 0.837 & 0.837 & 0.996 & 0.496 & 0.907 & 0.910 & 0.910 \\ 
   & 100 & 20 & 0.996 & 0.496 & 0.931 & 0.933 & 0.933 & 1.000 & 0.495 & 0.980 & 0.980 & 0.980 \\ 
  \midrule\multirow{6}{*}{\textbf{DGP 3}} & 50 & 5 & 0.862 & 0.505 & 0.699 & 0.707 & 0.708 & 0.951 & 0.501 & 0.798 & 0.802 & 0.803 \\ 
   & 50 & 10 & 0.945 & 0.499 & 0.806 & 0.814 & 0.816 & 0.991 & 0.498 & 0.879 & 0.886 & 0.887 \\ 
   & 50 & 20 & 0.992 & 0.500 & 0.904 & 0.907 & 0.906 & 1.000 & 0.500 & 0.955 & 0.957 & 0.958 \\ 
   & 100 & 5 & 0.848 & 0.502 & 0.700 & 0.707 & 0.707 & 0.947 & 0.499 & 0.788 & 0.792 & 0.792 \\ 
   & 100 & 10 & 0.948 & 0.502 & 0.794 & 0.804 & 0.803 & 0.991 & 0.500 & 0.888 & 0.892 & 0.893 \\ 
   & 100 & 20 & 0.992 & 0.502 & 0.905 & 0.910 & 0.910 & 1.000 & 0.500 & 0.952 & 0.956 & 0.956 \\ 
  \midrule\multirow{6}{*}{\textbf{DGP 4}} & 50 & 5 & 0.548 & 0.831 & 0.828 & 0.772 & 0.766 & 0.793 & 0.945 & 0.947 & 0.885 & 0.885 \\ 
   & 50 & 10 & 0.613 & 0.954 & 0.953 & 0.942 & 0.940 & 0.921 & 0.992 & 0.994 & 0.983 & 0.983 \\ 
   & 50 & 20 & 0.700 & 0.995 & 0.995 & 0.994 & 0.994 & 0.983 & 1.000 & 1.000 & 1.000 & 1.000 \\ 
   & 100 & 5 & 0.555 & 0.842 & 0.838 & 0.786 & 0.787 & 0.806 & 0.951 & 0.950 & 0.891 & 0.891 \\ 
   & 100 & 10 & 0.605 & 0.954 & 0.955 & 0.943 & 0.942 & 0.928 & 0.994 & 0.994 & 0.985 & 0.984 \\ 
   & 100 & 20 & 0.703 & 0.996 & 0.995 & 0.995 & 0.995 & 0.988 & 1.000 & 1.000 & 1.000 & 1.000 \\ 
   \bottomrule 
\end{tabular}

\begin{tablenotes}
	\item Note: The value in each cell denotes the mean of the Rand index across 100 simulations, with 100 starting group memberships randomly generated for each estimate.
\end{tablenotes}
\end{threeparttable}
}
\end{center}
\end{table}

Table \ref{tab:ri_table} displays the average RI of our estimators under the above DGPs.
For TGFE, we consider two specifications of first-stage heterogeneity, $K=2$ and $K=N/4$. These are denoted as TGFE$_{2}$ and TGFE$_{N/4}$, respectively.
We omit the reduced-form approach because it yields estimates identical to the 2SLS estimates in just-identified cases.\footnote{The second stage in the TSLS estimation minimizes $\sum_{i=1}^N \sum_{t=1}^T (y_{it} - z_{it}' (\hat \Pi ' \beta_{g_i}))^2$. In just-identified cases, $\hat \Pi$ is invertible when the IVs are relevant. Thus, setting $\hat \Pi' \beta_{g_i} =\xi_{l_i}$ makes the minimization problem equivalent to minimizing $\sum_{i=1}^N \sum_{t=1}^T (y_{it} - z_{it}'\xi_{l_i})^2$.}
Every estimator performs well when predicted to be so by asymptotic theory.
Overall, an increase in $N$ does not contribute much to classification accuracy as opposed to $T$.
The 2SLS estimator displays high classification performance in DGPs 1 and 4, both of which assume a homogeneous first stage. In particular, in DGP 4, 2SLS  far outperforms the other competing estimators when $T$ is small.
However, it suffers from the weak instrument problem in DGPs 2--3, which 
indicates the importance of modeling the first stage as exhibiting a group structure.
The TGFE and UGFE estimators perform well in all specifications.
A notable observation is that the TGFE$_2$ estimator performs similarly to the UGFE estimator even under DGP 3, where the first stage exhibits full heterogeneity.
The result indicates that the TGFE estimators may be employed as an approximation tool for the UGFE estimator.
Indeed, the TGFE$_{N/4}$ estimator performs very similarly to the UGFE estimator in all cases.
The IG estimator performs very well in DGPs 1--3, but poorly in DGP 4, especially when $\sigma=0.5$ even after we increase $N$ and $T$.
Intuitively, this is because the true group structure $\gamma^0$ and the grouped pattern $\{\rho_i\}_{i=1}^N$ in the correlation between the two error terms are tangled up, and the IG estimator is unable to disentangle them.

\bigskip\begin{table}[hp]
\begin{center}
\resizebox{0.9\textwidth}{!}{
\begin{threeparttable}
\caption{Average Hausdorff distance (Pre-estimates)}
\label{tab:hd_table}
\begin{tabular}{lllcccccccc}
  \toprule 
 \multicolumn{3}{c}{} & \multicolumn{4}{c}{$\sigma=0.5$} & \multicolumn{4}{c}{$\sigma=0.75$} \\
 \cmidrule(l){4-7} \cmidrule(l){8-11}
  & $N$ & $T$ & TSLS & TGFE$_2$ & TGFE$_{N/4}$ & UGFE & TSLS & TGFE$_2$ & TGFE$_{N/4}$ & UGFE \\
 \midrule 
 \midrule\multirow{6}{*}{\textbf{DGP 1}} & 50 & 5 & 0.438 & 0.334 & 0.328 & 0.326 & 0.266 & 0.200 & 0.190 & 0.190 \\ 
   & 50 & 10 & 0.251 & 0.213 & 0.212 & 0.212 & 0.137 & 0.127 & 0.120 & 0.120 \\ 
   & 50 & 20 & 0.129 & 0.130 & 0.132 & 0.132 & 0.081 & 0.078 & 0.079 & 0.080 \\ 
   & 100 & 5 & 0.394 & 0.276 & 0.267 & 0.266 & 0.240 & 0.174 & 0.180 & 0.181 \\ 
   & 100 & 10 & 0.240 & 0.198 & 0.192 & 0.192 & 0.103 & 0.101 & 0.103 & 0.103 \\ 
   & 100 & 20 & 0.101 & 0.105 & 0.107 & 0.107 & 0.047 & 0.052 & 0.054 & 0.054 \\ 
  \midrule\multirow{6}{*}{\textbf{DGP 2}} & 50 & 5 & 115.135 & 0.445 & 0.306 & 0.301 & 61.235 & 0.257 & 0.198 & 0.201 \\ 
   & 50 & 10 & 73.606 & 0.211 & 0.180 & 0.182 & 41.249 & 0.149 & 0.130 & 0.131 \\ 
   & 50 & 20 & 51.571 & 0.118 & 0.116 & 0.116 & 86.899 & 0.075 & 0.071 & 0.071 \\ 
   & 100 & 5 & 91.446 & 0.425 & 0.285 & 0.287 & 69.831 & 0.230 & 0.178 & 0.179 \\ 
   & 100 & 10 & 109.139 & 0.202 & 0.176 & 0.176 & 185.538 & 0.125 & 0.115 & 0.115 \\ 
   & 100 & 20 & 93.052 & 0.098 & 0.097 & 0.099 & 544.140 & 0.053 & 0.058 & 0.058 \\ 
  \midrule\multirow{6}{*}{\textbf{DGP 3}} & 50 & 5 & 73.604 & 0.463 & 0.315 & 0.310 & 161.554 & 0.253 & 0.171 & 0.171 \\ 
   & 50 & 10 & 138.895 & 0.232 & 0.176 & 0.177 & 106.551 & 0.152 & 0.115 & 0.116 \\ 
   & 50 & 20 & 83.750 & 0.145 & 0.119 & 0.119 & 67.040 & 0.086 & 0.073 & 0.074 \\ 
   & 100 & 5 & 87.780 & 0.419 & 0.260 & 0.260 & 168.426 & 0.252 & 0.163 & 0.164 \\ 
   & 100 & 10 & 111.934 & 0.264 & 0.169 & 0.169 & 111.968 & 0.118 & 0.086 & 0.087 \\ 
   & 100 & 20 & 169.544 & 0.120 & 0.093 & 0.093 & 139.661 & 0.070 & 0.061 & 0.061 \\ 
  \midrule\multirow{6}{*}{\textbf{DGP 4}} & 50 & 5 & 0.185 & 0.250 & 0.319 & 0.319 & 0.115 & 0.171 & 0.215 & 0.218 \\ 
   & 50 & 10 & 0.134 & 0.153 & 0.189 & 0.190 & 0.079 & 0.108 & 0.139 & 0.139 \\ 
   & 50 & 20 & 0.092 & 0.107 & 0.125 & 0.126 & 0.054 & 0.074 & 0.088 & 0.089 \\ 
   & 100 & 5 & 0.145 & 0.185 & 0.260 & 0.262 & 0.081 & 0.157 & 0.209 & 0.210 \\ 
   & 100 & 10 & 0.089 & 0.126 & 0.178 & 0.179 & 0.058 & 0.088 & 0.117 & 0.118 \\ 
   & 100 & 20 & 0.064 & 0.087 & 0.109 & 0.110 & 0.039 & 0.054 & 0.068 & 0.068 \\ 
   \bottomrule 
\end{tabular}

\begin{tablenotes}
	\item Note: The value in each cell denotes the mean of the Hausdorff distance across 100 simulations, with 100 starting group memberships randomly generated for each estimate.
\end{tablenotes}
\end{threeparttable}
}
\end{center}
\end{table}

Table \ref{tab:hd_table} presents the average Hausdorff distance between the estimates and the true coefficient parameter $\beta^0$.
The IG estimation is omitted because it only provides estimates for the group memberships.
The results overall follow those in Table \ref{tab:ri_table}.
As an estimator attains higher classification accuracy, its estimate also approaches the true coefficient value.

\begin{table}[hp]
\begin{center}
\resizebox{\textwidth}{!}{
\begin{threeparttable}
\caption{Average Hausdorff distance (Post-estimates)}
\label{tab:hd_reest_table}
\begin{tabular}{lllcccccccccc}
  \toprule 
 \multicolumn{3}{c}{} & \multicolumn{5}{c}{$\sigma=0.5$} & \multicolumn{5}{c}{$\sigma=0.75$} \\
 \cmidrule(l){4-8} \cmidrule(l){9-13}
  & $N$ & $T$ & IG & 2SLS & TGFE$_2$ & TGFE$_{N/4}$ & UGFE & IG & 2SLS & TGFE$_2$ & TGFE$_{N/4}$ & UGFE \\
 \midrule 
 \midrule\multirow{6}{*}{\textbf{DGP 1}} & 50 & 5 & 0.204 & 0.319 & 0.312 & 0.281 & 0.279 & 0.146 & 0.187 & 0.180 & 0.159 & 0.157 \\ 
   & 50 & 10 & 0.138 & 0.187 & 0.179 & 0.184 & 0.183 & 0.083 & 0.097 & 0.095 & 0.094 & 0.094 \\ 
   & 50 & 20 & 0.100 & 0.116 & 0.113 & 0.114 & 0.114 & 0.068 & 0.068 & 0.068 & 0.068 & 0.068 \\ 
   & 100 & 5 & 0.156 & 0.283 & 0.275 & 0.225 & 0.223 & 0.106 & 0.157 & 0.152 & 0.134 & 0.134 \\ 
   & 100 & 10 & 0.103 & 0.174 & 0.165 & 0.161 & 0.161 & 0.065 & 0.076 & 0.074 & 0.073 & 0.073 \\ 
   & 100 & 20 & 0.075 & 0.085 & 0.084 & 0.084 & 0.084 & 0.043 & 0.041 & 0.041 & 0.041 & 0.041 \\ 
  \midrule\multirow{6}{*}{\textbf{DGP 2}} & 50 & 5 & 7.825 & 27.009 & 5.349 & 4.829 & 5.070 & 4.324 & 45.869 & 11.318 & 4.829 & 5.160 \\ 
   & 50 & 10 & 3.518 & 52.256 & 6.476 & 5.172 & 9.523 & 3.921 & 83.225 & 4.600 & 14.287 & 14.309 \\ 
   & 50 & 20 & 39.294 & 64.611 & 44.333 & 44.502 & 44.502 & 4.887 & 2411.873 & 4.852 & 4.691 & 4.695 \\ 
   & 100 & 5 & 47.885 & 157.277 & 7.541 & 10.174 & 10.802 & 3.549 & 253.861 & 10.258 & 4.781 & 4.906 \\ 
   & 100 & 10 & 7.073 & 161.123 & 6.398 & 6.171 & 4.338 & 32.893 & 78.121 & 3.476 & 6.383 & 6.372 \\ 
   & 100 & 20 & 9.156 & 210.235 & 5.259 & 5.533 & 5.485 & 21.470 & 255.334 & 20.574 & 20.745 & 20.733 \\ 
  \midrule\multirow{6}{*}{\textbf{DGP 3}} & 50 & 5 & 0.249 & 41.228 & 0.408 & 0.367 & 0.359 & 0.136 & 47.176 & 0.172 & 0.164 & 0.160 \\ 
   & 50 & 10 & 0.141 & 63.842 & 0.197 & 0.186 & 0.185 & 0.088 & 55.036 & 0.105 & 0.102 & 0.101 \\ 
   & 50 & 20 & 0.104 & 51.546 & 0.117 & 0.117 & 0.117 & 0.069 & 35.139 & 0.073 & 0.073 & 0.073 \\ 
   & 100 & 5 & 0.193 & 263.935 & 0.337 & 0.316 & 0.317 & 0.100 & 178.029 & 0.161 & 0.153 & 0.153 \\ 
   & 100 & 10 & 0.110 & 55.298 & 0.206 & 0.195 & 0.195 & 0.066 & 202.919 & 0.080 & 0.078 & 0.078 \\ 
   & 100 & 20 & 0.075 & 57.106 & 0.101 & 0.099 & 0.099 & 0.049 & 84.641 & 0.052 & 0.051 & 0.051 \\ 
  \midrule\multirow{6}{*}{\textbf{DGP 4}} & 50 & 5 & 0.431 & 0.213 & 0.232 & 0.229 & 0.225 & 0.191 & 0.146 & 0.150 & 0.144 & 0.145 \\ 
   & 50 & 10 & 0.384 & 0.155 & 0.158 & 0.154 & 0.152 & 0.115 & 0.098 & 0.097 & 0.098 & 0.098 \\ 
   & 50 & 20 & 0.299 & 0.108 & 0.108 & 0.107 & 0.107 & 0.072 & 0.070 & 0.070 & 0.070 & 0.070 \\ 
   & 100 & 5 & 0.374 & 0.167 & 0.177 & 0.159 & 0.157 & 0.150 & 0.098 & 0.101 & 0.110 & 0.110 \\ 
   & 100 & 10 & 0.371 & 0.104 & 0.108 & 0.111 & 0.110 & 0.084 & 0.069 & 0.069 & 0.071 & 0.071 \\ 
   & 100 & 20 & 0.288 & 0.075 & 0.076 & 0.076 & 0.076 & 0.050 & 0.048 & 0.048 & 0.048 & 0.048 \\ 
   \bottomrule 
\end{tabular}

\begin{tablenotes}
	\item Note: The value in each cell denotes the mean of the Hausdorff distance across 100 simulations, with 100 starting group memberships randomly generated for each estimate.
\end{tablenotes}
\end{threeparttable}
}
\end{center}
\end{table}

Table \ref{tab:hd_reest_table} includes estimates obtained by conducting separate IV estimations for each estimated group.
Again, the estimates largely correspond to those in Table \ref{tab:ri_table}.
The estimators achieve small biases in situations under which the estimators attain high classification accuracy.
An exception is DGP 2, where every estimator, including the TGFE and UGFE estimators, suffers from the weak instrument problem.
This is because these estimates obtained ex-post do not utilize the first-stage information.
This result exemplifies the importance of taking into account possible heterogeneity in the first stage, which constitutes the distinct characteristic of our procedures. 

\section{Empirical example} \label{sec:empirical}

In this section, we apply our estimators to revisit the relationship between democracy and income studied by \textcite{AcemogluJohnsonRobinsonYared08}.
The authors use linear dynamic models and find that once we control for the country fixed effects, the positive effect of income on the level of democracy disappears.

However, because their model assumes that the effect of income growth is homogeneous across countries, the literature has pointed out that we might reach different conclusions when the potential heterogeneity is considered.
For example, in their comment, \textcite{CervellatiJungSundeVischer14} report opposing estimation results for former colonies and noncolonies.
Meanwhile, by using GFE estimation, \textcite[Table S16]{BonhommeManresa2015Supp} show that the effect varies across groups.
We contribute to this literature by reporting new results on heterogeneity in the effect of income using our estimators.

The key deviation from \textcite{BonhommeManresa2015Supp} is that we take into account the potential endogeneity of lagged income growth due to the reverse causality.
\textcite{AcemogluJohnsonRobinsonYared08} manage this issue by employing the savings rate and trade-weighted world income as instruments for income growth.
By contrast, \textcite{BonhommeManresa2015Supp} implicitly assume that the potential correlation between the income growth and unobserved factors is entirely subsumed by the group-specific fixed effects.
Here, our interest is in the consequences of simultaneously allowing the endogeneity of income and the heterogeneity in its coefficient.

We consider various models of the relationship between the level of democracy and income growth.
Let $d_{it}$ and $y_{it-1}$ denote the level of democracy measured by the Freedom House score and the lagged value of log income per capita, respectively.
Let $\mathbb G =\{1,2\}$, and denote the group-specific fixed effects by $\alpha_{g}$.
The $u_{it}$ in each equation is the error term.
\begin{description}
	\item[Model 1a:] $d_{it}=\beta_{2,g_i} y_{it-1} + \alpha + u_{it}$
	\item[Model 1b:] $d_{it}=\beta_{2,g_i} y_{it-1} + \alpha_{g_i} + u_{it}$
	\item[Model 2a:] $d_{it}=\beta_1 d_{it-1} + \beta_{2,g_i} y_{it-1} + \alpha + u_{it}$
	\item[Model 2b:] $d_{it}=\beta_1 d_{it-1} + \beta_{2,g_i} y_{it-1} + \alpha_{g_i} + u_{it}$
\end{description}

Models 1a and 1b are static in the sense that they do not contain the lagged term for the level of democracy.
They differ by whether the group-specific fixed effects are included.
Models 2a and 2b are dynamic, where the term $d_{it-1}$ captures the mean-reverting dynamics of democracy.
In Model 2a, only the coefficient of income growth exhibits a grouped pattern, while in Model 2b the constant also does so.
As in \textcite{CervellatiJungSundeVischer14}, we employ trade-weighted world income as an instrument.

Similar to the simulation section, we report two types of estimates for $\beta_{2,g_i}$, which captures the effect of income growth on democracy. The first ones, which we call the ``pre-estimates,'' are directly generated from our procedures, and the second ones, which we call the ``post-estimates,'' are from separate IV regressions post-run for each estimated group.

\begin{table}[htp]
\begin{center}
\resizebox{0.8\columnwidth}{!}{
\begin{threeparttable}
\caption{Estimation results for the models}
\label{tab:m_table}
\begin{tabular}{llccccc}
  \toprule 
  Model & Method & Pre-estimates & SE & Group size & Post-estimates & SE \\
 \midrule 
 \midrule\multirow{1}{*}{\textbf{No groups}} & \multirow{1}{*}{2SLS} &  &  &  & 0.112 & 0.126 \\ 
  \midrule\multirow{10}{*}{\textbf{Model 1a}} & \multirow{2}{*}{IG} &  &  & 47 & 0.180 & 0.036 \\ 
   &  &  &  & 32 & 0.134 & 0.040 \\ 
   & \multirow{2}{*}{2SLS} & 0.166 & 0.083 & 37 & 0.150 & 0.053 \\ 
   &  & 0.232 & 0.083 & 42 & 0.192 & 0.045 \\ 
   & \multirow{2}{*}{TGFE$_2$} & 0.170 & 0.022 & 35 & 0.138 & 0.054 \\ 
   &  & 0.220 & 0.019 & 44 & 0.180 & 0.048 \\ 
   & \multirow{2}{*}{TGFE$_{N/4}$} & 0.210 & 0.012 & 48 & 0.156 & 0.036 \\ 
   &  & 0.170 & 0.015 & 31 & 0.108 & 0.038 \\ 
   & \multirow{2}{*}{UGFE} & 0.218 & 0.012 & 52 & 0.200 & 0.036 \\ 
   &  & 0.178 & 0.015 & 27 & 0.155 & 0.039 \\ 
  \midrule\multirow{10}{*}{\textbf{Model 1b}} & \multirow{2}{*}{IG} &  &  & 47 & 0.276 & 0.102 \\ 
   &  &  &  & 32 & 0.092 & 0.028 \\ 
   & \multirow{2}{*}{2SLS} & 0.141 & 0.069 & 34 & 0.113 & 0.037 \\ 
   &  & 0.632 & 0.148 & 45 & 0.445 & 0.234 \\ 
   & \multirow{2}{*}{TGFE$_2$} & 0.236 & 0.026 & 36 & 0.109 & 0.050 \\ 
   &  & 0.167 & 0.020 & 43 & 0.303 & 0.267 \\ 
   & \multirow{2}{*}{TGFE$_{N/4}$} & 0.181 & 0.026 & 33 & 0.082 & 0.038 \\ 
   &  & 0.197 & 0.015 & 46 & 0.261 & 0.078 \\ 
   & \multirow{2}{*}{UGFE} & 0.187 & 0.029 & 27 & 0.101 & 0.031 \\ 
   &  & 0.215 & 0.014 & 52 & 0.271 & 0.071 \\ 
   \bottomrule 
\end{tabular}

\smallskip
\begin{tabular}{llccccc}
  \toprule 
  Model & Method & Pre-estimates & SE & Group size & Post-estimates & SE \\
 \midrule 
 \midrule\multirow{1}{*}{\textbf{No groups}} & \multirow{1}{*}{2SLS} &  &  &  & -0.006 & 0.053 \\ 
  \midrule\multirow{10}{*}{\textbf{Model 2a}} & \multirow{2}{*}{IG} &  &  & 27 & 0.098 & 0.046 \\ 
   &  &  &  & 52 & 0.125 & 0.047 \\ 
   & \multirow{2}{*}{2SLS} & 0.113 & 0.050 & 45 & 0.109 & 0.039 \\ 
   &  & 0.083 & 0.049 & 34 & 0.081 & 0.040 \\ 
   & \multirow{2}{*}{TGFE$_2$} & 0.088 & 0.017 & 48 & 0.125 & 0.041 \\ 
   &  & 0.059 & 0.017 & 31 & 0.096 & 0.041 \\ 
   & \multirow{2}{*}{TGFE$_{N/4}$} & 0.083 & 0.012 & 27 & 0.098 & 0.046 \\ 
   &  & 0.107 & 0.012 & 52 & 0.125 & 0.047 \\ 
   & \multirow{2}{*}{UGFE} & 0.088 & 0.011 & 27 & 0.098 & 0.046 \\ 
   &  & 0.112 & 0.011 & 52 & 0.125 & 0.047 \\ 
  \midrule\multirow{10}{*}{\textbf{Model 2b}} & \multirow{2}{*}{IG} &  &  & 27 & 0.089 & 0.049 \\ 
   &  &  &  & 52 & 0.219 & 0.089 \\ 
   & \multirow{2}{*}{2SLS} & 0.326 & 0.061 & 54 & 0.337 & 0.146 \\ 
   &  & 0.116 & 0.049 & 25 & 0.103 & 0.051 \\ 
   & \multirow{2}{*}{TGFE$_2$} & 0.099 & 0.019 & 48 & 0.222 & 0.171 \\ 
   &  & 0.033 & 0.023 & 31 & 0.059 & 0.046 \\ 
   & \multirow{2}{*}{TGFE$_{N/4}$} & 0.105 & 0.013 & 52 & 0.219 & 0.089 \\ 
   &  & 0.091 & 0.019 & 27 & 0.089 & 0.049 \\ 
   & \multirow{2}{*}{UGFE} & 0.122 & 0.012 & 54 & 0.213 & 0.077 \\ 
   &  & 0.121 & 0.020 & 25 & 0.093 & 0.052 \\ 
   \bottomrule 
\end{tabular}

\begin{tablenotes}
	\item Note: 10,000 starting group memberships are randomly generated for each estimate. Standard errors clustered at country level.
\end{tablenotes}
\end{threeparttable}
}
\end{center}
\end{table}

Table \ref{tab:m_table} presents our estimation results. In Models 1a and 2a, the pre- and post-estimates are positive and mostly statistically significant. They are similar to each other, while the pre-estimates tend to show smaller standard errors. The results are overall consistent across procedures. The group with more units displays larger estimates---around 0.4 to 0.6 in Model 1a and around 0.3 in Model 2a. 

Likewise, in Models 1b and 2b, the estimates are positive and statistically significant. However, in Model 1b, some pre- and post-estimates are markedly different from each other. Their difference is not prevalent in Model 2b, but there, the pre-estimates show some variation by procedure. For instance, the 2SLS estimator reports a large difference between groups---around 0.2---whereas the UGFE estimator reports a negligible difference. This illustrates the situation where modeling the first-stage heterogeneity requires careful attention.

\section{Conclusion} \label{sec:conclusion}
This paper considers the estimation of linear panel data models with endogenous regressors and a latent group structure in the coefficients.
In particular, it provides a relatively comprehensive view of possible estimation methods and clarifies the conditions under which the estimators work.
We first argue that a na\"ive approach of minimizing the GMM objective function should not be used to estimate group membership parameters because at least asymptotically, the minimizer of the objective function is not uniquely determined. 
Then, we consider various alternative estimation strategies.
The first three procedures are 2SLS estimations which differ in the first-stage modeling.
The 2SLS estimator is based on the homogeneity of the first-stage coefficients.
The TGFE estimator assumes that the first-stage coefficients exhibit a group structure.
The UGFE estimator allows complete heterogeneity across units.
We provide conditions under which these procedures yield consistent estimates for the true coefficient and group memberships.
Furthermore, for the 2SLS estimators, we provide data-driven procedures that select the number of groups.
The next two approaches are more heuristic.
The ignoring endogeneity approach estimates the true memberships by applying the GFE estimation directly to the endogenous variables.
The reduced-form approach applies the GFE estimation directly to the instrument.
The conditions that ensure their consistency can be obtained from the literature, but it is difficult to test or verify them in practice.
The results from the Monte Carlo simulations suggest that estimating the group memberships using the TGFE/UGFE estimators performs well in terms of classification performance; the IG estimator also shows good performance, but it might perform poorly in cases where the first- and second-stage errors are highly correlated with heterogeneous correlation coefficients.


\printbibliography[notcategory={onlyappx}]

\end{mtchideinmaintoc}

\newpage\appendix

\setstretch{1.35}
\setcounter{page}{1}
\setcounter{assumption}{0}
\setcounter{equation}{0}
\setcounter{table}{0}
\setcounter{theorem}{0}
\renewcommand{\theassumption}{\thesection\arabic{assumption}}
\renewcommand{\theequation}{\thesection\arabic{equation}}
\counterwithin{table}{section}
\renewcommand{\thetable}{\thesection\arabic{table}}
\renewcommand{\thetheorem}{\thesection\arabic{theorem}}
\renewcommand{\thelemma}{\thesection\arabic{lemma}}

\title{Online appendix to ``Latent group structure in linear panel data models with endogenous regressors''}
\author{\setcounter{footnote}{1} Junho Choi\thanks{Department of Economics, University of Wisconsin-Madison. Email: \href{mailto: jamesjunho@wisc.edu}{jamesjunho@wisc.edu}} \ and Ryo Okui\thanks{Graduate School of Economics and Faculty of Economics, University of Tokyo. Email: \href{mailto:okuiryo@e.u-tokyo.ac.jp}{okuiryo@e.u-tokyo.ac.jp}}}
\date{May 2024}
\maketitle

\begin{abstract}
This online appendix includes the proofs for the theoretical results in the main text and additional simulation/empirical results.
In Section \ref{sec-appdx:proofs}, after establishing several useful lemmas, we provide the proofs for the theorems in Sections \ref{sec:naive}--\ref{sec:extensions}, although some detailed steps are included in Section \ref{sec-appdx:supp} for the sake of conciseness.
In Section \ref{sec-appdx:num-of-groups}, we provide a theoretical background for the information criteria suggested in Section \ref{sec:select-num-grps}. Under certain high-level conditions, we establish the required rate for the penalty term to ensure the consistency of the selection procedure. In Section \ref{sec-appdx:sim}, using Monte Carlo simulations, we evaluate the performance of our information criteria. Furthermore, we illustrate the bias-variance tradeoff encountered in deciding the number of groups in the first stage. In Section \ref{sec-appdx:emp}, we provide the details of our empirical example, including the functional forms of the first-stage models and how we calculate the standard errors for the pre-estimates. We also examine an extended version of the baseline model and discuss its estimation results.
\end{abstract}

\newpage
\tableofcontents

\newpage
\section{Proofs of theorems} \label{sec-appdx:proofs}
\subsection{Lemmas}
We consider the following baseline linear model: 
\begin{align*}
	\underset{1\times d_y}{y_{it}'} = \underset{\mathclap{1\times d_{x,N}}}{x_{it}'}\overset{\mathclap{d_{x,N}\times d_y}}{\beta_{g_i^0}^0} + \underset{1\times d_y}{\alpha_{g_i^0t}^{0\prime}} + \underset{1\times d_y}{e_{it}'}, \label{eq-sineq}
\end{align*}
where the dimensions of the variables are documented either below or above each one. The dimension $d_{x,N}$ of the regressor $x_{it}$ may vary by the total number $N$ of units. This aims to make Lemma \ref{lem-sineq} applicable to the unit-specific first-stage scenario in the main text.

We make the following assumptions about the variables in this model. For some notations, the dependence on $N$ and $T$ are omitted to simplify the presentation. Unless otherwise stated, we adhere to the notation defined in the main text.
\begin{assumption} \label{ass-sineq} 
For some constant $M>0$,
\begin{enumerate}[label=(\alph*),ref=(\alph*)]
	\item \label{ass-sineq:cpt-param} Let $\mathcal B_N$ and $\mathcal A$ be compact sets in $\mathbb R^{d_{x,N}\times d_y}$ and $\mathbb R^{d_y}$ respectively. For all $g\in\mathbb G$, $\beta_g^0\in\mathcal B_N$ and $\{\alpha_{gt}^0\}_{t=1}^T\in\mathcal A^T$. Moreover, $\mathcal B_N$'s are uniformly bounded.
	\item \label{ass-sineq:wkdep-xe} As $N$ and $T$ tend to infinity,
	\begin{align*}
		\max_{\gamma\in\mathbb G^N}\max_{(g,\tilde g)\in\mathbb G^2}\biggl\|\frac{1}{N}\sum_{i=1}^N\mathbf 1\{g_i^0=g\}\mathbf 1\{g_i=\tilde g\}\frac{1}{T}\sum_{t=1}^Tx_{it}e_{it}'\biggr\|=o_p(1).
	\end{align*}
	\item \label{ass-sineq:wkdep-e} As $N$ and $T$ tend to infinity, 
	\begin{align*}
		\max_{\gamma\in\mathbb G^N}\max_{(g,\tilde g)\in\mathbb G^2}\frac{1}{T}\sum_{t=1}^T\biggl\|\frac{1}{N}\sum_{i=1}^N\mathbf 1\{g_i^0=g\}\mathbf 1\{g_i=\tilde g\}e_{it}\biggr\|=o_p(1)	
	\end{align*}
	\item \label{ass-sineq:relev-ss} Let $\mathbf x_{it}\equiv (x_{it}',\sqrt T d1_t,\dots,\sqrt T dT_t)$, where $d\tau_t \equiv \mathbf 1\{t=\tau\}$, so that
	\begin{align*}
		\frac{1}{T}\sum_{t=1}^T\mathbf x_{it}\mathbf x_{it}'
		= \begin{pmatrix}
			\frac{1}{T}\sum_{t=1}^Tx_{it}x_{it}' & \frac{1}{\sqrt T} x_{i1} & \frac{1}{\sqrt T} x_{i2} & \cdots & \frac{1}{\sqrt T} x_{iT}\\
			\frac{1}{\sqrt T} x_{i1}' & 1 & 0 & \cdots & 0\\
			\frac{1}{\sqrt T} x_{i2}' & 0 & 1 & \cdots & 0\\
			\vdots & \vdots & \vdots & \ddots & \vdots\\
			\frac{1}{\sqrt T} x_{iT}' & 0 & 0 & \cdots & 1
		\end{pmatrix}.
	\end{align*}
	For every $(g,\tilde g)\in\mathbb G^2$, define a $(d_{x,N}+T)\times (d_{x,N} + T)$ matrix
	\begin{align*}
		M(\gamma,g,\tilde g)
	 	\equiv \frac{1}{N}\sum_{i=1}^N\mathbf 1\{g_i^0=g\}\mathbf 1 \{g_i=\tilde g\}\frac{1}{T}\sum_{t=1}^T\mathbf x_{it}\mathbf x_{it}'.
	\end{align*}
	There exists $\underline \rho$, which depends on data, such that
	\begin{enumerate}[label=(\roman*)]
		\item for all $g\in\mathbb G$, $\min_{\gamma\in\mathbb G^N}\max_{\tilde g\in\mathbb G}\rho(\gamma,g,\tilde g)\geq \underline \rho$, where $\rho(\gamma,g,\tilde g)$ is a non-negative scalar that satisfies: for all $v\in\mathbb R^{d_{x,N}+T}$, 
		\begin{align*}
			v'M(\gamma,g,\tilde g)v\geq \rho(\gamma,g,\tilde g)\|v_J\|^2	
		\end{align*}
		where $v_J$ denotes a subvector of $v$ consisting only of the components indexed by a subset $J\subseteq\{1,\dots,d_{x,N},d_{x,N}+1,\dots,d_{x,N}+T\}$.\footnote{This formulation is to encompass the cases where $M(\gamma,g,\tilde g)$ is singular. If $M(\gamma,g,\tilde g)$ is positive definite, we can take $\rho(\gamma,g,\tilde g)$ as its smallest eigenvalue and set $J=\{1,\dots,d_{x,N},d_{x,N}+1,\dots,d_{x,N}+T\}$.}
		\item as $N$ and $T$ tend to infinity, $\underline \rho\rightarrow_p \rho^*>0$ 
	\end{enumerate}
	\item \label{ass-sineq:grpsep-ss} There exists a constant $c>0$ such that 
	\begin{align*}
		\plim_{N,T\rightarrow \infty}\frac{1}{N}\sum_{i=1}^N\min_{g\neq \tilde g} D_{i,g,\tilde g} >c,
	\end{align*}
	where $D_{i,g,\tilde g} \equiv T^{-1}\sum_{t=1}^T\|x_{it}'(\beta_g^0-\beta_{\tilde g}^0) + (\alpha_{gt}^0-\alpha_{\tilde gt}^0)\|^2$.
	\item \label{ass-sineq:finmt-z}$\Exp[\|(NT)^{-1}\sum_{i=1}^N\sum_{t=1}^Tx_{it}x_{it}'\|_2] \leq M$.
\end{enumerate}	
\end{assumption}

Under these assumptions, we establish the following lemma:
\begin{lemma} \label{lem-sineq}
Let $(\hat\beta,\hat\alpha,\hat\gamma)$ be any estimator for $(\beta^0,\alpha^0,\gamma^0)$. For each $g\in\mathbb G$, we define
\begin{align*}
	\sigma(g)\equiv \arg\min_{\tilde g\in\mathbb G}\biggl(\|\beta_g^0-\hat\beta_{\tilde g}\|^2 + \frac{1}{T}\sum_{t=1}^T\|\alpha_{gt}^0-\hat\alpha_{\tilde gt}\|^2\biggr).
\end{align*}
Then, we have the following results:
\begin{enumerate}[label=(\alph*),ref=\ref{lem-sineq}(\alph*)]
	\item \label{lem-sineq:fit} 	Suppose that Assumptions \ref{ass-sineq}\ref{ass-sineq:cpt-param}--\ref{ass-sineq:wkdep-e} hold, and that $(\hat\beta,\hat\alpha,\hat\gamma)$ satisfies:
	\begin{equation}
		\inf_{(\beta,\alpha,\gamma)\in\mathcal B_N^{G}\times\mathcal A^{TG}\times \mathbb G^N}\frac{1}{NT}\sum_{i=1}^N\sum_{t=1}^T\|y_{it}'-x_{it}'\beta_g - \alpha_{gt}'\|^2 + o_p(1) \geq \frac{1}{NT}\sum_{i=1}^N\sum_{t=1}^T\|y_{it}'-x_{it}'\hat\beta_{\hat g_i}-\hat\alpha_{\hat g_i}'\|^2, \label{eq-sineq:min}
	\end{equation}
	as $N$ and $T$ tend to infinity. Then, we have:
	\begin{equation}
		\frac{1}{NT}\sum_{i=1}^N\sum_{t=1}^T\|x_{it}'(\beta_{g_i^0}^0-\hat\beta_{\hat g_i}) + (\alpha_{g_i^0 t}^{0\prime} - \hat\alpha_{\hat g_i t}')\|^2\rightarrow_p 0. \label{eq-sineq:fit}
	\end{equation}
	In addition, if the coefficient parameter does not exhibit group heterogeneity, that is, if $\beta_g=\beta_{\tilde g}$ for every $g\neq \tilde g$, we can restrict the parameter space $\mathcal B_N^G$ to $\{(\beta,\cdots,\beta):\beta\in\mathcal B_N\}$ accordingly and replace Assumption \ref{ass-sineq}\ref{ass-sineq:wkdep-xe} by a weaker $(NT)^{-1}
	\sum_{i=1}^N\sum_{t=1}^Tx_{it}e_{it}'=o_p(1)$.
	\item \label{lem-sineq:beta-cnst} Suppose that $(\hat\beta,\hat\alpha,\hat\gamma)$ satisfies equation \eqref{eq-sineq:fit}. Then, as $N$ and $T$ tend to infinity,
	\begin{enumerate}[label=(\roman*)]
		\item if Assumption \ref{ass-sineq}\ref{ass-sineq:relev-ss} holds with $J_{\gamma,g,\tilde g}=\{1,\dots,d_{x,N},d_{x,N}+1,\dots d_{x,N}+T\}$,
		\begin{align*}
			\max_{g\in\mathbb G}\biggl\{\|\beta_g^0-\hat\beta_{\sigma(g)}\|^2 + \frac{1}{T}\sum_{t=1}^T\|\alpha_{gt}^0-\hat\alpha_{\sigma(g) t}\|^2\biggr\}
		= o_p(1).
		\end{align*}
		If Assumptions \ref{ass-sineq}\ref{ass-sineq:grpsep-ss}--\ref{ass-sineq:finmt-z} additionally hold, $(\hat\beta,\hat\alpha)$ is consistent for $(\beta^0,\alpha^0)$.\footnote{Specifically, their Hausdorff distance tends to zero in probability, i.e.,
		\begin{align*}
			\max\biggl\{\max_{g\in\mathbb G}\min_{\tilde g\in\mathbb G}\biggl\{\|\beta_g^0-\hat\beta_{\tilde g}\|^2 + \frac{1}{T}\sum_{t=1}^T\|\alpha_{gt}^0-\hat\alpha_{\tilde g t}\|^2\biggr\},\max_{\tilde g\in\mathbb G}\min_{g\in\mathbb G}\biggl\{\|\beta_g^0-\hat\beta_{\tilde g}\|^2 + \frac{1}{T}\sum_{t=1}^T\|\alpha_{gt}^0-\hat\alpha_{\tilde g t}\|^2\biggr\}\biggr\}\rightarrow_p 0.	
		\end{align*}}
		\item if Assumption \ref{ass-sineq}\ref{ass-sineq:relev-ss} holds with just $J_{\gamma,g,\tilde g}\supseteq \{d_{x,N}+1,\dots,d_{x,N}+T\}$,
		\begin{align*}
			\max_{g\in\mathbb G}\frac{1}{T}\sum_{t=1}^T\|\alpha_{gt}^0-\hat\alpha_{\sigma(g)t}\|^2 = o_p(1).
		\end{align*}
		If Assumption \ref{ass-sineq}\ref{ass-sineq:grpsep-ss} additionally holds, $\hat\alpha$ is consistent for $\alpha^0$.\footnote{Likewise, we mean
		\begin{align*}
			\max\biggl\{\max_{g\in\mathbb G}\min_{\tilde g\in\mathbb G}\frac{1}{T}\sum_{t=1}^T\|\alpha_{gt}^0-\hat\alpha_{\tilde g t}\|^2, \max_{\tilde g\in\mathbb G}\min_{g\in\mathbb G}\frac{1}{T}\sum_{t=1}^T\|\alpha_{gt}^0-\hat\alpha_{\tilde g t}\|^2\biggr\}\rightarrow_p 0.
		\end{align*}}
	\end{enumerate}
	\item \label{lem-sineq:gamma-cnst} Suppose that Assumptions \ref{ass-sineq}\ref{ass-sineq:cpt-param} and \ref{ass-sineq:grpsep-ss}--\ref{ass-sineq:finmt-z} hold. Furthermore, suppose that $(\hat\beta,\hat\alpha)$ is consistent for $(\beta^0,\alpha^0)$. Then, as $N$ and $T$ tend to infinity, 
	\begin{enumerate}[label=(\roman*)]
		\item with probability approaching one, $\sigma$ is bijective. Thus, without loss of generality, we may assume that $\sigma$ is an identity by relabeling $(\hat\beta,\hat\alpha)$.
		\item the probability of misclassification is bounded by	
		\begin{align*}
			&\Pr\biggl(\max_{i\in\{1,\dots,N\}}|\hat g_i - g_i^0|>0\biggr)\\
			\leq& \Pr\biggl(\max_{i\in\{1,\dots,N\}}\biggl(\frac{1}{T}\sum_{t=1}^T\|y_{it}'-x_{it}'\hat\beta_{\hat g_i}-\hat\alpha_{\hat g_i t}'\|^2-\min_{g\in\mathbb G}\frac{1}{T}\sum_{t=1}^T\|y_{it}'-x_{it}'\hat\beta_g-\hat\alpha_{gt}'\|^2\biggr)\gtrsim d\biggr)\\
			& + \Pr\biggl(\min_{i\in\{1,\dots,N\}}\min_{g\neq \tilde g} \frac{1}{T}\sum_{t=1}^T\|x_{it}'(\beta_g^0-\beta_{\tilde g}^0) + (\alpha_{g t}^{0\prime} - \alpha_{\tilde g t}^{0\prime})\|^2<\frac{d}{2}\biggr)\\
			& + \Pr\biggl(\max_{i\in\{1,\dots,N\}}\max_{g\neq \tilde g}\biggl|\frac{1}{T}\sum_{t=1}^Te_{it}'(\alpha_{g t}^{0\prime} - \alpha_{\tilde g t}^{0\prime})'\biggr|\gtrsim d\biggr)\\
			& + 2\Pr\biggl(\max_{i\in\{1,\dots,N\}}\biggl\|\frac{1}{T}\sum_{t=1}^Tx_{it}x_{it}'\biggr\|_2> L\biggr) + \Pr\biggl(\max_{i\in\{1,\dots,N\}}\biggl\|\frac{1}{T}\sum_{t=1}^Tx_{it}e_{it}'\biggr\|\gtrsim d \biggr)\\
			& + \Pr\biggl(\biggl(\max_{g\in\mathbb G}\frac{1}{T}\sum_{t=1}^T\|\hat\alpha_{gt}-\alpha_{gt}^0\|^2\biggr)^{\frac{1}{2}}\biggl(\max_{i\in\{1,\dots,N\}}\frac{1}{T}\sum_{t=1}^T\|e_{it}\|^2\biggr)^{\frac{1}{2}}\gtrsim d\biggr) + o(1),
		\end{align*}
		for any positive constants $d$ and $L$.
	\end{enumerate} 
\end{enumerate}
\end{lemma}
In Lemma \ref{lem-sineq}, the group membership estimates $\hat g_i$'s do not necessarily minimize unit-specific sum of squared errors, that is, we allow $\hat g_i\neq \arg\min_{g\in\mathbb G}T^{-1}\sum_{t=1}^T\|y_{it}'-x_{it}'\hat\beta_g-\hat\alpha_{gt}'\|^2$. This slackness turns out to be useful for extending the results of Lemma \ref{lem-sineq} to the case of the TGFE estimator. Meanwhile, for (ii) of Lemma \ref{lem-sineq:gamma-cnst}, since $\hat\alpha$ is consistent for $\alpha^0$ by assumption, the sixth term can be bounded by
\begin{align*}
	o(1) + \Pr\biggl(\max_{i\in\{1,\dots,N\}}\frac{1}{T}\sum_{t=1}^T\|e_{it}\|^2\geq L\biggr).
\end{align*}
However, we did not take this additional step, because keeping the term $\max_gT^{-1}\sum_{t=1}^T\|\hat\alpha_{gt}-\alpha_{gt}^0\|^2$ enables us to obtain a sharper bound when
applied to models without fixed effects. 

Now, we proceed to examine the two-stage linear model as presented in the main text:
\begin{align*}
	\underset{1\times 1}{y_{it}} =& \underset{\mathclap{1\times d}}{x_{it}'}\overset{\mathclap{d\times 1}}{\beta_{g_i^0}^0} + \underset{1\times 1}{\alpha_{g_i^0t}^0} + \underset{1\times 1}{u_{it}}\\
	\underset{d\times 1}{x_{it}} =& \overset{d\times m_N}{\Pi_{k_i^0}^{0\prime}}\underset{\mathclap{m_N\times 1}}{z_{it}} + \underset{d\times 1}{\mu_{k_i^0 t}^0}  + \underset{d\times 1}{v_{it}}.
\end{align*}
Here, the dimension of instruments can increase as $N$ diverges. This allows us to incorporate the two separate scenarios---group-specific and unit-specific first stages---in the main text into a unified framework. Specifically, the unit-specific case can be understood as a homogeneous linear model via the following reformulation:
\begin{align*}
	x_{it} = \begin{pmatrix}
		\Pi_1^0 & \cdots & \Pi_N^0		
	\end{pmatrix}'\begin{pmatrix}
		\mathbf 1\{1=i\}z_{it}\\
		\vdots\\
		\mathbf 1\{N=i\}z_{it}
	\end{pmatrix} + \mu_{k_i^0 t}^0 + v_{it}.
\end{align*}

We make the following assumptions in terms of the variables forming this model:
\begin{assumption} \label{ass-ts} 
For some constant $M>0$,
\begin{enumerate}[label=(\alph*)]
	\item \label{ass-ts:cpt-param} Let $\mathbf \Pi_N$, $\mathcal M$, $\mathcal B$, and $\mathcal A$ be compact sets in $\mathbb R^{m_N\times d}$, $\mathbb R^d$, $\mathbb R^d$, and $\mathbb R$ respectively. For all $k\in\mathbb K$, $\Pi_k^0\in\mathbf \Pi_N$ and $\{\mu_{kt}^0\}_{t=1}^T\in \mathcal M^T$, and for all $g\in\mathbb G$, $\beta_g^0\in\mathcal B$ and $\{\alpha_{gt}\}_{t=1}^T\in\mathcal A^T$.  Moreover, $\mathbf \Pi_N$'s are uniformly bounded.
	\item \label{ass-ts:wkdep-zv} $\Exp[(NT)^{-1}\sum_{i=1}^N\|\sum_{t=1}^Tz_{it}v_{it}'\|^2]\leq M$.
	\item \label{ass-ts:wkcdep-v} $(NT)^{-1}\sum_{i=1}^N\sum_{j=1}^N\sum_{t=1}^T|\Exp[v_{it}v_{jt}]|\leq M$
	\item \label{ass-ts:wkdep-v} $|(N^2T)^{-1}\sum_{i=1}^N\sum_{j=1}^N\sum_{t=1}^T\sum_{s=1}^T\cov(v_{it}v_{jt},v_{is}v_{js})|\leq M$.
	\item \label{ass-ts:wkdep-zu} $\Exp[(NT)^{-1}\sum_{i=1}^N\|\sum_{t=1}^T(\Pi_{k_i^0}^{0\prime}z_{it})u_{it}\|^2]\lesssim M$.
	\item \label{ass-ts:wkcdep-u} $(NT)^{-1}\sum_{i=1}^N\sum_{j=1}^N\sum_{t=1}^T|\Exp[u_{it}u_{jt}]|\leq M$
	\item \label{ass-ts:wkdep-u} $|(N^2T)^{-1}\sum_{i=1}^N\sum_{j=1}^N\sum_{t=1}^T\sum_{s=1}^T\cov(u_{it}u_{jt},u_{is}u_{js})|\leq M$.
	\item \label{ass-ts:relev-ss} Let $\mathbf{\tilde x}_{it}\equiv ((\Pi_{k_i^0}^{0\prime}z_{it} + \mu_{k_i^0t}^0)',\sqrt T d1_t,\dots,\sqrt T dT_t)$, where $d\tau_t \equiv \mathbf 1\{t=\tau\}$, so that
	\begin{align*}
		&\frac{1}{T}\sum_{t=1}^T\mathbf{\tilde x}_{it}\mathbf{\tilde x}_{it}'\\
		=& \begin{pmatrix}
	 		\frac{1}{T}\sum_{t=1}^T(\Pi_{k_i^0}^{0\prime}z_{it} + \mu_{k_i^0t}^0)(\Pi_{k_i^0}^{0\prime}z_{it} + \mu_{k_i^0t}^0)'	 & 	\frac{1}{\sqrt T}(\Pi_{k_i^0}^{0\prime}z_{i0} + \mu_{k_i^01}^0) & \cdots & \frac{1}{\sqrt T}(\Pi_{k_i^0}^{0\prime}z_{iT} + \mu_{k_i^0T}^0)\\
	 		\frac{1}{\sqrt T}(\Pi_{k_i^0}^{0\prime}z_{i1} + \mu_{k_i^01}^0)' & 1 & \cdots & 0\\
	 		\vdots & \vdots & \ddots & \vdots\\
	 		\frac{1}{\sqrt T}(\Pi_{k_i^0}^{0\prime}z_{iT} + \mu_{k_i^0T}^0)' & 0 & \cdots & 1\\
		\end{pmatrix}.
	\end{align*}
	There exists $\underline \rho^\tss$, which depends on data, such that $\min_{\gamma\in\mathbb G^N}\max_{\tilde g\in\mathbb G}\rho^\tss(\gamma,g,\tilde g)\geq \underline \rho^\tss $, for all $g\in\mathbb G$, where $\rho^\tss(\gamma,g,\tilde g)$ is the smallest eigenvalue of
	\begin{align*}
		M^\tss(\gamma,g,\tilde g)\equiv\frac{1}{N}\sum_{i=1}^N\mathbf 1\{g_i^0=g\}\mathbf 1\{g_i=\tilde g\}\frac{1}{T}\sum_{t=1}^T\mathbf{\tilde x}_{it}\mathbf{\tilde x}_{it}',
	\end{align*}
	and $\underline \rho^\tss\rightarrow \rho^{\tss*}>0$ as $N$ and $T$ tend to infinity.
	\item \label{ass-ts:grpsep-ss} There exists $c^\tss>0$ such that 
	\begin{align*}
		\plim_{N,T\rightarrow \infty}\frac{1}{N}\sum_{i=1}^N\min_{g\neq \tilde g}D_{i,g,\tilde g}^\tss >c^\tss,
	\end{align*}
	where $D_{i,g,\tilde g}^\tss \equiv T^{-1}\sum_{t=1}^T((z_{it}'\Pi_{k_i^0}^0 + \mu_{k_i^0t}^0)(\beta_g^0-\beta_{\tilde g}^0) + (\alpha_{gt}^0-\alpha_{\tilde gt}^0))^2$.
	\item \label{ass-ts:finmt-z} $\Exp[\|(NT)^{-1}\sum_{i=1}^N\sum_{t=1}^T(\Pi_{k_i^0}^{0\prime}z_{it})(\Pi_{k_i^0}^{0\prime}z_{it})'\|_2] \lesssim M$.
\end{enumerate}	
\end{assumption}

\begin{lemma}\label{lem-ts} Suppose that Assumption \ref{ass-ts} holds. Then, as $N$ and $T$ tend to infinity, the following results hold.
\begin{enumerate}[label=(\alph*),ref=\ref{lem-ts}(\alph*)]
	\item \label{lem-ts:beta-cnst} The TGFE estimator is consistent, that is, $d_H((\hat\beta^\tgfe,\hat\alpha^\tgfe),(\beta^0,\alpha^0))\rightarrow_p 0$. In addition, if the first-stage coefficient parameter does not exhibit group heterogeneity, that is, if $\Pi_k^0=\Pi_{\tilde k}^0$ for every $k\neq \tilde k$, we can restrict the parameter space $\mathbf\Pi_N^{K}$ to $\{(\Pi,\cdots,\Pi):\Pi\in\mathbf \Pi_N\}$ accordingly and replace Assumption \ref{ass-ts}\ref{ass-ts:wkdep-zv} by a weaker set of assumptions: $\Exp[\|(NT)^{-1}\sum_{i=1}^N\sum_{t=1}^Tz_{it}v_{it}'\|^2]\leq M$ and $\Exp[(NT)^{-1}\sum_{i=1}^N\|\sum_{t=1}^T(\Pi_{k_i^0}^{0\prime}z_{it})v_{it}'\|^2]\leq M$.
	\item \label{lem-ts:gamma-cnst} For small enough $d>0$, and for large enough $\tilde L>0$, the probability of misclassification is bounded by
	\begin{align}
		&\Pr\biggl(\max_{i\in\{1,\dots,N\}}|\hat g_i^\tgfe-g_i^0|>0\biggr) \nonumber\\
		\leq &3\Pr\biggl(\max_{i\in\{1,\dots,N\}}\biggl\|\frac{1}{T}\sum_{t=1}^Tz_{it}z_{it}'\biggr\|_2\gtrsim \tilde L\biggr) \nonumber \\
		& + 3\Pr\biggl(\max_{i\in\{1,\dots,N\}}\frac{1}{T}\sum_{t=1}^T\|(\Pi_{k_i^0}^0-\hat\Pi_{\hat k_i})'z_{it}+(\mu_{k_i^0t}^0-\hat\mu_{\hat k_it})\|^2\gtrsim d\biggr) \nonumber\\ 
		& + \Pr\biggl(\max_{i\in\{1,\dots,N\}}\biggl\|\frac{1}{T}\sum_{t=1}^T(v_{it}'\beta_{g_i^0}^0+u_{it})\mathbf{\tilde x}_{it}\biggr\|_2\gtrsim \tilde L\biggr) \nonumber\\
		& + \Pr\biggl(\max_{i\in\{1,\dots,N\}}\biggl\|\frac{1}{T}\sum_{t=1}^T(v_{it}'\beta_{g_i^0}^0+u_{it})((\Pi_{k_i^0}^0-\hat\Pi_{\hat k_i})'z_{it}+(\mu_{k_i^0t}^0-\hat\mu_{\hat k_it}))'\biggr\|\gtrsim d\biggr) \nonumber\\ 
		& + \Pr\biggl(\min_{i\in\{1,\dots,N\}}\min_{g\neq \tilde g}\frac{1}{T}\sum_{t=1}^T\|(\Pi_{k_i^0}^{0\prime}z_{it}+\mu_{k_i^0t}^0)'(\beta_g^0-\beta_{\tilde g}^0) + (\alpha_{gt}^0-\alpha_{\tilde gt}^0)\|^2<\frac{d}{2}\biggr) \nonumber\\
		& + \Pr\biggl(\max_{i\in\{1,\dots,N\}}\max_{g\neq g}\biggl|\frac{1}{T}\sum_{t=1}^Tv_{it}(\alpha_{g t}^0-\alpha_{\tilde g t}^0)\biggr|\gtrsim d \biggr) \nonumber \\
		&+ \Pr\biggl(\max_{i\in\{1,\dots,N\}}\max_{g\neq g}\biggl|\frac{1}{T}\sum_{t=1}^Tu_{it}(\alpha_{g t}^0-\alpha_{\tilde g t}^0)\biggr|\gtrsim d \biggr) \nonumber\\
		& + \Pr\biggl(\max_{i\in\{1,\dots,N\}}\biggl\|\frac{1}{T}\sum_{t=1}^Tz_{it}v_{it}'\biggr\|\gtrsim d\biggr)
		+ \Pr\biggl(\max_{i\in\{1,\dots,N\}}\biggl\|\frac{1}{T}\sum_{t=1}^Tz_{it}u_{it}\biggr\|\gtrsim d\biggr) \nonumber\\
		& + \Pr\biggl(\biggl(\min_{g\in\mathbb G}\frac{1}{T}\sum_{t=1}^T\|\tilde\alpha_{gt}^\tgfe-\alpha_{gt}^0\|^2\biggr)^{\frac{1}{2}}\biggl(\max_{i\in\{1,\dots,N\}}\frac{1}{T}\sum_{t=1}^T\|v_{it}\|^2\biggr)^{\frac{1}{2}}\gtrsim d\biggr) \nonumber\\
		& + \Pr\biggl(\biggl(\min_{g\in\mathbb G}\frac{1}{T}\sum_{t=1}^T\|\tilde\alpha_{gt}^\tgfe-\alpha_{gt}^0\|^2\biggr)^{\frac{1}{2}}\biggl(\max_{i\in\{1,\dots,N\}}\frac{1}{T}\sum_{t=1}^T|u_{it}|^2\biggr)^{\frac{1}{2}}\gtrsim d\biggr) \nonumber\\
		& + \Pr\biggl(\max_{i\in\{1,\dots,N\}}\biggl\|\frac{1}{T}\sum_{t=1}^T\mu_{k_i^0t}^0(v_{it}'\beta_{g_i^0}^0+u_{it})\biggr\|\gtrsim d\biggr) + o(1),
		\label{eq-ts:gamma-cnst}
	\end{align}
	where $\tilde\alpha_{gt}^{\tgfe}$ denotes an infeasible estimator for $\alpha_{gt}^0$, which we will define in the proof.
\end{enumerate}
\end{lemma}

Lastly, we restate the crucial lemma of \textcite{BonhommeManresa2015}, which was key to determining the orders of several concentration probabilities that were needed to determine the overall order of the probability of misclassification. This lemma is a direct consequence of equation (1.7) of \textcite{MerlevedePeligradRio2011}.
\begin{lemma}[Lemma B.5 of \textcite{BonhommeManresa2015}] \label{lem-bm}
	Let $(w_t)_t$ be a strongly mixing process with zero mean, with strong mixing coefficient $\alpha[t]\leq \exp(-at^{d_1})$, and with exponentially decaying tail mass $\Pr(|w_t|>w)=\exp(1-(w/b)^{d_2})$, where $a$, $b$, $d_1$, and $d_2$ are positive constants. Then, for all $w,\delta>0$, as $T$ tends to infinity, 
	\begin{align*}
		T^{\delta}\Pr\biggl(\biggl|\frac{1}{T}\sum_{t=1}^Tw_t\biggr|\geq w\biggr)\rightarrow 0.
	\end{align*}
\end{lemma}

\subsection{Proofs of lemmas}
\subsubsection{Proof of Lemma \ref{lem-sineq:fit}}
Let $Q(\beta,\alpha,\gamma) \equiv (NT)^{-1}\sum_{i=1}^N\sum_{t=1}^T\|y_{it}'-x_{it}'\beta_{g_i}-\alpha_{g_i t}'\|^2$ be the minimand of the GFE estimation, and $\tilde Q(\beta,\gamma) \equiv (NT)^{-1}\sum_{i=1}^N\sum_{t=1}^T\|x_{it}'(\beta_{g_i^0}^0-\beta_{g_i})+(\alpha_{g_i^0}^{0\prime}-\alpha_{g_i}')\|^2 + (NT)^{-1}\sum_{i=1}^N\sum_{t=1}^T\|e_{it}\|^2$ be its auxiliary counterpart. The difference between these two objectives is bounded by
\begin{align*}
	&|Q(\beta,\alpha,\gamma)-\tilde Q(\beta,\alpha,\gamma)|\\ 
	=& \biggl|\frac{2}{NT}\sum_{i=1}^N\sum_{t=1}^Te_{it}'(x_{it}'(\beta_{g_i^0}^0-\beta_{g_i})+(\alpha_{g_i^0t}^{0\prime}-\alpha_{g_it}'))'\biggr|\\
	\leq& 	\biggl|\tr\biggl(\biggl(\frac{1}{N}\sum_{i=1}^N(\beta_{g_i^0}^0-\beta_{g_i})\biggr)'\frac{2}{NT}\sum_{i=1}^N\sum_{t=1}^Tx_{it}e_{it}'\biggr)\biggr|\\
	&+ \left|\sum_{g=1}^G\sum_{\tilde g=1}^G\tr\left(\begin{pmatrix}
		(\beta_g^0-\beta_{\tilde g})-\frac{1}{N}\sum_{i=1}^N(\beta_{g_i^0}^0-\beta_{g_i})\\
		\frac{1}{\sqrt T}(\alpha_{g1}^{0\prime}-\alpha_{\tilde g1}')\\
		\vdots\\
		\frac{1}{\sqrt T}(\alpha_{gT}^{0\prime}-\alpha_{\tilde gT}')
	\end{pmatrix}'\frac{2}{N}\sum_{i=1}^N\mathbf 1\{g_i^0=g\}\mathbf 1\{g_i=\tilde g\}\sum_{t=1}^T\mathbf x_{it}e_{it}'\right)\right|\\
	\leq& \biggl\|\frac{1}{NT}\sum_{i=1}^N\sum_{t=1}^Tx_{it}e_{it}'\biggr\| \cdot\max_{\gamma\in\mathbb G^N}\max_{\beta\in\mathcal B_N^G}\biggl\|\frac{1}{N}\sum_{i=1}^N(\beta_{g_i^0}^0-\beta_{g_i})\biggr\|\\
	& + \max_{\gamma\in\mathbb G^N}\max_{(g,\tilde g)\in\mathbb G^2}\biggl\|\frac{1}{NT}\sum_{i=1}^N\mathbf 1\{g_i^0=g\}\mathbf 1\{g_i=\tilde g\}\sum_{t=1}^Tx_{it}e_{it}'\biggr\| \\
	& \quad \times
		 \max_{\gamma\in\mathbb G^N}\max_{\beta\in\mathcal B_N^G}\sum_{g=1}^G\sum_{\tilde g=1}^G\biggl\|(\beta_g^0-\beta_{\tilde g})-\frac{1}{N}\sum_{i=1}^N(\beta_{g_i^0}^0-\beta_{g_i})\biggr\|\\
	& + \max_{\gamma\in\mathbb G^N}\max_{(g,\tilde g)\in\mathbb G^2}\left\|\frac{1}{NT}\sum_{i=1}^N\mathbf 1\{g_i^0=g\}\mathbf 1\{g_i=\tilde g\}\sum_{t=1}^T\begin{pmatrix}
			\sqrt T d1_t\\
			\vdots\\
			\sqrt T dT_t
		\end{pmatrix}e_{it}'\right\| \\
		& \quad \times \max_{\gamma\in\mathbb G^N}\max_{\alpha\in\mathcal A^{TG}}\sum_{g=1}^G\sum_{\tilde g=1}^G\left\|\begin{pmatrix}
		\frac{1}{\sqrt T}(\alpha_{g1}^{0\prime}-\alpha_{\tilde g1}')\\
		\vdots\\
		\frac{1}{\sqrt T}(\alpha_{gT}^{0\prime}-\alpha_{\tilde gT}')
	\end{pmatrix}\right\|.
\end{align*} 
Note that
\begin{align*}
&	\left\|\frac{1}{NT}\sum_{i=1}^N\mathbf 1\{g_i^0=g\}\mathbf 1\{g_i=\tilde g\}\sum_{t=1}^T\begin{pmatrix}
		\sqrt T d1_t\\
		\vdots\\
		\sqrt T dT_t
	\end{pmatrix}e_{it}'\right\| \\
	=& \left\|\frac{1}{\sqrt T}\begin{pmatrix}
		\frac{1}{N}\sum_{i=1}^N\mathbf 1\{g_i^0=g\}\mathbf 1\{g_i=\tilde g\}e_{i1}\\
		\vdots\\
		\frac{1}{N}\sum_{i=1}^N\mathbf 1\{g_i^0=g\}\mathbf 1\{g_i=\tilde g\}e_{iT}
	\end{pmatrix}\right\|
	\leq \biggl(\frac{1}{T}\sum_{t=1}^T\biggl\|\frac{1}{N}\sum_{i=1}^N\mathbf 1\{g_i^0=g\}\mathbf 1\{g_i=\tilde g\}e_{it}\biggr\|^2\biggr)^{\frac{1}{2}}.
\end{align*}
It follows from Assumption \ref{ass-sineq}\ref{ass-sineq:cpt-param}--\ref{ass-sineq:wkdep-e} that the last equation is $o_p(1)$. Since it does not depend on any parameter, we have the uniform convergence result:
\begin{align*}
	\sup_{(\beta,\alpha,\gamma)\in\mathcal B\times\mathcal A\times \mathbb G^N}|Q(\beta,\alpha,\gamma)-\tilde Q(\beta,\alpha,\gamma)|=o_p(1).
\end{align*}
Combining this with equation \eqref{eq-sineq:min} yields:
\begin{align*}
	&\frac{1}{NT}\sum_{i=1}^N\sum_{t=1}^T\|x_{it}'(\beta_{g_i^0}^0-\hat\beta_{\hat g_i}) + (\alpha_{g_i^0 t}^{0\prime}-\hat\alpha_{\hat g_i t}')\|^2\\
	=& \tilde Q(\hat\beta,\hat\alpha,\hat\gamma) - \tilde Q(\beta^0,\alpha^0,\gamma^0)
	\vphantom{\inf_{(\beta,\alpha,\gamma)\in\mathcal B^G\times\mathcal A^G\times \mathbb G^N}}\\
	=& \tilde Q(\hat\beta,\hat\alpha,\hat\gamma) -  Q(\hat\beta,\hat\alpha,\hat\gamma) + Q(\hat\beta,\hat\alpha,\hat\gamma) - \inf_{(\beta,\alpha,\gamma)\in\mathcal B^G\times\mathcal A^G\times \mathbb G^N}Q(\beta,\alpha,\gamma)\\
	& + \inf_{(\beta,\alpha,\gamma)\in\mathcal B^G\times\mathcal A^G\times \mathbb G^N} Q(\beta,\alpha,\gamma) - Q(\beta^0,\alpha^0,\gamma^0) + Q(\beta^0,\alpha^0,\gamma^0) - \tilde Q(\beta^0,\alpha^0,\gamma^0)\\
	\leq& 2\sup_{(\beta,\alpha,\gamma)\in\mathcal B\times\mathcal A\times \mathbb G^N}|Q(\beta,\alpha,\gamma)-\tilde Q(\beta,\alpha,\gamma)| + o_p(1) = o_p(1).
\end{align*}

If the coefficient parameter is assumed to be homogeneous across groups, and if we have restricted its space to $\{(\beta,\cdots,\beta):\beta\in\mathcal B_N\}$, then in our previous bound analysis, the term 
\begin{align*}
	 \sum_{g=1}^G\sum_{\tilde g=1}^G\biggl\|(\beta_g^0-\beta_{\tilde g})-\frac{1}{N}\sum_{i=1}^N(\beta_{g_i^0}^0-\beta_{\tilde g_i})\biggr\|	
\end{align*}
reduces to zero, and thus $\|(NT)^{-1}\sum_{i=1}^N\sum_{t=1}^Tx_{it}e_{it}'\|=o_p(1)$ will suffice.

\subsubsection{Proof of Lemma \ref{lem-sineq:beta-cnst}}
By Assumption \ref{ass-sineq}\ref{ass-sineq:relev-ss}, as $N$ and $T$ tend to infinity,
\begin{align}
	&\tilde Q(\beta,\alpha,\gamma) - \tilde Q(\beta^0,\alpha^0,\gamma^0) \nonumber\\ 
	=& \frac{1}{NT}\sum_{i=1}^N\sum_{t=1}^T\|x_{it}'(\beta_{g_i^0}^0-\beta_{g_i})+(\alpha_{g_i^0}^{0\prime}-\alpha_{g_i}')\|^2 \nonumber\\
	=& \sum_{g=1}^G\sum_{\tilde g=1}^G\tr\left(\begin{pmatrix}
		\beta_g^0-\beta_{\tilde g}\\ \frac{1}{\sqrt T}(\alpha_{g1}^{0\prime}-\alpha_{\tilde g1}')\\ \vdots\\ \frac{1}{\sqrt T}(\alpha_{gT}^{0\prime}-\alpha_{\tilde gT}')
	\end{pmatrix}\begin{pmatrix}
		\beta_g^0-\beta_{\tilde g}\\ \frac{1}{\sqrt T}(\alpha_{g1}^{0\prime}-\alpha_{\tilde g1}')\\ \vdots\\ \frac{1}{\sqrt T}(\alpha_{gT}^{0\prime}-\alpha_{\tilde gT}')
	\end{pmatrix}'\biggl(\frac{1}{N}\sum_{i=1}^N\mathbf 1\{g_i^0=g\}\mathbf 1\{g_i=\tilde g\}\frac{1}{T}\sum_{t=1}^T\mathbf x_{it}\mathbf x_{it}'\biggr)\right) \nonumber\\
	\geq& \sum_{g=1}^G\sum_{\tilde g=1}^G\rho(\gamma,g,\tilde g)\biggl(\|\beta_g^0-\beta_{\tilde g}\|^2 + \frac{1}{T}\sum_{t=1}^T\|\alpha_{gt}^0-\alpha_{\tilde g t}\|^2\biggr) \label{eq-sineq:beta-cnst:hd-tfe}\\
	\geq& \sum_{g=1}^G\max_{\tilde g\in\mathbb G}\rho(\gamma,g,\tilde g) \min_{\tilde g\in\mathbb G}\biggl(\|\beta_g^0-\beta_{\tilde g}\|^2 + \frac{1}{T}\sum_{t=1}^T\|\alpha_{gt}^0-\alpha_{\tilde g t}\|^2\biggr) \nonumber\\
	\geq& \sum_{g=1}^G\min_{\gamma\in\mathbb G^N}\max_{\tilde g\in\mathbb G}\rho(\gamma,g,\tilde g)\min_{\tilde g\in\mathbb G}\biggl(\|\beta_g^0-\beta_{\tilde g}\|^2 + \frac{1}{T}\sum_{t=1}^T\|\alpha_{gt}^0-\alpha_{\tilde g t}\|^2\biggr) \nonumber\\
	\geq& \underline\rho\max_{g\in\mathbb G}\min_{\tilde g\in\mathbb G}\biggl(\|\beta_g^0-\beta_{\tilde g}\|^2 + \frac{1}{T}\sum_{t=1}^T\|\alpha_{gt}^0-\alpha_{\tilde g t}\|^2\biggr) \nonumber
\end{align}
with probability approaching one. It follows from Lemma \ref{ass-sineq}\ref{ass-sineq:cpt-param} that
\begin{align*}
	&\underline\rho \max_{g\in\mathbb G}\min_{\tilde g\in\mathbb G}\biggl(\|\beta_g^0-\hat\beta_{\tilde g}\|^2 + \frac{1}{T}\sum_{t=1}^T\|\alpha_{gt}^0-\hat\alpha_{\tilde g t}\|^2\biggr)\\
	&\leq \tilde Q(\hat\beta,\hat\alpha,\hat\gamma) - \tilde Q(\beta^0,\alpha^0,\gamma^0)
	= \frac{1}{NT}\sum_{i=1}^N\sum_{t=1}^T\|x_{it}'(\beta_{g_i^0}^0-\hat\beta_{\hat g_i}) + (\alpha_{g_i^0 t}^{0\prime} - \hat\alpha_{\hat g_i t}')\|^2 = o_p(1).
\end{align*}
Invoking Assumption \ref{ass-sineq}\ref{ass-sineq:relev-ss} once more, we obtain 
\begin{align}
	\max_{g\in\mathbb G}\min_{\tilde g\in\mathbb G}\biggl(\|\beta_g^0-\hat\beta_{\tilde g}\|^2 + \frac{1}{T}\sum_{t=1}^T\|\alpha_{gt}^0-\hat\alpha_{\tilde g t}\|^2\biggr)
	= o_p(1).\label{eq-sineq:beta-cnst:hd}
\end{align}

For any $g\neq \tilde g$,
\begin{align}
	&\biggl(\frac{1}{NT}\sum_{i=1}^N\sum_{t=1}^T\|x_{it}'(\hat\beta_{\sigma (g)}-\hat\beta_{\sigma(\tilde g)} ) + (\hat\alpha_{\sigma(g)t}' - \hat\alpha_{\sigma(\tilde g) t}')\|^2\biggr)^{\frac{1}{2}} \nonumber\\
	\geq&  - \biggl(\frac{1}{NT}\sum_{i=1}^N\sum_{t=1}^T\|x_{it}'(\beta_{g}^0-\hat\beta_{\sigma(g)})+ (\alpha_{g t}^{0\prime} - \hat\alpha_{\sigma(g) t}')\|^2\biggr)^{\frac{1}{2}} \nonumber\\
	&- \biggl(\frac{1}{NT}\sum_{i=1}^N\sum_{t=1}^T\|x_{it}'(\beta_{\tilde g}^0-\hat\beta_{\sigma(\tilde g)}) + (\alpha_{\tilde g t}^{0\prime} - \hat\alpha_{\sigma(\tilde g) t}')\|^2\biggr)^{\frac{1}{2}} \nonumber\\
	&+ \biggl(\frac{1}{NT}\sum_{i=1}^N\sum_{t=1}^T\|x_{it}'(\beta_g^0-\beta_{\tilde g}^0) + (\alpha_{g t}^{0\prime} - \alpha_{\tilde g t}^{0\prime})\|^2\biggr)^{\frac{1}{2}} \nonumber\\
	\geq& -2\biggl(\max_{g\in\mathbb G}\frac{1}{NT}\sum_{i=1}^N\sum_{t=1}^T\|x_{it}'(\beta_{g}^0-\hat\beta_{\sigma(g)}) + (\alpha_{g t}^{0\prime} - \hat\alpha_{\sigma(g) t}')\|^2\biggr)^{\frac{1}{2}} \nonumber\\
	&+ \biggl(\frac{1}{NT}\sum_{i=1}^N\sum_{t=1}^T\min_{g\neq g}\|x_{it}'(\beta_g^0-\beta_{\tilde g}^0)+ (\alpha_{g t}^{0\prime} - \alpha_{\tilde g t}^{0\prime})\|^2\biggr)^{\frac{1}{2}}, 
	\label{eq-sineq:beta-cnst:bijection}
\end{align}
where the first inequality follows from the Minkowski inequality. The two terms in the last equation neither depend on $g$ nor $\tilde g$. Combining equation \eqref{eq-sineq:beta-cnst:hd} with Assumption \ref{ass-sineq}\ref{ass-sineq:finmt-z},
\begin{align*}
	&\max_{g\in\mathbb G}\frac{1}{NT}\sum_{i=1}^N\sum_{t=1}^T\|x_{it}'(\beta_{g}^0-\hat\beta_{\sigma(g)}) + (\alpha_{g t}^{0\prime} - \hat\alpha_{\sigma(g) t}')\|^2\\
	=& \max_{g\in\mathbb G}\tr\left(\begin{pmatrix}
		\beta_g^0-\beta_{\tilde g}\\ \frac{1}{\sqrt T}(\alpha_{g1}^{0\prime}-\alpha_{\tilde g1}')\\ \vdots\\ \frac{1}{\sqrt T}(\alpha_{gT}^{0\prime}-\alpha_{\tilde gT}')
	\end{pmatrix}'\biggl(\frac{1}{NT}\sum_{i=1}^N\sum_{t=1}^T\mathbf x_{it}\mathbf x_{it}'\biggr)\begin{pmatrix}
		\beta_g^0-\beta_{\tilde g}\\ \frac{1}{\sqrt T}(\alpha_{g1}^{0\prime}-\alpha_{\tilde g1}')\\ \vdots\\ \frac{1}{\sqrt T}(\alpha_{gT}^{0\prime}-\alpha_{\tilde gT}')
	\end{pmatrix}\right)\\
	\leq& \max_{g\in\mathbb G}\left\|\begin{pmatrix}
		\beta_g^0 - \hat\beta_{\sigma(g)}\\
		\frac{1}{\sqrt T}(\alpha_{g 1}^{0\prime} - \hat\alpha_{\sigma(g) 1}')\\
		\vdots\\
		\frac{1}{\sqrt T}(\alpha_{g T}^{0\prime} - \hat\alpha_{\sigma(g) T}')
	\end{pmatrix}\right\|^2\biggl\|\frac{1}{NT}\sum_{i=1}^N\sum_{t=1}^T\mathbf x_{it}\mathbf x_{it}'\biggr\|_2
	= o_p(1)O_p(1)=o_p(1).
\end{align*}
Consequently, by Assumption \ref{ass-sineq}\ref{ass-sineq:grpsep-ss}, as $N$ and $T$ tends to infinity, with probability approaching one, $\sigma$ is a bijection. Hence, for all $\tilde g\in\mathbb G$, with probability approaching one, 
\begin{align*}
	&\min_{g\in\mathbb G}\biggl(\|\beta_g^0-\hat\beta_{\tilde g}\|^2 + \frac{1}{T}\sum_{t=1}^T\|\alpha_{gt}^0-\hat\alpha_{\tilde g t}\|^2\biggr)\\
	\leq& \|\beta_{\sigma^{-1}(\tilde g)}^0-\hat\beta_{\tilde g}\|^2 + \frac{1}{T}\sum_{t=1}^T\|\alpha_{gt}^0-\hat\alpha_{\tilde g t}\|^2 
	= \min_{h\in\mathbb G}\biggl(\|\beta_{\sigma^{-1}(\tilde g)}^0 -\hat\beta_h\|^2 + \frac{1}{T}\sum_{t=1}^T\|\alpha_{\sigma^{-1}(\tilde g)t}^0-\hat\alpha_{h t}\|^2\biggr).
\end{align*}
That is, $\max_{\tilde g\in\mathbb G}\min_{g\in\mathbb G}(\|\beta_g^0-\hat\beta_{\tilde g}\|^2+T^{-1}\sum_{t=1}^T\|\alpha_{gt}^0-\hat\alpha_{\tilde g t}\|^2)=o_p(1)$.

Meanwhile, if Assumption \ref{ass-sineq}\ref{ass-sineq:relev-ss} holds only with $J\supseteq \{d_{x,N}+1,\dots,d_{x,N}+T\}$, then equation \eqref{eq-sineq:beta-cnst:hd-tfe} has to be adjusted to
\begin{align*}
	\sum_{g=1}^G\sum_{\tilde g=1}^G\rho(\gamma, g,\tilde g)\frac{1}{T}\sum_{t=1}^T\|\alpha_{gt}^0-\alpha_{\tilde g t}\|^2.
\end{align*}
Following steps similar to those outlined above, we can establish the consistency of $\hat\alpha$.

\subsubsection{Proof of Lemma \ref{lem-sineq:gamma-cnst}}
We begin with showing that $\sigma$ is bijective. Based on equation \eqref{eq-sineq:beta-cnst:bijection} in the proof of Lemma \ref{lem-sineq:beta-cnst}, it is sufficient to check if the consistency of $(\hat\beta,\hat\alpha)$ implies
\begin{align*}
	\max_{g\in\mathbb G}\frac{1}{NT}\sum_{i=1}^N\sum_{t=1}^T\|x_{it}'(\beta_g^0-\hat\beta_{\sigma(g)})+(\alpha_{gt}^{0\prime}-\hat\alpha_{\sigma(g)t}')\|^2=o_p(1).	
\end{align*}
This directly follows from our previous observation: for all $g\in\mathbb G$,
\begin{align*}
	&\frac{1}{NT}\sum_{i=1}^N\sum_{t=1}^T\|x_{it}'(\beta_g^0-\hat\beta_{\sigma(g)})+(\alpha_{gt}^{0\prime}-\hat\alpha_{\sigma(g)t}')\|^2\\
	\leq& \underbrace{\max_{g\in\mathbb G}\biggl\{\|\beta_g^0-\hat\beta_{\sigma(g)}\|^2 + \frac{1}{T}\sum_{t=1}^T\|\alpha_{gt}^0-\hat\alpha_{\sigma(g)t}\|^2\biggr\}}_{\leq d_H((\hat\beta,\hat\alpha),(\beta^0,\alpha^0))\text{ by definition}}\underbrace{\biggl\|\frac{1}{NT}\sum_{i=1}^N\sum_{t=1}^T\mathbf x_{it}\mathbf x_{it}'\biggr\|_2.}_{\mathclap{=O_p(1)\text{ by Assumption \ref{ass-sineq}\ref{ass-sineq:finmt-z}}}}
\end{align*} 
For this reason, without loss of generality, we henceforth assume that $\sigma$ is set to be an identity by relabeling $(\hat\beta,\hat\alpha)$. In particular, for all $g\in\mathbb G$, with probability approaching one, 
\begin{align*}
	d_H((\hat\beta,\hat\alpha),(\beta^0,\alpha^0))^2	 
	\geq \frac{1}{T}\sum_{t=1}^T\|\alpha_{gt}^0-\hat\alpha_{gt}\|^2.
\end{align*}

Next, we make an observation:
\begin{align*}
	&\frac{1}{T}\sum_{t=1}^T\|y_{it}'-x_{it}'\hat\beta_{g_i^0}-\hat\alpha_{g_i^0 t}'\|^2 - \frac{1}{T}\sum_{t=1}^T\|y_{it}'-x_{it}'\hat\beta_{\hat g_i}-\hat\alpha_{\hat g_it}'\|^2\\
	=& \frac{1}{T}\sum_{t=1}^T\|x_{it}'(\beta_{g_i^0}^0-\hat\beta_{g_i^0}) + (\alpha_{g_i^0 t}^0-\hat\alpha_{g_i^0 t})+e_{it}'\|^2\\
	& - \frac{1}{T}\sum_{t=1}^T\|x_{it}'(\beta_{g_i^0}^0 - \hat\beta_{g_i^0}) + (\alpha_{g_i^0 t}^{0\prime} - \hat\alpha_{g_i^0 t}')+ e_{it}' + x_{it}'(\hat\beta_{g_i^0}-\hat\beta_{\hat g_i}) + (\hat\alpha_{g_i^0 t}' - \hat\alpha_{\hat g_i t}')\|^2\\
	=& -\frac{2}{T}\sum_{t=1}^T(x_{it}'(\beta_{g_i^0}^0-\hat\beta_{g_i^0}) + (\alpha_{g_i^0 t}^{0\prime} - \hat\alpha_{g_i^0 t}') + e_{it}')(x_{it}'(\hat\beta_{g_i^0}-\hat\beta_{\hat g_i}) + (\hat\alpha_{g_i^0 t}' - \hat\alpha_{\hat g_i t}'))'\\
	& - \frac{1}{T}\sum_{t=1}^T\|x_{it}'(\hat\beta_{g_i^0}-\hat\beta_{\hat g_i}) + (\hat\alpha_{g_i^0 t}' - \hat\alpha_{\hat g_i t}')\|^2\\
	=& -\frac{2}{T}\sum_{t=1}^T(x_{it}'(\beta_{g_i^0}^0-\hat\beta_{g_i^0}) + (\alpha_{g_i^0 t}^{0\prime} - \hat\alpha_{g_i^0 t}'))(x_{it}'(\hat\beta_{g_i^0}-\hat\beta_{\hat g_i}) + (\hat\alpha_{g_i^0 t}' - \hat\alpha_{\hat g_i t}'))'\\
	& - \frac{2}{T}\sum_{t=1}^Te_{it}'(x_{it}'(\hat\beta_{g_i^0}-\hat\beta_{\hat g_i}) + (\hat\alpha_{g_i^0 t}' - \hat\alpha_{\hat g_i t}'))'\\
	& - \mathbf 1\{\hat g_i\neq g_i^0\}\frac{1}{T}\sum_{t=1}^T\|x_{it}'(\hat\beta_{g_i^0}-\hat\beta_{\hat g_i}) + (\hat\alpha_{g_i^0 t}' - \hat\alpha_{\hat g_i t}')\|^2.\\
	\leq& \max_{i\in\{1,\dots,N\}} \biggl|\frac{2}{T}\sum_{t=1}^T(x_{it}'(\beta_{g_i^0}^0-\hat\beta_{g_i^0}) + (\alpha_{g_i^0 t}^0 - \hat\alpha_{g_i^0 t}))(x_{it}'(\hat\beta_{g_i^0}-\hat\beta_{\hat g_i}) + (\hat\alpha_{g_i^0 t} - \hat\alpha_{\hat g_i t}))'\biggr|\\
	& + \max_{i\in\{1,\dots,N\}} \biggl|\frac{2}{T}\sum_{t=1}^Te_{it}'(x_{it}'(\hat\beta_{g_i^0}-\hat\beta_{\hat g_i}))'\biggr| 
	+ \max_{i\in\{1,\dots,N\}} \biggl|\frac{2}{T}\sum_{t=1}^Te_{it}'((\hat\alpha_{g_i^0 t}' - \hat\alpha_{\hat g_i t}')-(\alpha_{g_i^0 t}^{0\prime} - \alpha_{\hat g_i t}^{0\prime}))'\biggr|\\
	& + \mathbf 1\{\hat g_i\neq g_i^0\}\max_{g\neq \tilde g}\biggl|\frac{2}{T}\sum_{t=1}^Te_{it}'(\alpha_{g t}^{0\prime}-\alpha_{\tilde g t}^{0\prime})'\biggr| \\
	& - \mathbf 1\{\hat g_i\neq g_i^0\}\min_{g\neq \tilde g} \frac{1}{T}\sum_{t=1}^T\|x_{it}'(\hat\beta_g-\hat\beta_{\tilde g}) + (\hat\alpha_{g t}' - \hat\alpha_{\tilde g t}')\|^2.
\end{align*}
We examine the terms that appear in the last equation. Specifically, we establish bounds for the probabilities of the terms being concentrated around zero so that the bounds closely correspond to the assumptions in the main text. Let $d$ and $L$ be any positive constants.
\begin{itemize}
	\item The first term concentrates around zero with the probability of
	\begin{align*}
		&\Pr\biggl(\max_{i\in\{1,\dots,N\}}\biggl|\frac{2}{T}\sum_{t=1}^T(x_{it}'(\beta_{g_i^0}^0-\hat\beta_{g_i^0}) + (\alpha_{g_i^0 t}^{0\prime} - \hat\alpha_{g_i^0 t}'))(x_{it}'(\hat\beta_{g_i^0}-\hat\beta_{\hat g_i})+ (\hat\alpha_{g_i^0 t}' - \hat\alpha_{\hat g_i t}'))'\biggr|\gtrsim d\biggr)\\
		\leq&\Pr\biggl(\max_{i\in\{1,\dots,N\}}\biggl\|\frac{1}{T}\sum_{t=1}^T\mathbf x_{it}\mathbf x_{it}'\biggr\|_2\biggl(\|\beta_{g_i^0}^0-\hat\beta_{g_i^0}\|^2 + \frac{1}{T}\sum_{t=1}^T\|\alpha_{g_i^0t}^0-\hat\alpha_{g_i^0t}\|^2\biggr)^{\frac{1}{2}}\gtrsim d\biggr)\\
		\leq& \Pr\biggl(\max_{i\in\{1,\dots,N\}}\biggl\|\frac{1}{T}\sum_{t=1}^Tx_{it}x_{it}'\biggr\|_2 > L \biggr) + o(1),
	\end{align*}
	where Assumption \ref{ass-sineq}\ref{ass-sineq:cpt-param} and the consistency of $(\hat\beta,\hat\alpha)$ are used in the second inequality.	
	\item By Assumption \ref{ass-sineq}\ref{ass-sineq:cpt-param}, the second term concentrates around zero with the probability of
	\begin{align*}
		\Pr\biggl(\max_{i\in\{1,\dots,N\}}\biggl|\frac{1}{T}\sum_{t=1}^Te_{it}'(x_{it}'(\hat\beta_{g_i^0}-\hat\beta_{\hat g_i}))'\biggr|\gtrsim d \biggr)
		\leq \Pr\biggl(\max_{i\in\{1,\dots,N\}}\biggl\|\frac{1}{T}\sum_{t=1}^Tx_{it}e_{it}'\biggr\|\gtrsim d\biggr).
	\end{align*}
	\item The third term concentrates around zero with the probability of
	\begin{align*}
		&\Pr\biggl(\max_{i\in\{1,\dots,N\}}\biggl|\frac{1}{T}\sum_{t=1}^Te_{it}'((\hat\alpha_{g_i^0 t}' - \hat\alpha_{\hat g_i t}')-(\alpha_{g_i^0 t}^{0\prime} - \alpha_{\hat g_i t}^{0\prime}))'\biggr|\gtrsim d\biggr)	\\
		\leq& \Pr\biggl(\max_{i\in\{1,\dots,N\}}\biggl(\biggl(\frac{1}{T}\sum_{t=1}^T\|\hat\alpha_{g_i^0t}-\alpha_{g_i^0t}^0\|^2\biggr)^{\frac{1}{2}}+\biggl(\frac{1}{T}\sum_{t=1}^T\|\hat\alpha_{\hat g_it}-\alpha_{\hat g_it}^0\|^2\biggr)^{\frac{1}{2}}\biggr)\biggl(\frac{1}{T}\sum_{t=1}^T\|e_{it}\|^2\biggr)^{\frac{1}{2}}\gtrsim d\biggr)\\
		\leq& \Pr\biggl(\biggl(\max_{g\in\mathbb G}\frac{1}{T}\sum_{t=1}^T\|\hat\alpha_{gt}-\alpha_{gt}^0\|^2\biggr)^{\frac{1}{2}}\biggl(\max_{i\in\{1,\dots,N\}}\frac{1}{T}\sum_{t=1}^T\|e_{it}\|^2\biggr)^{\frac{1}{2}}\gtrsim d\biggr).
	\end{align*}
	\item Lastly, we examine the last term. For a $(d_{x,N}+T)\times (d_{x,N}+T)$ matrix $A$, we define
	\begin{align*}
		h_A:(\beta,\alpha)\mapsto \min_{g\neq \tilde g}\tr\left(\begin{pmatrix}
		\beta_g-\beta_{\tilde g}\\
		\frac{1}{\sqrt T}(\alpha_{g1}' - \alpha_{\tilde g1}')\\
		\vdots\\
		\frac{1}{\sqrt T}(\alpha_{gT}' - \alpha_{\tilde gT}')
	\end{pmatrix}'A\begin{pmatrix}
		\beta_g-\beta_{\tilde g}\\
		\frac{1}{\sqrt T}(\alpha_{g1}' - \alpha_{\tilde g1}')\\
		\vdots\\
		\frac{1}{\sqrt T}(\alpha_{gT}' - \alpha_{\tilde gT}')
	\end{pmatrix}\right),
	\end{align*}
	Then, for any two elements $(\beta,\alpha)$ and $(\tilde\beta,\tilde\alpha)$ of the parameter space $\mathcal B^{G}\times\mathcal A^{TG}$,
	\begin{align*}
		&h_A(\beta,\alpha)\\ 
		\geq& \min_{g\neq \tilde g}	\tr\left(\begin{pmatrix}
		(\beta_g-\beta_{\tilde g})-(\tilde \beta_g-\tilde \beta_{\tilde g})\\
		\frac{1}{\sqrt T}(\alpha_{g1}' - \alpha_{\tilde g1}') - \frac{1}{\sqrt T}(\tilde\alpha_{g1}' - \tilde \alpha_{\tilde g1}')\\
		\vdots\\
		\frac{1}{\sqrt T}(\alpha_{gT}' - \alpha_{\tilde gT}') - \frac{1}{\sqrt T}(\tilde\alpha_{gT}' - \tilde\alpha_{\tilde gT}'))
		\end{pmatrix}'A\begin{pmatrix}
		(\beta_g-\beta_{\tilde g})-(\tilde\beta_g-\tilde\beta_{\tilde g})\\
		\frac{1}{\sqrt T}(\alpha_{g1}' - \alpha_{\tilde g1}') - \frac{1}{\sqrt T}(\tilde\alpha_{g1}' - \tilde\alpha_{\tilde g1}')\\
		\vdots\\
		\frac{1}{\sqrt T}(\alpha_{gT}' - \alpha_{\tilde gT}') - \frac{1}{\sqrt T}(\tilde\alpha_{gT}' - \tilde\alpha_{\tilde gT}')
		\end{pmatrix}\right. \\
		&\left.\hphantom{\min_{g\neq \tilde g}	\tr\left( \right.} +  2\begin{pmatrix}
		(\beta_g-\beta_{\tilde g})-(\tilde\beta_g-\tilde\beta_{\tilde g})\\
		\frac{1}{\sqrt T}(\alpha_{g1}' - \alpha_{\tilde g1}') - \frac{1}{\sqrt T}(\tilde\alpha_{g1}' - \tilde\alpha_{\tilde g1}')\\
		\vdots\\
		\frac{1}{\sqrt T}(\alpha_{gT}' - \alpha_{\tilde gT}') - \frac{1}{\sqrt T}(\tilde\alpha_{gT}' - \tilde\alpha_{\tilde gT}')
		\end{pmatrix}'A\begin{pmatrix}
		(\beta_g-\beta_{\tilde g})\\
		\frac{1}{\sqrt T}(\alpha_{g1}' - \alpha_{\tilde g1}')\\
		\vdots\\
		\frac{1}{\sqrt T}(\alpha_{gT}' - \alpha_{\tilde gT}')
		\end{pmatrix}\right) + h_A(\tilde\beta,\tilde\alpha).
	\end{align*}
	Consequently, by Assumption \ref{ass-sineq}\ref{ass-sineq:cpt-param},
	\begin{align*}
		|h_A(\beta,\alpha)-h_A(\beta^0,\alpha^0)|
		\lesssim \|A\|_2\max_{g\in\mathbb G}\biggl(\|\beta_g-\beta_g^0\|^2 + \frac{1}{T}\sum_{t=1}^T\|\alpha_{gt} - \alpha_{gt}^0\|^2\biggr)^{\frac{1}{2}}.
	\end{align*}
	Substituting $A=T^{-1}\sum_{t=1}^T\mathbf x_{it}\mathbf x_{it}'$, we have
	\begin{align}
		&\min_{g\neq \tilde g} \frac{1}{T}\sum_{t=1}^T\|x_{it}'(\hat\beta_g-\hat\beta_{\tilde g}) + (\hat\alpha_{g t} - \hat\alpha_{\tilde g t})\|^2 \nonumber\\
		=&h_{\frac{1}{T}\sum_{t=1}^T\mathbf x_{it}\mathbf x_{it}'}(\hat\beta,\hat\alpha)\nonumber\\
		\geq& h_{\frac{1}{T}\sum_{t=1}^T\mathbf x_{it}\mathbf x_{it}'}(\beta^0,\alpha^0)
		- c\biggl\|\frac{1}{T}\sum_{t=1}^T\mathbf x_{it}\mathbf x_{it}'\biggr\|_2\max_{g\in\mathbb G}\biggl(\|\hat\beta_g-\beta_g^0\|^2 + \frac{1}{T}\sum_{t=1}^T\|\hat\alpha_{gt} - \alpha_{gt}^0\|^2\biggr)^{\frac{1}{2}}\nonumber\\
		\geq& \min_{g\neq \tilde g} \frac{1}{T}\sum_{t=1}^T\|x_{it}'(\beta_g^0-\beta_{\tilde g}^0) + (\alpha_{g t}^0 - \alpha_{\tilde g t}^0)\|^2
		- c\biggl\|\frac{1}{T}\sum_{t=1}^T\mathbf x_{it}\mathbf x_{it}'\biggr\|_2 d_H((\hat\beta,\hat\alpha),(\beta^0,\alpha^0)) \label{eq-sineq:gamma-cnst:grpsep-ss}
	\end{align}
	with probability approaching one, where $c$ is some positive constant.
\end{itemize}
Combining results, we have
\begin{align*}
	&\Pr\biggl(\min_{i\in\{1,\dots,N\}}\biggl\{\frac{1}{T}\sum_{t=1}^T\|y_{it}'-x_{it}'\hat\beta_{g_i^0}-\hat\alpha_{g_i^0 t}'\|^2 - \frac{1}{T}\sum_{t=1}^T\|y_{it}'-x_{it}'\hat\beta_{\hat g_i}-\hat\alpha_{\hat g_it}'\|^2\biggr\}\geq -\frac{d}{8},\\
	&\phantom{\Pr\biggl(} \max_{i\in\{1,\dots,N\}}|\hat g_i-g_i^0|>0\biggr)\\
	\leq& \Pr\biggl(\max_{i\in\{1,\dots,N\}} \biggl|\frac{2}{T}\sum_{t=1}^T(x_{it}'(\beta_{g_i^0}^0-\hat\beta_{g_i^0}) + (\alpha_{g_i^0 t}^0 - \hat\alpha_{g_i^0 t}))(x_{it}'(\hat\beta_{g_i^0}-\hat\beta_{\hat g_i}) + (\hat\alpha_{g_i^0 t} - \hat\alpha_{\hat g_i t}))'\biggr|\\
	&\phantom{\Pr\biggl(} + \max_{i\in\{1,\dots,N\}} \biggl|\frac{2}{T}\sum_{t=1}^Te_{it}'(x_{it}'(\hat\beta_{g_i^0}-\hat\beta_{\hat g_i}))'\biggr| 
	+ \max_{i\in\{1,\dots,N\}} \biggl|\frac{2}{T}\sum_{t=1}^Te_{it}'((\hat\alpha_{g_i^0 t}' - \hat\alpha_{\hat g_i t}')-(\alpha_{g_i^0 t}^{0\prime} - \alpha_{\hat g_i t}^{0\prime}))'\biggr| + \frac{d}{8}\\
	&\phantom{\Pr\biggl(} \geq \max_{i\in\{1,\dots,N\}}\mathbf 1\{\hat g_i\neq g_i^0\}\biggl\{\min_{g\neq \tilde g} \frac{1}{T}\sum_{t=1}^T\|x_{it}'(\hat\beta_g-\hat\beta_{\tilde g}) + (\hat\alpha_{g t}' - \hat\alpha_{\tilde g t}')\|^2 - \max_{g\neq \tilde g}\biggl|\frac{2}{T}\sum_{t=1}^Te_{it}'(\alpha_{g_i^0t}^{0\prime}-\alpha_{\hat g_i^0t}^{0\prime})'\biggr|\biggr\},\\
	&\phantom{\Pr\biggl(} \max_{i\in\{1,\dots,N\}}|\hat g_i-g_i^0|>0\biggr)\\
	\leq& \Pr\biggl(\overbrace{
	\min_{i\in\{1,\dots,N\}}\biggl\{\min_{g\neq \tilde g} \frac{1}{T}\sum_{t=1}^T\|x_{it}'(\hat\beta_g-\hat\beta_{\tilde g}) + (\hat\alpha_{g t}' - \hat\alpha_{\tilde g t}')\|^2 - \max_{g\neq \tilde g}\biggl|\frac{2}{T}\sum_{t=1}^Te_{it}'(\alpha_{g t}^{0\prime} - \alpha_{\tilde g t}^{0\prime})'\biggr|\biggr\} < \frac{d}{4}
	}^{\text{implies } \min_i\min_{g\neq \tilde g} \frac{1}{T}\sum_{t=1}^T\|x_{it}'(\hat\beta_g-\hat\beta_{\tilde g}) + (\hat\alpha_{g t}' - \hat\alpha_{\tilde g t}')\|^2< -\min_i-\max_{g\neq \tilde g}|\frac{2}{T}\sum_{t=1}^Te_{it}'(\alpha_{g t}^{0\prime} - \alpha_{\tilde g t}^{0\prime})'|+\frac{d}{4}}\biggr)\\
	& + \Pr\biggl(\max_{i\in\{1,\dots,N\}} \biggl|\frac{2}{T}\sum_{t=1}^T(x_{it}'(\beta_{g_i^0}^0-\hat\beta_{g_i^0}) + (\alpha_{g_i^0 t}^0 - \hat\alpha_{g_i^0 t}))(x_{it}'(\hat\beta_{g_i^0}-\hat\beta_{\hat g_i}) + (\hat\alpha_{g_i^0 t} - \hat\alpha_{\hat g_i t}))'\biggr|\geq \frac{1}{3}\biggl(\frac{d}{4}-\frac{d}{8}\biggr)\biggr)\\
	& + \Pr\biggl(\max_{i\in\{1,\dots,N\}} \biggl|\frac{2}{T}\sum_{t=1}^Te_{it}'(x_{it}'(\hat\beta_{g_i^0}-\hat\beta_{\hat g_i}))'\biggr| \geq \frac{d}{24}\biggr)\\ 
	& + \Pr\biggl(\max_{i\in\{1,\dots,N\}} \biggl|\frac{2}{T}\sum_{t=1}^Te_{it}'((\hat\alpha_{g_i^0 t}' - \hat\alpha_{\hat g_i t}')-(\alpha_{g_i^0 t}^{0\prime} - \alpha_{\hat g_i t}^{0\prime}))'\biggr| \geq \frac{d}{24}\biggr)\\
	\leq& \biggl\{\Pr\biggl(\overbrace{\min_{i\in\{1,\dots,N\}}\min_{g\neq \tilde g}\frac{1}{T}\sum_{t=1}^T\|x_{it}'(\hat\beta_g-\hat\beta_{\tilde g})+(\hat\alpha_{gt}'-\hat\alpha_{\tilde gt}')\|^2<\frac{3d}{8}}^{\hspace{2in}\mathclap{\text{implies }\min_i\min_{g\neq \tilde g}\frac{1}{T}\sum_{t=1}^T\|x_{it}'(\beta_g^0-\beta_{\tilde g}^0)+(\alpha_{gt}^{0\prime}-\hat\alpha_{\tilde gt}^{0\prime})\|^2-c\max_i\|\frac{1}{T}\sum_{t=1}^T\mathbf x_{it}\mathbf x_{it}'\|_2d_H((\hat\beta,\hat\alpha),(\beta^0,\alpha^0))<\frac{3d}{8}\text{ by equation \eqref{eq-sineq:gamma-cnst:grpsep-ss}}}}\biggr)\\
	&\hphantom{\biggl\{} + \Pr\biggl(\max_{i\in\{1,\dots,N\}}\max_{g\neq \tilde g}\biggl|\frac{2}{T}\sum_{t=1}^Te_{it}'(\alpha_{g t}^{0\prime} - \alpha_{\tilde g t}^{0\prime})'\biggr| \geq \frac{d}{8}\biggr)\biggr\}\\
	& + \bigg\{\Pr\biggl(\max_{i\in\{1,\dots,N\}}\biggl\|\frac{1}{T}\sum_{t=1}^Tx_{it}x_{it}'\biggr\|_2>L\biggr) + o(1)\biggr\} + \Pr\biggl(\max_{i\in\{1,\dots,N\}}\biggl\|\frac{1}{T}\sum_{t=1}^Tx_{it}e_{it}'\biggr\|\gtrsim d\biggr)\\
	& + \Pr\biggl(\biggl(\max_{g\in\mathbb G}\frac{1}{T}\sum_{t=1}^T\|\hat\alpha_{gt}-\alpha_{gt}^0\|^2\biggr)^{\frac{1}{2}}\biggl(\max_{i\in\{1,\dots,N\}}\frac{1}{T}\sum_{t=1}^T\|e_{it}\|^2\biggr)^{\frac{1}{2}}\gtrsim d\biggr)\\
	\leq& \Pr\biggl(\min_{i\in\{1,\dots,N\}}\min_{g\neq \tilde g}\frac{1}{T}\sum_{t=1}^T\|x_{it}'(\beta_g^0-\beta_{\tilde g}^0)+(\alpha_{gt}^{0\prime}-\alpha_{\tilde gt}^{0\prime})\|^2<\frac{d}{2}\biggr)\\
	& + \Pr\biggl(\max_{i\in\{1,\dots,N\}}\max_{g\neq \tilde g}\biggl|\frac{1}{T}\sum_{t=1}^Te_{it}'(\alpha_{g t}^{0\prime} - \alpha_{\tilde g t}^{0\prime})'\biggr|\gtrsim d\biggr)\\
	& + 2\Pr\biggl(\max_{i\in\{1,\dots,N\}}\biggl\|\frac{1}{T}\sum_{t=1}^Tx_{it}x_{it}'\biggr\|_2>L\biggr) + \Pr\biggl(\max_{i\in\{1,\dots,N\}}\biggl\|\frac{1}{T}\sum_{t=1}^Tx_{it}e_{it}'\biggr\|\gtrsim d\biggr)\\
	& + \Pr\biggl(\biggl(\max_{g\in\mathbb G}\frac{1}{T}\sum_{t=1}^T\|\hat\alpha_{gt}-\alpha_{gt}^0\|^2\biggr)^{\frac{1}{2}}\biggl(\max_{i\in\{1,\dots,N\}}\frac{1}{T}\sum_{t=1}^T\|e_{it}\|^2\biggr)^{\frac{1}{2}}\gtrsim d\biggr) + o(1).
\end{align*}

Now, note that if
\begin{align*}
	\max_{i\in\{1,\dots,N\}}\biggl\{\frac{1}{T}\sum_{t=1}^T\|y_{it}'-x_{it}'\hat\beta_{\hat g_i}-\hat\alpha_{\hat g_i t}\|^2 
	- \min_{g\in\mathbb G}\frac{1}{T}\sum_{t=1}^T\|y_{it}'-x_{it}'\hat\beta_g-\hat\alpha_{g t}\|^2\biggr\} \leq \frac{d}{8},
\end{align*}
then for all $i\in\{1,\dots,N\}$,
\begin{align*}
	&\frac{1}{T}\sum_{t=1}^T\|y_{it}'-x_{it}'\hat\beta_{g_i^0}-\hat\alpha_{g_i^0 t}'\|^2 
	- \frac{1}{T}\sum_{t=1}^T\|y_{it}'-x_{it}'\hat\beta_{\hat g_i}-\hat\alpha_{\hat g_it}'\|^2\\
	=&\biggl(\frac{1}{T}\sum_{t=1}^T\|y_{it}'-x_{it}'\hat\beta_{g_i^0}-\hat\alpha_{g_i^0 t}'\|^2
	- \min_{g\in\mathbb G}\frac{1}{T}\sum_{t=1}^T\|y_{it}'-x_{it}'\hat\beta_g-\hat\alpha_{g t}\|^2\biggr)\\
	& + \biggl(\min_{g\in\mathbb G}\frac{1}{T}\sum_{t=1}^T\|y_{it}'-x_{it}'\hat\beta_g-\hat\alpha_{g t}\|^2
	- \frac{1}{T}\sum_{t=1}^T\|y_{it}'-x_{it}'\hat\beta_{\hat g_i}-\hat\alpha_{\hat g_it}'\|^2\biggr) \geq 0 + (-\frac{d}{8}) = -\frac{d}{8}.
\end{align*}
In particular, the minimum of the leftmost side is no smaller than $-d/8$, that is, 
\begin{align*}
	\min_{i\in\{1,\dots,N\}}\biggl\{\frac{1}{T}\sum_{t=1}^T\|y_{it}'-x_{it}'\hat\beta_{g_i^0}-\hat\alpha_{g_i^0 t}'\|^2 
	- \frac{1}{T}\sum_{t=1}^T\|y_{it}'-x_{it}'\hat\beta_{\hat g_i}-\hat\alpha_{\hat g_it}'\|^2\biggr\}\geq -\frac{d}{8}.
\end{align*}
The desired result then follows, because
\begin{align*}
	&\Pr\biggl(\max_{i\in\{1,\dots,N\}}|\hat g_i-g_i^0|>0\biggr)\\
	\leq& \Pr\biggl(\max_{i\in\{1,\dots,N\}}\biggl\{\frac{1}{T}\sum_{t=1}^T\|y_{it}'-x_{it}'\hat\beta_{\hat g_i}-\hat\alpha_{\hat g_i t}\|^2 - \min_{g\in\mathbb G}\frac{1}{T}\sum_{t=1}^T\|y_{it}'-x_{it}'\hat\beta_g-\hat\alpha_{g t}\|^2\biggr\} > \frac{d}{8} \biggr)\\
	& + \Pr\biggl(\min_{i\in\{1,\dots,N\}}\biggl\{\frac{1}{T}\sum_{t=1}^T\|y_{it}'-x_{it}'\hat\beta_{g_i^0}-\hat\alpha_{g_i^0 t}'\|^2 - \frac{1}{T}\sum_{t=1}^T\|y_{it}'-x_{it}'\hat\beta_{\hat g_i}-\hat\alpha_{\hat g_it}'\|^2\biggr\}\geq -\frac{d}{8},\\
	& \phantom{+ \Pr\biggl(} \max_{i\in\{1,\dots,N\}}|\hat g_i-g_i^0|>0\biggr).
\end{align*}
Recall that we have just obtained a bound on the second term of the last equation.

\subsubsection{Proof of Lemma \ref{lem-ts:beta-cnst}}
Recast the given two-stage linear model into a single equation:
\begin{align*}
	y_{it} = \smash[tb]{(\underbrace{\Pi_{k_i^0}^{0\prime}z_{it} + \mu_{k_i^0t}^0}_{x_{it}})'\beta_{g_i^0}^0 + \alpha_{g_i^0t}^0 + (\overbrace{v_{it}'\beta_{g_i^0} + u_{it}}^{e_{it}})}.
\end{align*}
Under Assumptions \ref{ass-ts}\ref{ass-ts:cpt-param}--\ref{ass-ts:wkdep-u}, $\Pi_{k_i^0}^{0\prime}z_{it} + \mu_{k_i^0 t}^0$, $\alpha_{g_i^0t}^0$, and $v_{it}'\beta_{g_i^0}^0+u_{it}$, when considered as $x_{it}$, $\alpha_{g_i^0t}^0$, and $e_{it}$ in the baseline model, satisfy Assumptions \ref{ass-sineq}\ref{ass-sineq:wkdep-xe}--\ref{ass-sineq:wkdep-e}.
Thus, once the TGFE estimator satisfies the translated version of equation \eqref{eq-sineq:min}:
\begin{align}
	&\min_{(\beta,\alpha,\gamma)\in \mathcal B^G\times \mathcal A^{TG}\times \mathbb G^N}\frac{1}{NT}\sum_{i=1}^N\sum_{t=1}^T(y_{it}-(\Pi_{k_i^0}^{0\prime}z_{it}+\mu_{k_i^0 t}^0)'\beta_g - \alpha_{gt})^2 + o_p(1) \nonumber\\ 
	\geq& \frac{1}{NT}\sum_{i=1}^N\sum_{t=1}^T(y_{it}-(\Pi_{k_i^0}^{0\prime}z_{it}+\mu_{k_i^0 t}^0)'\hat\beta_{\hat g_i^\tgfe}^\tgfe - \hat\alpha_{\hat g_i^\tgfe t}^\tgfe)^2, \label{eq-ts:beta-cnst:min}
\end{align}
by Lemma \ref{lem-sineq:fit}, the corresponding version of equation \eqref{eq-sineq:fit} follows:
\begin{align}
	\frac{1}{NT}\sum_{i=1}^N\sum_{t=1}^T((z_{it}'\Pi_{k_i^0}^0 + \mu_{k_i^0t}^0)(\beta_{g_i^0}^0-\hat\beta_{\hat g_i^\tgfe}^\tgfe) + (\alpha_{g_i^0}^{0\prime}-\hat\alpha_{\hat g_i^\tgfe t}^{\tgfe\prime}))^2 \rightarrow_p 0. \label{eq-ts:beta-cnst:fit-ss}
\end{align}
The desired consistency of $(\hat\beta^\tgfe,\hat\alpha^\tgfe)$ then becomes a direct outcome of (i) of Lemma \ref{lem-sineq:beta-cnst}.\footnote{Under Assumption \ref{ass-ts}\ref{ass-ts:relev-ss}--\ref{ass-ts:finmt-z}, Assumptions \ref{lem-sineq}\ref{ass-sineq:relev-ss}--\ref{ass-sineq:finmt-z} holds with $J_{\gamma,g,\tilde g}=\{1,\dots,d+T\}$.} Hence, we proceed to show that equation \eqref{eq-ts:beta-cnst:min} indeed holds.

It is sufficient to establish the uniform convergence between the sums of squared residuals, each generated from the regressions on feasible and infeasible first-stage fitted values:
\begin{align*}
	&\sup_{(\beta,\alpha,\gamma)\in\mathcal B^G\times \mathcal A^G\times  \mathbb G^N} \biggl|\frac{1}{NT}\sum_{i=1}^N\sum_{t=1}^T(y_{it}-(\hat \Pi_{\hat k_i^\tgfe}^{\tgfe\prime}z_{it}+\hat\mu_{\hat k_i^\tgfe t}^\tgfe)'\beta_{g_i}-\alpha_{g_i t})^2\\
	&\hphantom{\sup_{(\beta,\alpha,\gamma)\in\mathcal B^G\times \mathcal A^G\times  \mathbb G^N}\biggl|}- \frac{1}{NT}\sum_{i=1}^N\sum_{t=1}^T(y_{it}-(\Pi_{k_i^0}^{0\prime}z_{it}+\mu_{k_i^0 t}^0)'\beta_{g_i}-\alpha_{g_it})^2\biggr| = o_p(1).
\end{align*}
Note that given Assumptions \ref{ass-ts}\ref{ass-ts:cpt-param}--\ref{ass-ts:wkdep-v}, the variables $z_{it}$, $\mu_{k_i^0t}^0$, and $v_{it}$, when considered as $x_{it}$, $\alpha_{g_i^0}^0$, and $e_{it}$ in the baseline model, satisfy Assumptions \ref{ass-sineq}\ref{ass-sineq:wkdep-xe}--\ref{ass-sineq:wkdep-e}. By Lemma \ref{lem-sineq:fit}, we thus obtain equation \eqref{eq-sineq:fit} in terms of the first-stage GFE estimate $(\hat\Pi_{\hat k_i^\tgfe}^\tgfe,\hat\mu_{\hat k_i^\tgfe}^\tgfe)$:
\begin{align}
	\frac{1}{NT}\sum_{i=1}^N\sum_{t=1}^T\|(\Pi_{k_i^0}^0-\hat\Pi_{\hat k_i^\tgfe}^\tgfe)'z_{it} + (\mu_{k_i^0t}^0-\hat\mu_{\hat k_i^\tgfe t}^\tgfe)\|^2 = o_p(1). \label{eq-ts:beta-cnst:fit-fs}
\end{align}
Combining this with Assumptions \ref{ass-ts}\ref{ass-ts:finmt-z}, the above supremand is then bounded by
\begin{align*}
	&\biggl|\frac{2}{NT}\sum_{i=1}^N\sum_{t=1}^T\bigl((\Pi_{k_i^0}^{0\prime}z_{it} + \mu_{k_i^0t}^0)'(\beta_{g_i^0}^0-\beta_{g_i}) + (\alpha_{g_i^0t}^0-\alpha_{g_it})\bigr)
	\bigl((\Pi_{k_i^0}^0-\hat\Pi_{\hat k_i^\tgfe}^\tgfe)'z_{it} + (\mu_{k_i^0t}^0-\hat\mu_{\hat k_i^\tgfe t}^\tgfe)\bigr)'\beta_{g_i}\\
	&+ \frac{1}{NT}\sum_{i=1}^N\sum_{t=1}^T\bigl(\bigl((\Pi_{k_i^0}^0-\hat\Pi_{\hat k_i^\tgfe}^\tgfe)'z_{it} + (\mu_{k_i^0t}^0-\hat\mu_{\hat k_i^\tgfe t}^\tgfe)\bigr)'\beta_{g_i}\bigr)^2\biggr|\\
	\leq& 2\biggl(\frac{1}{NT}\sum_{i=1}^N\sum_{t=1}^T\bigl((\Pi_{k_i^0}^{0\prime}z_{it} + \mu_{k_i^0t}^0)'(\beta_{g_i^0}^0-\beta_{g_i}) + (\alpha_{g_i^0t}^0-\alpha_{g_it})\bigr)^2\biggr)^{\frac{1}{2}}\\
	&\phantom{2}\times \biggl(\frac{1}{NT}\sum_{i=1}^N\sum_{t=1}^T\big\|(\Pi_{k_i^0}^0-\hat\Pi_{\hat k_i^\tgfe}^\tgfe)'z_{it} + (\mu_{k_i^0t}^0-\hat\mu_{\hat k_i^\tgfe t}^\tgfe)\|^2\|\beta_{g_i}\|^2\biggr)^{\frac{1}{2}}\\
	&+ \frac{1}{NT}\sum_{i=1}^N\sum_{t=1}^T\|(\Pi_{k_i^0}^0-\hat\Pi_{\hat k_i^\tgfe}^\tgfe)'z_{it} + (\mu_{k_i^0t}^0-\hat\mu_{\hat k_i^\tgfe t}^\tgfe)\|^2 \leq O_p(1)o_p(1) + o_p(1),
\end{align*}
which does not depend on any parameter.

Recall that if the coefficient parameter is homogeneous, instead of Assumption \ref{ass-sineq}\ref{ass-sineq:wkdep-xe}, $(NT)^{-1}\sum_{i=1}^N\sum_{t=1}^Tx_{it}e_{it}' = o_p(1)$ is sufficient for Lemma \ref{lem-sineq:fit}. This implies that  Assumption \ref{ass-ts}\ref{ass-ts:wkdep-zv} is stronger than necessary for equation \eqref{eq-ts:beta-cnst:fit-fs}. Specifically, it can be weakened into 
\begin{align*}
	\Exp\biggl[\biggl\|\frac{1}{NT}\sum_{i=1}^N\sum_{t=1}^Tz_{it}v_{it}'\biggr\|^2\biggr]\leq M,
\end{align*}
where the exogeneity arises from both cross-sectional and intertemporal dimensions. Nevertheless, for equation \eqref{eq-ts:beta-cnst:fit-ss}, we need to impose
\begin{align*}
	\Exp\biggl[\frac{1}{NT}\sum_{i=1}^N\biggl\|\sum_{t=1}^T(\Pi_{k_i^0}^{0\prime}z_{it})v_{it}'\biggr\|^2\biggr]\leq M,
\end{align*} 
where the exogeneity along the intertemporal dimension is required. Still, note that the combination of these two conditions is weaker than Assumption \ref{ass-ts}\ref{ass-ts:wkdep-zv}.

\subsubsection{Proof of Lemma \ref{lem-ts:gamma-cnst}}
Let $(\tilde\beta^\tgfe,\tilde\alpha^\tgfe,\tilde\gamma^\tgfe)\equiv \arg\min_{(\beta,\alpha,\gamma)\in\mathcal B^G\times\mathcal A^G\times \mathbb G^N} \sum_{i=1}^N\sum_{t=1}^T(y_{it}-(\Pi_{k_i^0}^{0\prime}z_{it} + \mu_{k_i^0 t}^0)'\beta_{g_i} - \alpha_{g_i})^2$ be an infeasible GFE estimator using the population first-stage fitted values $\Pi_{k_i^0}^{0\prime}z_{it} + \mu_{k_i^0 t}^0$ as regressors. As noted in the proof of Lemma \ref{lem-ts:beta-cnst}, under Assumptions \ref{ass-ts}\ref{ass-ts:cpt-param}--\ref{ass-ts:finmt-z}, $\Pi_{k_i^0}^{0\prime}z_{it}+\mu_{k_i^0t}^0$, $\alpha_{g_i^0t}^0$, and $v_{it}'\beta_{g_i^0}^0+u_{it}$, if considered as $x_{it}$, $\alpha_{g_i^0}^0$, and $e_{it}$ in the baseline model, satisfy Assumptions \ref{ass-sineq}\ref{ass-sineq:wkdep-xe}--\ref{ass-sineq:finmt-z}. Since equation \eqref{eq-sineq:min} automatically holds in terms of the infeasible GFE estimator, it follows from (i) of Lemma \ref{lem-sineq:beta-cnst} that $(\tilde\beta^\tgfe,\tilde\alpha^\tgfe)$ is consistent for $(\beta^0,\alpha^0)$.

Now, we apply Lemma \ref{lem-sineq:gamma-cnst} to the combined estimator $(\smash[b]{\underbrace{\tilde\beta^\tgfe,\tilde\alpha^\tgfe}_{\text{infeasible}}},\smash[b]{\underbrace{\hat\gamma^\tgfe}_{\mathclap{\text{feasible}}}})$ to obtain:
\begin{align}
	&\Pr\biggl(\max_{i\in\{1,\dots,N\}}|\hat g_i^\tgfe-g_i^0|>0\biggr)\nonumber\\
	\leq& \Pr\biggl(\max_{i\in\{1,\dots,N\}}\biggl(\frac{1}{T}\sum_{t=1}^T(y_{it}-(\Pi_{k_i^0}^{0\prime}z_{it}+\mu_{k_i^0 t}^0)'\tilde\beta_{\hat g_i^\tgfe}^\tgfe-\tilde\alpha_{\hat g_i^\tgfe t}^\tgfe)^2\nonumber\\
	&\hphantom{\Pr\biggl(\max_{i\in\{1,\dots,N\}}\biggl(} - \min_{g\in\mathbb G}\frac{1}{T}\sum_{t=1}^T(y_{it}-(\Pi_{k_i^0}^{0\prime}z_{it}+\mu_{k_i^0t}^0)'\tilde\beta_g^\tgfe-\tilde\alpha_{g t}^\tgfe)^2\biggr)\gtrsim d\biggr)\nonumber\\
	& + \Pr\biggl(\min_{i\in\{1,\dots,N\}}\min_{g\neq \tilde g}\frac{1}{T}\sum_{t=1}^T\|(\Pi_{k_i^0}^{0\prime}z_{it}+\mu_{k_i^0t}^0)'(\beta_g^0-\beta_{\tilde g}^0) + (\alpha_{gt}^0-\alpha_{\tilde gt}^0)\|^2<\frac{d}{2}\biggr)\nonumber\\
	& + \Pr\biggl(\max_{i\in\{1,\dots,N\}}\min_{g\neq g}\biggl|\frac{1}{T}\sum_{t=1}^T(v_{it}'\beta_{g_i^0}^0 + u_{it})(\alpha_{g t}^0-\alpha_{\tilde g t}^0)\biggr|\gtrsim d \biggr)\nonumber\\ 
	& + 2\Pr\biggl(\max_{i\in\{1,\dots,N\}}\biggl\|\frac{1}{T}\sum_{t=1}^T(\Pi_{k_i^0}^{0\prime}z_{it}+\mu_{k_i^0t}^0)(\Pi_{k_i^0}^{0\prime}z_{it}+\mu_{k_i^0t}^0)'\biggr\|_2>L\biggr)\nonumber\\
	& + \Pr\biggl(\max_{i\in\{1,\dots,N\}}\biggl\|\frac{1}{T}\sum_{t=1}^T(\Pi_{k_i^0}^{0\prime}z_{it}+\mu_{k_i^0t}^0)(v_{it}'\beta_{g_i^0}^0+u_{it})\biggr\|\gtrsim d\biggr)\nonumber\\
	& + \Pr\biggl(\biggl(\min_{g\in\mathbb G}\frac{1}{T}\sum_{t=1}^T\|\tilde\alpha_{gt}^\tgfe-\alpha_{gt}^0\|^2\biggr)^{\frac{1}{2}}\biggl(\max_{i\in\{1,\dots,N\}}\frac{1}{T}\sum_{t=1}^T|v_{it}'\beta_{g_i^0}^0+u_{it}|^2\biggr)^{\frac{1}{2}}\gtrsim d\biggr) + o(1). \label{eq-ts:gamma-cnst:bound-ss}
\end{align}
The third to sixth terms can be readily bounded using the concentration inequalities regarding $z_{it}z_{it}'$, $z_{it}v_{it}'$, $z_{it}u_{it}$, $v_{it}(\alpha_{g t}^0-\alpha_{\tilde g t}^0)$, and $u_{it}(\alpha_{g t}^0-\alpha_{\tilde g t}^0)$. Consequently, the subsequent part of the proof will be dedicated to deriving a bound for the first term. For the sake of notational conciseness, we will henceforth drop the superscript ``tgfe.''

Similar to our approach in the proof of Lemma \ref{lem-sineq:beta-cnst}, we establish the order of uniform convergence between the unit-wise sums of squared errors obtained by using population and sample fitted values respectively. Specifically, we derive an upper bound for the probability of the event:
\begin{align*}
	\max_{i\in\{1,\dots,N\}}\bigg|\frac{1}{T}\sum_{t=1}^T(y_{it}-(\hat\Pi_{\hat k_i}'z_{it}+\hat\mu_{\hat k_i t})'\hat\beta_g-\hat\alpha_{gt})^2
	- \frac{1}{T}\sum_{t=1}^T(y_{it}-(\Pi_{k_i^0}^{0\prime}z_{it}+\mu_{k_i^0 t}^0)'\tilde\beta_g-\tilde\alpha_{gt})^2\bigg|
	>\eta,
\end{align*}
where $\eta$ is any positive constant. If $\eta$ is small enough, from the definition of $\hat g_i$, the first term of equation \eqref{eq-ts:gamma-cnst:bound-ss} collapses to zero under the complementary event.

The difference between the two sums inside the absolute value expands as
\begin{align*}
	&- \frac{2}{T}\sum_{t=1}^T\biggl(\bigl((\Pi_{k_i^0}^{0\prime}z_{it}+\mu_{k_i^0t}^0)'(\beta_{g_i^0}^0-\tilde\beta_g) + (\alpha_{g_i^0 t}^0-\tilde\alpha_{g t})\bigr) + \bigl(v_{it}'\beta_{g_i^0}^0+u_{it}\bigr)\biggr)\\
	&\phantom{- \frac{1}{T}\sum_{t=1}^T} \times \biggl(\bigl((\Pi_{k_i^0}^{0\prime}z_{it} + \mu_{k_i^0t}^0)'(\tilde\beta_g - \hat\beta_g) + (\tilde\alpha_{gt} - \hat\alpha_{gt})\bigr) + \bigl((\Pi_{k_i^0}^0-\hat\Pi_{\hat k_i})'z_{it}+(\mu_{k_i^0t}^0-\hat\mu_{\hat k_it})\bigr)'\hat\beta_g\biggr)\\
	& + \frac{1}{T}\sum_{t=1}^T\biggl(\bigl((\Pi_{k_i^0}^{0\prime}z_{it} + \mu_{k_i^0t}^0)'(\tilde\beta_g - \hat\beta_g) + (\tilde\alpha_{gt} - \hat\alpha_{gt})\bigr)
	+ \bigl((\Pi_{k_i^0}^0-\hat\Pi_{\hat k_i})'z_{it}+(\mu_{k_i^0t}^0-\hat\mu_{\hat k_it})\bigr)'\hat\beta_g\biggr)^2.
\end{align*}
By Assumption \ref{ass-ts}\ref{ass-ts:cpt-param}, for large enough $L$, there exists some $\tilde L>0$ such that
\begin{align*}
	\Pr\biggl(\max_{i\in\{1,\dots,N\}}\biggl\|\frac{1}{T}\sum_{t=1}^T(\Pi_{k_i^0}^{0\prime}z_{it}+\mu_{k_i^0t}^0)(\Pi_{k_i^0}^{0\prime}z_{it}+\mu_{k_i^0t}^0)'\biggr\|_2> L\biggr)
	\leq \Pr\biggl(\max_{i\in\{1,\dots,N\}}\biggl\|\frac{1}{T}\sum_{t=1}^Tz_{it}z_{it}'\biggr\|_2\gtrsim \tilde L\biggr),
\end{align*}
and so the probability associated with the above event can be bounded as follows:
\begin{align*}
	&\Pr\biggl(\max_{i\in\{1,\dots,N\}}\biggl\|\frac{1}{T}\sum_{t=1}^T(\Pi_{k_i^0}^{0\prime}z_{it}+\mu_{k_i^0t}^0)(\Pi_{k_i^0}^{0\prime}z_{it}+\mu_{k_i^0t}^0)'\biggr\|_2> L\biggr)\\
	& + \biggl\{\Pr\biggl((L^{\frac{1}{2}}+1)^2\biggl(\|\tilde\beta_g - \hat\beta_g\|^2 + \frac{1}{T}\sum_{t=1}^T\|\tilde\alpha_{gt}-\hat\alpha_{gt}\|^2\biggr)^{\frac{1}{2}}\gtrsim \eta\biggr)\\
	& \hphantom{+ \biggl\{} + \Pr\biggl((L^{\frac{1}{2}}+1)\biggl(\max_{i\in\{1,\dots,N\}}\frac{1}{T}\sum_{t=1}^T\|(\Pi_{k_i^0}^0-\hat\Pi_{\hat k_i})'z_{it}+(\mu_{k_i^0t}^0-\hat\mu_{\hat k_it})\|^2\biggr)^{\frac{1}{2}}\gtrsim \eta\biggr)\\
	& \hphantom{+ \biggl\{} + \Pr\biggl(\max_{i\in\{1,\dots,N\}}\biggl\|\frac{1}{T}\sum_{t=1}^T(v_{it}'\beta_{g_i^0}^0+u_{it})\mathbf{\tilde x}_{it}\biggr\|_2\biggl(\|\tilde\beta_g - \hat\beta_g\|^2 + \frac{1}{T}\sum_{t=1}^T\|\tilde\alpha_{gt}-\hat\alpha_{gt}\|^2\biggr)^{\frac{1}{2}}\gtrsim \eta\biggr)\\
	& \hphantom{+ \biggl\{} + \Pr\biggl(\max_{i\in\{1,\dots,N\}}\biggl\|\frac{1}{T}\sum_{t=1}^T(v_{it}'\beta_{g_i^0}^0+u_{it})((\Pi_{k_i^0}^0-\hat\Pi_{\hat k_i})'z_{it}+(\mu_{k_i^0t}^0-\hat\mu_{\hat k_it}))'\biggr\|\gtrsim \eta\biggr)\biggr\}\\
	& + \biggl\{\Pr\biggl((L^{\frac{1}{2}}+1)^2\biggl(\|\tilde\beta_g - \hat\beta_g\|^2 + \frac{1}{T}\sum_{t=1}^T\|\tilde\alpha_{gt}-\hat\alpha_{gt}\|^2\biggr)\gtrsim \eta\biggr)\\
	& \hphantom{+ \biggl\{} + \Pr\biggl((L^{\frac{1}{2}}+1)\biggl(\|\tilde\beta_g - \hat\beta_g\|^2 + \frac{1}{T}\sum_{t=1}^T\|\tilde\alpha_{gt}-\hat\alpha_{gt}\|^2\biggr)^{\frac{1}{2}}\\
	& \hphantom{+ \biggl\{+ \Pr\biggl(} \times \biggl(\max_{i\in\{1,\dots,N\}}\frac{1}{T}\sum_{t=1}^T\|(\Pi_{k_i^0}^0-\hat\Pi_{\hat k_i})'z_{it}+(\mu_{k_i^0t}^0-\hat\mu_{\hat k_it})\|^2\biggr)^{\frac{1}{2}}\gtrsim \eta\biggr)\\
	& \hphantom{+ \biggl\{}+ \Pr\biggl(\max_{i\in\{1,\dots,N\}}\frac{1}{T}\sum_{t=1}^T\|(\Pi_{k_i^0}^0-\hat\Pi_{\hat k_i})'z_{it}+(\mu_{k_i^0t}^0-\hat\mu_{\hat k_it})\|^2\gtrsim \eta\biggr)\biggr\}\\
	\leq&\Pr\biggl(\max_{i\in\{1,\dots,N\}}\biggl\|\frac{1}{T}\sum_{t=1}^Tz_{it}z_{it}'\biggr\|_2\gtrsim \tilde L\biggr)
	+ 3\Pr\biggl(\max_{i\in\{1,\dots,N\}}\frac{1}{T}\sum_{t=1}^T\|(\Pi_{k_i^0}^0-\hat\Pi_{\hat k_i})'z_{it}+(\mu_{k_i^0t}^0-\hat\mu_{\hat k_it})\|^2\gtrsim \eta\biggr)\\
	& + \Pr\biggl(\max_{i\in\{1,\dots,N\}}\biggl\|\frac{1}{T}\sum_{t=1}^T(v_{it}'\beta_{g_i^0}^0+u_{it})\mathbf{\tilde x}_{it}\biggr\|_2\gtrsim \tilde L\biggr)\\
	& + \Pr\biggl(\max_{i\in\{1,\dots,N\}}\biggl\|\frac{1}{T}\sum_{t=1}^T(v_{it}'\beta_{g_i^0}^0+u_{it})((\Pi_{k_i^0}^0-\hat\Pi_{\hat k_i})'z_{it}+(\mu_{k_i^0t}^0-\hat\mu_{\hat k_it}))'\biggr\|\gtrsim \eta\biggr) + o(1).
\end{align*}
Note that this upper bound neither depends on $i$ nor $g$.

Now, note that the sum of the previous third to sixth terms in \eqref{eq-ts:gamma-cnst:bound-ss} can be bounded by
\begin{align*}
	&\bigg\{\Pr\biggl(\max_{i\in\{1,\dots,N\}}\min_{g\neq g}\biggl|\frac{1}{T}\sum_{t=1}^Tv_{it}(\alpha_{g t}^0-\alpha_{\tilde g t}^0)\biggr|\gtrsim d \biggr)
	+ \Pr\biggl(\max_{i\in\{1,\dots,N\}}\min_{g\neq g}\biggl|\frac{1}{T}\sum_{t=1}^Tu_{it}(\alpha_{g t}^0-\alpha_{\tilde g t}^0)\biggr|\gtrsim d \biggr)\biggr\}\\ 
	& + 2\Pr\biggl(\max_{i\in\{1,\dots,N\}}\biggl\|\frac{1}{T}\sum_{t=1}^Tz_{it}z_{it}'\biggr\|_2>\tilde L\biggr)\\
	& + \bigg\{\Pr\biggl(\max_{i\in\{1,\dots,N\}}\biggl\|\frac{1}{T}\sum_{t=1}^Tz_{it}v_{it}'\biggr\|\gtrsim d\biggr)
	+ \Pr\biggl(\max_{i\in\{1,\dots,N\}}\biggl\|\frac{1}{T}\sum_{t=1}^Tz_{it}u_{it}\biggr\|\gtrsim d\biggr)\\
	& \hphantom{ + \biggl\{} + \Pr\biggl(\max_{i\in\{1,\dots,N\}}\biggl\|\frac{1}{T}\sum_{t=1}^T\mu_{k_i^0t}^0(v_{it}'\beta_{g_i^0}^0+u_{it})\biggr\|\gtrsim d\biggr)\biggr\}\\
	& + \biggl\{\Pr\biggl(\biggl(\min_{g\in\mathbb G}\frac{1}{T}\sum_{t=1}^T\|\tilde\alpha_{gt}^\tgfe-\alpha_{gt}^0\|^2\biggr)^{\frac{1}{2}}\biggl(\max_{i\in\{1,\dots,N\}}\frac{1}{T}\sum_{t=1}^T\|v_{it}\|^2\biggr)^{\frac{1}{2}}\gtrsim d\biggr)\\
	& \hphantom{ + \biggl\{} + \Pr\biggl(\biggl(\min_{g\in\mathbb G}\frac{1}{T}\sum_{t=1}^T\|\tilde\alpha_{gt}^\tgfe-\alpha_{gt}^0\|^2\biggr)^{\frac{1}{2}}\biggl(\max_{i\in\{1,\dots,N\}}\frac{1}{T}\sum_{t=1}^T|u_{it}|^2\biggr)^{\frac{1}{2}}\gtrsim d\biggr)\biggr\},	
\end{align*}
our desired result follows once we set $d=\eta$.	 Recall that Lemma \ref{lem-sineq:gamma-cnst} holds for any positive $d>0$.

\subsection{Proof of Theorem \ref{th-unid-obj1}}

The sample analog of the moment vector can be expanded as:
\begin{align*}
	&\frac{1}{NT} \sum_{i=1}^N \sum_{t=1}^T z_{it} (y_{it} - x_{it}'\beta_{g_i} )\\
	=& \frac{1}{NT}  \sum_{g_i =1} \sum_{t=1}^T z_{it}x_{it}'(\beta_{g_i^0}^0-\beta_1) 
	+ \frac{1}{NT}  \sum_{g_i =2} \sum_{t=1}^T z_{it}x_{it}'(\beta_{g_i^0}^0-\beta_2)
	+ \frac{1}{NT} \sum_{i=1}^N\sum_{t=1}^Tz_{it}u_{it}\\
	=& \frac{1}{NT}  \sum_{g_i =1}\sum_{g_i^0=1} \sum_{t=1}^T z_{it} x_{it}'(\beta_1^0-\beta_1 ) 
	+ \frac{1}{NT}  \sum_{g_i =1}\sum_{g_i^0=2} \sum_{t=1}^T z_{it} x_{it}'(\beta_2^0-\beta_1 )\\ 
	&+ \frac{1}{NT}  \sum_{g_i =2}\sum_{g_i^0=1} \sum_{t=1}^T z_{it}x_{it}'(\beta_1^0-\beta_2 )
	+ \frac{1}{NT}  \sum_{g_i =2}\sum_{g_i^0=2} \sum_{t=1}^T z_{it} x_{it}'(\beta_2^0-\beta_2 )
	+ \frac{1}{NT} \sum_{i=1}^N\sum_{t=1}^T z_{it} u_{it}.
\end{align*}
By Assumption \ref{assumption:basic} and \ref{assumption:basic-sample}\ref{assumption:basic-sample:lambda}--\ref{assumption:basic-sample:M}, as $N$ and $T$ tend to infinity, it converges in probability to
\begin{align*}
	m(\beta,\gamma)\equiv& \lambda_{11} (\gamma) M_{11}(\gamma) (\beta_1^0 - \beta_1 )  + (\lambda_2^0 - \lambda_{22}(\gamma) ) M_{21}(\gamma) (\beta_2^0 - \beta_1 ) \\
	&+ (\lambda_1^0 - \lambda_{11}(\gamma) ) M_{12}(\gamma) (\beta_1^0 - \beta_2 )
	+  \lambda_{22} (\gamma) M_{22}(\gamma) (\beta_2^0 - \beta_2 ).
\end{align*}

From
\begin{align*}
	\beta_1^0 - \beta_1(\gamma) =& -\frac{\lambda_2^0 - \lambda_{22}(\gamma)}{\lambda_{11}(\gamma) + \lambda_2^0 - \lambda_{22}(\gamma)}(\beta_2^0-\beta_1^0),\\
	\beta_1^0 - \beta_2(\gamma) =& -\frac{\lambda_{22}(\gamma)}{\lambda_{22}(\gamma) + \lambda_1^0 - \lambda_{11}(\gamma)}(\beta_2^0-\beta_1^0),\\
	\beta_2^0 - \beta_1(\gamma) =& \frac{\lambda_{11}(\gamma)}{\lambda_{11}(\gamma) + \lambda_2^0 - \lambda_{22}(\gamma)}(\beta_2^0-\beta_1^0), \text{ and}\\
	\beta_2^0 - \beta_2(\gamma) =& \frac{\lambda_1^0 - \lambda_{11}(\gamma)}{\lambda_{22}(\gamma) + \lambda_1^0 - \lambda_{11}(\gamma)}(\beta_2^0-\beta_1^0),
\end{align*}
it follows that
\begin{align*}
	&m(\beta(\gamma),\gamma)\\ 
	=& \Biggl(\frac{\lambda_{11}(\gamma)(\lambda_2^0 - \lambda_{22}(\gamma))}{\lambda_{11}(\gamma) + \lambda_2^0 - \lambda_{22}(\gamma)}(M_{21}(\gamma)-M_{11}(\gamma)) 
	+ \frac{\lambda_{22}(\gamma)(\lambda_1^0 - \lambda_{11}(\gamma))}{\lambda_{22}(\gamma) + \lambda_1^0 - \lambda_{11}(\gamma)}(M_{22}(\gamma)-M_{12}(\gamma))	\Biggr)
	(\beta_2^0-\beta_1^0).
\end{align*}
By Assumption \ref{assumption:basic-sample}\ref{assumption:basic-sample:M}, for $g\in\mathbb G$, $M_{g1}(\gamma)=M_{g2}(\gamma)\equiv M_g$. Thus, we have
\begin{align*}
	m(\beta(\gamma),\gamma)=& \Biggl(\frac{\lambda_{11}(\gamma)(\lambda_2^0 - \lambda_{22}(\gamma))}{\lambda_{11}(\gamma) + \lambda_2^0 - \lambda_{22}(\gamma)} + \frac{\lambda_{22}(\gamma)(\lambda_1^0 - \lambda_{11}(\gamma))}{\lambda_{22}(\gamma) + \lambda_1^0 - \lambda_{11}(\gamma)}\Biggr)(M_2-M_1)(\beta_2^0-\beta_1^0).
\end{align*}

By the CMT, as $N$ and $T$ tend to infinity, the objective function
\begin{align*}
	&\Biggl(\frac{1}{NT} \sum_{i=1}^N \sum_{t=1}^T z_{it} (y_{it} - x_{it}'\beta_{g_i} ) \Biggr)' \hat W \ \Biggl(\frac{1}{NT} \sum_{i=1}^N \sum_{t=1}^T z_{it} (y_{it} - x_{it}'\beta_{g_i} ) \Biggr)
\end{align*}
converges in probability to
\begin{align*}
	&m(\beta(\gamma),\gamma)'Wm(\beta(\gamma),\gamma)\\
	=& \Biggl(	 \frac{\lambda_{11} (\gamma)(\lambda_2^0 - \lambda_{22}(\gamma) ) }{\lambda_{11} (\gamma)+ (\lambda_2^0 - \lambda_{22}(\gamma) ) } +  \frac{\lambda_{22}(\gamma)(\lambda_1^0 - \lambda_{11}(\gamma) ) }{\lambda_{22} (\gamma)+ (\lambda_1^0 - \lambda_{11}(\gamma) } \Biggr)^2
	(\beta_2^0 - \beta_1^0)' (M_2 - M_1)'W(M_2 - M_1)(\beta_2^0 - \beta_1^0).
\end{align*}
Now, note that this limit reduces to zero if $\beta_2^0 - \beta_1^0 $ is included in the null space of $M_2- M_1$.

\subsection{Proof of Theorem \ref{th-unid-obj2}}
For each $g\in\mathbb G$, let $\lambda_g(\gamma)\equiv (NT)^{-1}\sum_{i=1}^N\mathbf 1\{g_i=g\}$.
Then, for any given $\gamma\in\mathbb G^{\mathbb N}$, as $N$ and $T$ tend to infinity, by Assumptions \ref{assumption:basic} and \ref{assumption:basic-sample},
\begin{align*}
	&\frac{1}{NT} \sum_{g_i =1}\sum_{t=1}^T z_{it} (y_{it} - x_{it}'\beta_1 )\\
	=& \frac{1}{NT}  \sum_{g_i =1} \sum_{t=1}^T z_{it}x_{it}'(\beta_{g_i^0}^0-\beta_1) 
	+ \frac{1}{NT} \sum_{g_i =1}\sum_{t=1}^Tz_{it}u_{it}\\
	=& \frac{1}{NT}  \sum_{g_i =1}\sum_{g_i^0=1} \sum_{t=1}^T z_{it} x_{it}'(\beta_1^0-\beta_1 ) 
	+ \frac{1}{NT}  \sum_{g_i =1}\sum_{g_i^0=2} \sum_{t=1}^T z_{it} x_{it}'(\beta_2^0-\beta_1 )
	+ \frac{1}{NT} \sum_{g_i=1}\sum_{t=1}^T z_{it} u_{it}
\end{align*}
converges in probability to
\begin{align*}
	m_1(\beta_1,\gamma)\equiv 
	\lambda_{11}(\gamma)M_{11}(\gamma)(\beta_1^0-\beta_1) 
	+ (\lambda_2^0-\lambda_{22}(\gamma))M_{21}(\gamma)(\beta_2^0-\beta_1)
	+ \lambda_1(\gamma)L_1(\gamma).
\end{align*}
Similarly, we have
\begin{align*}
	&\frac{1}{NT} \sum_{g_i =2}\sum_{t=1}^T z_{it} (y_{it} - x_{it}'\beta_2 )\\
	=& \frac{1}{NT}  \sum_{g_i =2} \sum_{t=1}^T z_{it}x_{it}'(\beta_{g_i^0}^0-\beta_1) 
	+ \frac{1}{NT} \sum_{g_i =2}\sum_{t=1}^Tz_{it}u_{it}\\
	=& \frac{1}{NT}  \sum_{g_i =2}\sum_{g_i^0=1} \sum_{t=1}^T z_{it} x_{it}'(\beta_1^0-\beta_2 ) 
	+ \frac{1}{NT}  \sum_{g_i =2}\sum_{g_i^0=2} \sum_{t=1}^T z_{it} x_{it}'(\beta_2^0-\beta_2 )
	+ \frac{1}{NT} \sum_{g_i=1}\sum_{t=1}^T z_{it} u_{it}
\end{align*}
converges in probability to
\begin{align*}
	m_2(\beta_2,\gamma)\equiv 
	(\lambda_1^0-\lambda_{11}(\gamma))M_{12}(\gamma)(\beta_2^0-\beta_2)
	+ \lambda_{22}(\gamma)M_{22}(\gamma)(\beta_1^0-\beta_2) 
	+ \lambda_2(\gamma)L_2(\gamma).
\end{align*}

By Assumption \ref{assumption:basic-sample}\ref{assumption:basic-sample:M}, for each $g\in\mathbb G$, $M_{g1}(\gamma)=M_{g2}(\gamma)\equiv M_g$. If $\gamma$ additionally satisfies $L_1(\gamma)=L_2(\gamma)=0$, then
\begin{align*}
	m_1(\beta_1(\gamma),\gamma) =& \frac{\lambda_{11}(\gamma)(\lambda_2^0-\lambda_{22}(\gamma))}{\lambda_{11}(\gamma) + \lambda_2^0 - \lambda_{22}(\gamma)}(M_2-M_1)(\beta_2^0-\beta_1^0) \text{ and}\\
	m_2(\beta_2(\gamma),\gamma) =& \frac{\lambda_{22}(\gamma)(\lambda_1^0-\lambda_{11}(\gamma))}{\lambda_{22}(\gamma) + \lambda_1^0 - \lambda_{11}(\gamma)}(M_2-M_1)(\beta_2^0-\beta_1^0).
\end{align*}
Accordingly, the objective function
\begin{align*}
	\sum_{g\in \mathbb{G}} \Biggl(\frac{1}{NT}  \sum_{g_i =g} \sum_{t=1}^T z_{it} (y_{it} - x_{it}'\beta_g ) \Biggr)' \hat W_g  \Biggl(\frac{1}{NT} \sum_{g_i=g} \sum_{t=1}^T z_{it} (y_{it} - x_{it}'\beta_g ) \Biggr)
\end{align*}
converges in probability to
\begin{align*}
	&m_1(\beta_1(\gamma),\gamma)'W_1 m_1(\beta_1(\gamma),\gamma) 
	+ m_2(\beta_2(\gamma),\gamma)'W_2 m_2(\beta_2(\gamma),\gamma)\\
	=&\Biggl(\frac{\lambda_{11}(\gamma)(\lambda_2^0-\lambda_{22}(\gamma))}{\lambda_{11}(\gamma) + \lambda_2^0 - \lambda_{22}(\gamma)}\Biggr)^2	(\beta_2^0-\beta_1^0)'(M_2-M_1)'W_1(M_2-M_1)(\beta_2^0-\beta_1^0)\\
	&+\Biggl(\frac{\lambda_{22}(\gamma)(\lambda_1^0-\lambda_{11}(\gamma))}{\lambda_{22}(\gamma) + \lambda_1^0 - \lambda_{11}(\gamma)}\Biggr)^2
	(\beta_2^0-\beta_1^0)(M_2-M_1)'W_2(M_2-M_1)(\beta_2^0-\beta_1^0).
\end{align*}
Note that this limit is zero if $\beta_2^0-\beta_1^0$ is included in the null space of $M_2-M_1$.

\subsection{Proofs of Theorems \ref{thm-gfs:beta-cnst}--\ref{thm-ufs:gamma-cnst}}
A model without time-varying group-specific fixed effects is a special case of the model with them (by restricting the fixed effects to be zero). Thus, by restricting the parameter space accordingly, that is, by letting $\mathcal A=\{0\}$ and $\mathcal M=\{\mathbf 0_d\}$, Theorems \ref{thm-gfs:beta-cnst}--\ref{thm-ufs:gamma-cnst} can be established effectively as direct outcomes of Theorems \ref{thm-gfe-gfs:beta-cnst}--\ref{thm-gfe-ufs:gamma-cnst}.\footnote{Note that under this formulation, $|\tilde\alpha_{gt}^{\tgfe}-\alpha_{gt}^0|=0$ and $\|\hat\mu_{kt}-\mu_{kt}^0\|=0$. Thus, the relevant terms in equation \eqref{eq-ts:gamma-cnst} have to be adjusted accordingly.}

\subsection{Proof of Theorem \ref{thm-gfe-gfs:beta-cnst}}
This theorem is a direct consequence of Lemma \ref{lem-ts:beta-cnst}.

\subsection{Proof of Theorem \ref{thm-gfe-gfs:gamma-cnst}}
If Assumptions \ref{ass-gfs:beta-cnst}\ref{ass-gfs:cpt-param}--\ref{ass-gfs:finmt-z} and \ref{ass-gfe-gfs:beta-cnst} are satisfied with respect to the two-stage model with a group-specific first stage, then the probability of misclassification, that is, $\Pr(\max_i|g_i^\tgfe-g_i^0|>0)$, can be bounded as described in equation \eqref{eq-ts:gamma-cnst} of Lemma \ref{lem-ts:gamma-cnst}. Henceforth, we derive the order of each term of \eqref{eq-ts:gamma-cnst}  using Lemma \ref{lem-bm}, which is applicable due to Assumptions \ref{ass-gfe-gfs:pi-cnst}--\ref{ass-gfe-gfs:gamma-cnst}.

Except for the second and fourth terms, the orders of the probabilities in \eqref{eq-ts:gamma-cnst} can be directly obtained using Lemma \ref{lem-bm}. For this reason, our subsequent part mainly focuses on deriving the orders of these two terms. To enhance notational conciseness, the superscript ``tgfe'' is omitted. Below, we assume $d<\min\{c^\tfs,c^\tss\}$.

First, letting $\mathbf z_{it}\equiv (z_{it}',\sqrt T d1_t,\dots, \sqrt T dT_t)'$, the second term of \eqref{eq-ts:gamma-cnst} can be bounded by
\begin{align*}
	&3\bigg\{\Pr\biggl(\max_{i\in\{1,\dots,N\}}|\hat k_i-k_i^0|>0\biggr)\\
	&\hphantom{3\bigg\{} + \Pr\biggl(\max_{i\in\{1,\dots,N\}}\biggl\|\frac{1}{T}\sum_{t=1}^T\mathbf z_{it}\mathbf z_{it}'\biggr\|_2\smash[t]{\overbrace{\max_{k\in\mathbb K}\biggl(\|\Pi_k^0-\hat\Pi_k\|^2+\frac{1}{T}\sum_{t=1}^T\|\mu_{kt}^0-\hat\mu_{kt}\|^2\biggr)}^{\leq d_H((\hat\Pi,\hat\mu),(\Pi^0,\mu^0))\text{ with probability approaching one}}}\gtrsim d\biggr)\biggr\}.
\end{align*}
As Assumptions \ref{ass-gfs:beta-cnst}\ref{ass-gfs:cpt-param}--\ref{ass-gfs:finmt-z}, \ref{ass-gfe-gfs:beta-cnst}\ref{ass-gfe-gfs:cpt-param},\ref{ass-gfe-gfs:wkcdep-v}--\ref{ass-gfe-gfs:wkdep-v}, and \ref{ass-gfe-gfs:pi-cnst} are satisfied for the first stage,  it follows from (i) of Lemma \ref{lem-sineq:beta-cnst} that $d_H((\hat\Pi,\hat\mu),(\Pi^0,\mu^0))\rightarrow_p 0$, and then from (ii) of Lemma \ref{lem-sineq:gamma-cnst} that the probability of misclassification in the first stage, $\Pr(\max_i|\hat k_i-k_i^0|>0)$, is bounded by
\begin{align}
	\begin{split}
	&\overbrace{\Pr\biggl(\max_{i\in\{1,\dots,N\}}\biggl(\frac{1}{T}\sum_{t=1}^T\|x_{it}'-z_{it}'\hat\Pi_{\hat k_i}-\hat\mu_{\hat k_i t}'\|^2-\min_{k\in\mathbb K}\frac{1}{T}\sum_{t=1}^T\|x_{it}'-z_{it}'\hat\Pi_k-\hat\mu_{kt}'\|^2\biggr)\gtrsim d\biggr)}^{=0\text{ by the definition of $\hat\kappa$}}\\
	& + \Pr\biggl(\min_{i\in\{1,\dots,N\}}\min_{k\neq \tilde k} \frac{1}{T}\sum_{t=1}^T\|z_{it}'(\Pi_k^0-\Pi_{\tilde k}^0) + (\mu_{k t}^{0\prime} - \mu_{\tilde k t}^{0\prime})\|^2<\frac{d}{2}\biggr)\\
	& + \Pr\biggl(\max_{i\in\{1,\dots,N\}}\max_{k\neq \tilde k}\biggl|\frac{1}{T}\sum_{t=1}^Tv_{it}'(\mu_{k t}^{0\prime} - \mu_{\tilde k t}^{0\prime})'\biggr|\gtrsim d\biggr)\\
	& + 2\Pr\biggl(\max_{i\in\{1,\dots,N\}}\biggl\|\frac{1}{T}\sum_{t=1}^Tz_{it}z_{it}'\biggr\|_2> L\biggr) + \Pr\biggl(\max_{i\in\{1,\dots,N\}}\biggl\|\frac{1}{T}\sum_{t=1}^Tz_{it}v_{it}'\biggr\|\gtrsim d \biggr)\\
	& + \Pr\biggl(\biggl(\max_{k\in\mathbb K}\frac{1}{T}\sum_{t=1}^T\|\hat\mu_{kt}-\mu_{kt}^0\|^2\biggr)^{\frac{1}{2}}\biggl(\max_{i\in\{1,\dots,N\}}\frac{1}{T}\sum_{t=1}^T\|v_{it}\|^2\biggr)^{\frac{1}{2}}\gtrsim d\biggr) + o(1).
	\end{split}\label{eq-gfe-gfs:gamma-cnst:bound-fs}
\end{align}
The second term in this bound is $o(NT^{-\delta})$. 
To see this, define $s_{it}\equiv \min_{k\neq \tilde k}\|z_{it}'(\Pi_k^0-\Pi_{\tilde k}^0) + (\mu_{k t}^{0\prime} - \mu_{\tilde k t}^{0\prime})\|^2$. By Assumption \ref{ass-gfe-gfs:kappa-cnst}\ref{ass-gfe-gfs:mix-fs}--\ref{ass-gfe-gfs:tail-fs}, we can apply Lemma \ref{lem-bm} to the demeaned $s_{it}-\Exp[s_{it}]$, which yields
\begin{align*}
	\Pr\biggl(\biggl|\frac{1}{T}\sum_{t=1}^T(s_{it}-\Exp[s_{it}])\biggr|>\frac{d}{6}\biggr)=o(T^{-\delta}).
\end{align*}
Since the right-hand side does not depend on $i$, once combined with Assumption \ref{ass-gfe-gfs:kappa-cnst}\ref{ass-gfe-gfs:grpsep-fs-uni},
\begin{align*}
	&\Pr\biggl(\min_{i\in\{1,\dots,N\}}\frac{1}{T}\sum_{t=1}^Ts_{it}<\frac{d}{2}\biggr)\\
	=&\Pr\biggl(\min_{i\in\{1,\dots,N\}}\biggl\{\frac{1}{T}\sum_{t=1}^T(s_{it}-\Exp[s_{it}])+\frac{1}{T}\sum_{t=1}^T\Exp[s_{it}]\biggr\}<\frac{d}{2}\biggr)\\
	=&\Pr\biggl(\min_{i\in\{1,\dots,N\}}\biggl\{-\max_{i\in\{1,\dots,N\}}\biggl|\frac{1}{T}\sum_{t=1}^T(s_{it}-\Exp[s_{it}])\biggr|+\frac{1}{T}\sum_{t=1}^T\Exp[s_{it}]\biggr\}<\frac{d}{2}\biggr)\\
	\leq& \Pr\biggl(\max_{i\in\{1,\dots,N\}}\biggl|\frac{1}{T}\sum_{t=1}^T(s_{it}-\Exp[s_{it}])\biggr|>\frac{d}{6}\biggr) + \underbrace{\Pr\biggl(\min_{i\in\{1,\dots,N\}}\frac{1}{T}\sum_{t=1}^T\Exp[s_{it}]\leq \frac{2d}{3}\biggr)}_{=0 \text{ if $T$ is large enough }}
	=o(NT^{-\delta}).
\end{align*}
Since $\|v_{it}\|^2$ shares the mixing coefficients with $v_{it}$, through the similar process, by Assumption \ref{ass-gfe-gfs:pi-cnst}\ref{ass-gfe-gfs:mix-fs}, we can establish
\begin{align*}
	\Pr\biggl(\max_{i\in\{1,\dots,N\}}\frac{1}{T}\sum_{t=1}^T\|v_{it}\|^2\gtrsim \tilde L\biggr) = o(NT^{-\delta}).
\end{align*}
Combining this with the consistency of $\hat\mu$, the sixth term in the previous \eqref{eq-gfe-gfs:gamma-cnst:bound-fs} is $o(NT^{-\delta})+o(1)$. By applying Lemma \ref{lem-bm} to $v_{it}'(\mu_{kt}^{0\prime}-\mu_{\tilde kt}^{0\prime})'$, it can also be shown that the third term is $o(NT^{-\delta})$. The fourth and fifth terms are $o(NT^{-\delta})$ by Assumption \ref{ass-gfs:kappa-cnst}\ref{ass-gfs:meantail-z}--\ref{ass-gfs:meantail-zv}. Summing up, under our assumptions, the second term of \eqref{eq-ts:gamma-cnst} is $o(NT^{-\delta})+o(1)$.

Now, the order of the fourth term of \eqref{eq-ts:gamma-cnst} can be readily obtained from that of the second term. By the CS inequality,
\begin{align*}	
	&\biggl\|\frac{1}{T}\sum_{t=1}^T(v_{it}'\beta_{g_i^0}^0+u_{it})((\Pi_{k_i}^0-\hat\Pi_{\hat k_i})'z_{it}+(\mu_{k_i^0t}^0-\hat\mu_{\hat k_it}))'\biggr\|\\
	\leq& \biggl(\frac{1}{T}\sum_{t=1}^T\|v_{it}'\beta_{g_i^0}^0+u_{it}\|^2\biggr)^{\frac{1}{2}}
	\biggl(\frac{1}{T}\sum_{t=1}^T\|(\Pi_{k_i}^0-\hat\Pi_{\hat k_i})'z_{it}+(\mu_{k_i^0t}^0-\hat\mu_{\hat k_it})\|^2\biggr)^{\frac{1}{2}},
\end{align*}
and so we can bound the fourth term by
\begin{align*}
	\Pr\biggl(\max_{i\in\{1,\dots,N\}}\frac{1}{T}\sum_{t=1}^T\|v_{it}\|^2\gtrsim \tilde L\biggr)
	+ \Pr\biggl(\max_{i\in\{1,\dots,N\}}\frac{1}{T}\sum_{t=1}^T|u_{it}|^2\gtrsim \tilde L\biggr) + \smash[t]{\overbrace{o(NT^{-\delta})+o(1)}^{\mathclap{\text{the order of the second term}}}}.
\end{align*}
Recall that we already have shown that the first tail mass is of the order $o(NT^{-\delta})$. Using the same technique, under Assumption \ref{ass-gfe-gfs:gamma-cnst}\ref{ass-gfe-gfs:mix-ss}, we can show that the second one is also $o(NT^{-\delta})$. Hence, the fourth term is also $o(NT^{-\delta})+o(1)$. 

Combining Lemma \ref{lem-bm} with the relevant assumptions, we can find that the third, fifth to seventh, and twelfth terms in \eqref{eq-ts:gamma-cnst} are of the order $o(NT^{-\delta})(+o(1))$. In addition, the first, and the eighth to ninth terms are $o(NT^{-\delta})$ directly by assumption. Combining Lemma \ref{lem-bm} with the consistency of $\tilde\alpha$, we can show that the remaining tenth to eleventh terms are $o(NT^{-\delta})+o(1)$.

Summing our results thus far, we conclude that $\Pr(\max_i|\hat g_i-g_i^0|>0)$ is no larger than $o(NT^{-\delta})+o(1)$.

\subsection{Proof of Theorem \ref{thm-gfe-ufs:beta-cnst}}
Letting $\overline z_{it}\equiv (\sqrt N z_{it}'\mathbf 1\{1=i\},\dots,\sqrt N z_{it}'\mathbf 1\{N=i\})'$, observe that the first stage model can be recast into a homogeneous linear model:
\begin{align}
	x_{it}' = z_{it}'\Pi_i^0 + \mu_{k_i^0t}^{0\prime} + v_{it}'
	=& \sum_{j=1}^N\underbrace{z_{it}'\mathbf 1\{j=i\}}_{\text{ regressors}}\overbrace{\Pi_j^0}^{\mathclap{\text{coefficients}}} + \mu_{k_i^0 t}^{0\prime} + v_{it}'
	= \overline z_{it}'\smash[t]{\overbrace{\begin{pmatrix}
		\frac{1}{\sqrt N}\Pi_1^0 \\ \vdots \\ \frac{1}{\sqrt N}\Pi_N^0
	\end{pmatrix}}^{\equiv \overline \Pi^0}} + \mu_{k_i^0t}^{0\prime} + v_{it}',
	\label{eq-gfe-ufs:beta-cnst:homogeneous}
\end{align}
which can be fit into the two-stage model on which Lemma \ref{lem-ts} is based. Because the TGFE estimator under this formulation is the UGFE estimator, which is of current interest, it is sufficient to check if the required conditions for the consistency of the TGFE estimator are satisfied. We demonstrate that after replacing $z_{it}$ and $\Pi_{k_i^0}^0$ with $\overline z_{it}$ and $\overline \Pi^0$, Assumption \ref{ass-ts} still holds, thereby allowing us to invoke Lemma \ref{lem-ts:beta-cnst} for consistency.  

The space of the coefficient parameter is uniformly bounded, since for every $N$,
\begin{align*}
	\|\overline \Pi\|
	\leq \sqrt{\biggl\|\frac{1}{\sqrt N}\Pi_1\biggr\|^2 + \biggl\|\frac{1}{\sqrt N}\Pi_2\biggr\|^2 + \cdots + \biggl\|\frac{1}{\sqrt N}\Pi_N\biggr\|^2}\leq \sqrt{\max_{\Pi_i\in\mathbf \Pi}\|\Pi_i\|^2},
\end{align*}
which is finite by Assumption \ref{ass-ufs:beta-cnst}\ref{ass-ufs:cpt-param}. Thus, Assumption \ref{ass-ts}\ref{ass-ts:cpt-param} holds. 

Assumptions \ref{ass-ts}\ref{ass-ts:wkcdep-v}--\ref{ass-ts:finmt-z} are direct from the given assumptions, so we only need to examine whether Assumptions \ref{ass-ts}\ref{ass-ts:wkdep-zv} is satisfied. However, due to the homogeneity of the first stage, as noted in Lemma \ref{lem-ts:beta-cnst}, $\Exp[\|(NT)^{-1}\sum_{i=1}^N\sum_{t=1}^T\overline z_{it}v_{it}'\|^2]\leq M$ will suffice. Now, note that
\begin{align*}
	\biggl\|\frac{1}{NT}\sum_{i=1}^N\sum_{t=1}^T\overline z_{it}v_{it}'\biggr\|^2
	=&\left\|\frac{1}{NT}\sum_{i=1}^N\sum_{t=1}^T\begin{pmatrix}
		\sqrt N z_{it}\mathbf 1\{1=i\}\\
		\vdots\\
		\sqrt N z_{it}\mathbf 1\{N=i\}
	\end{pmatrix}v_{it}'\right\|^2\\
	=&\left\|\begin{pmatrix}
		\frac{1}{\sqrt NT}\sum_{t=1}^Tz_{1t}v_{1t}'\\
		\vdots\\
		\frac{1}{\sqrt NT}\sum_{t=1}^Tz_{Nt}v_{Nt}'
	\end{pmatrix}\right\|^2
	=\underbrace{\biggl\|\frac{1}{\sqrt NT}\sum_{t=1}^Tz_{1t}v_{1t}'\biggr\|^2 + \cdots + \biggl\|\frac{1}{\sqrt NT}\sum_{t=1}^Tz_{Nt}v_{Nt}'\biggr\|^2}_{\displaystyle\mathclap{=\frac{1}{T}\cdot\frac{1}{NT}\sum_{i=1}^N\biggl\|\sum_{t=1}^Tz_{it}v_{it}'\biggr\|^2\leq \frac{M}{T}\text{ by Assumption \ref{ass-gfs:beta-cnst}\ref{ass-gfs:wkdep-zv}.}}}	
\end{align*}

\subsection{Proof of Theorem \ref{thm-gfe-ufs:gamma-cnst}}
Recall that the UGFE estimator can be understood as a TGFE estimator once we recast the unit-specific first stage into a homogeneous counterpart as detailed in equation \eqref{eq-gfe-ufs:beta-cnst:homogeneous}. This viewpoint allows us to use Lemma \ref{lem-ts:gamma-cnst} and obtain the bound \eqref{eq-ts:gamma-cnst} for $\Pr(\max_i|\hat g_i^\ugfe - g_i^0|>0)$, with $\Pi_{k_i^0}^0$ $\hat\Pi_{\hat k_i}$, and $z_{it}$ replaced respectively by $\overline\Pi^0$, $\hat{\overline \Pi}$, and $\overline z_{it}$ introduced in \eqref{eq-gfe-ufs:beta-cnst:homogeneous}. As in the proof of Theorem \ref{thm-gfe-gfs:gamma-cnst}, the order of \eqref{eq-ts:gamma-cnst} can thus be derived by determining that of each of its components. However, to avoid redundancy, our focus hereafter will be on the second term:
\begin{align*}
	\Pr\biggl(\max_{i\in\{1,\dots,N\}}\frac{1}{T}\sum_{t=1}^T\|\smash[t]{\overbrace{(\Pi_i^0-\hat\Pi_i^\ugfe)z_{it}}^{(\overline \Pi^0-\hat{\overline \Pi})'\overline z_{it}}}+(\mu_{k_i^0t}^0-\hat\mu_{\hat k_i^\ugfe t}^\ugfe)\|^2\gtrsim d\biggr),
\end{align*}
where a major difference from the case of Theorem \ref{thm-gfe-gfs:gamma-cnst} occurs.

Given the first-stage group-specific fixed effects $\{\{\mu_{kt}\}_{t=1}^T\}_{k\in \mathbb K}$, define by 
\begin{align*}
	\Pi_{i,\{\mu_{kt}\}_{t=1}^T}\equiv \arg\min_{\Pi_i\in\mathbf \Pi}\frac{1}{T}\sum_{t=1}^T\|x_{it}-\Pi_i'z_{it}-\mu_{kt}\|^2
\end{align*} 
the coefficient parameter that minimizes the unit-specific sum of squared errors in case where the unit $i$ is assigned to group $k$. By the definition of the UGFE estimator, we have
\begin{align*}
	\hat\Pi_i^\ugfe = \Pi_{i,\{\hat\mu_{\hat k_i^\ugfe t}^\ugfe\}_{t=1}^T},
\end{align*}
and $\hat k_i^\ugfe = \arg\min_{k\in\mathbb K}T^{-1}\sum_{t=1}^T\|x_{it}-\Pi_{i,\{\hat\mu_{kt}^\ugfe\}_{t=1}^T}'z_{it}-\hat\mu_{kt}^\ugfe\|^2$.
Now, from the inequality:
\begin{align*}
	\frac{1}{T}\sum_{t=1}^T\|(\mu_{k_i^0t}^0-\hat\mu_{k_i^0t}^\ugfe) + v_{it}\|^2
	=&\frac{1}{T}\sum_{t=1}^T\|x_{it}-\Pi_i^{0\prime}z_{it}-\hat\mu_{k_i^0t}^\ugfe\|^2\\
	\geq& \frac{1}{T}\sum_{t=1}^T\|x_{it}-\Pi_{i,\{\hat\mu_{k_i^0t}^\ugfe\}_{t=1}^T}'z_{it}-\hat\mu_{k_i^0t}^\ugfe\|^2\\
	\geq& \frac{1}{T}\sum_{t=1}^T\|x_{it}-\Pi_{i,\{\hat\mu_{\hat k_i^\ugfe t}^\ugfe\}_{t=1}^T}'z_{it}-\hat\mu_{\hat k_i^\ugfe t}^\ugfe\|^2\\
	=& \frac{1}{T}\sum_{t=1}^T\|(\Pi_i^0-\hat\Pi_i^\ugfe)'z_{it}+(\mu_{k_i^0t}^0-\hat\mu_{\hat k_i^\ugfe t}^\ugfe) + v_{it}\|^2,
\end{align*}
we can derive an upper bound of
\begin{align*}
	&\frac{1}{T}\sum_{t=1}^T\|(\Pi_i^0-\hat\Pi_i^\ugfe)'z_{it}+(\mu_{k_i^0t}^0-\hat\mu_{\hat k_i^\ugfe t}^\ugfe)\|^2\\
	\leq& \frac{1}{T}\sum_{t=1}^T\|\mu_{k_i^0t}^0-\hat\mu_{k_i^0t}^\ugfe\|^2 + \frac{2}{T}\sum_{t=1}^T(\mu_{k_i^0t}^0-\hat\mu_{k_i^0t}^\ugfe)'v_{it}
	- \frac{2}{T}\sum_{t=1}^T((\Pi_i^0-\hat\Pi_i^\ugfe)'z_{it}+(\mu_{k_i^0t}^0-\hat\mu_{\hat k_i^\ugfe t}^\ugfe))'v_{it}\\
	=& \frac{1}{T}\sum_{t=1}^T\|\mu_{k_i^0t}^0-\hat\mu_{k_i^0t}^\ugfe\|^2 - \frac{2}{T}\sum_{t=1}^T((\Pi_i^0-\hat\Pi_i^\ugfe)'z_{it})'v_{it} 
	- \underbrace{\frac{2}{T}\sum_{t=1}^T(\hat\mu_{k_i^0t}^\ugfe - \hat\mu_{\hat k_i^\ugfe t}^\ugfe)'v_{it}}_{\mathclap{=\frac{2}{T}\sum_{t=1}^T((\hat\mu_{k_i^0t}^\ugfe-\mu_{k_i^0t}^0)-(\hat\mu_{\hat k_i^\ugfe t}^\ugfe-\mu_{\hat k_i^\ugfe t}^0))'v_{it} + \frac{2}{T}\sum_{t=1}^T(\mu_{k_i^0t}^0-\mu_{\hat k_i^\ugfe t}^0)'v_{it}}}\\
	\leq& \max_{k\in\mathbb K}\frac{1}{T}\sum_{t=1}^T\|\mu_{kt}^0-\hat\mu_{kt}^\ugfe\|^2 
	+ 4\max_{i\in\{1,\dots,N\}}\biggl\|\frac{1}{T}\sum_{t=1}^Tz_{it}v_{it}'\biggr\|\max_{\Pi_i\in\mathbf \Pi}\|\Pi_i\|\\
	&+ 4\biggl(\max_{k\in\mathbb K}\frac{1}{T}\sum_{t=1}^T\|\mu_{kt}^0-\hat\mu_{kt}^\ugfe\|^2\biggr)^{\frac{1}{2}}\biggl(\max_{i\in\{1,\dots,N\}}\frac{1}{T}\sum_{t=1}^T\|v_{it}\|^2\biggr)^{\frac{1}{2}}
	+ 2\max_{i\in\{1,\dots,N\}}\max_{k\neq\tilde k}\biggl|\frac{1}{T}\sum_{t=1}^T(\mu_{kt}^0 - \mu_{\tilde kt}^0)'v_{it}\biggr|,
\end{align*}
which neither depends on $i$ nor $k$. Once combined with the consistency of $\hat\mu^\ugfe$, which will be shown right after, the second term of \eqref{eq-ts:gamma-cnst} is then bounded by
\begin{align*}
	& o(1) + \Pr\biggl(\max_{i\in\{1,\dots,N\}}\biggl\|\frac{1}{T}\sum_{t=1}^Tz_{it}v_{it}'\biggr\|\gtrsim d\biggr) + \Pr\biggl(\max_{i\in\{1,\dots,N\}}\frac{1}{T}\sum_{t=1}^T\|v_{it}\|^2\gtrsim \tilde L\biggr)\\
	& + \Pr\biggl(\max_{i\in\{1,\dots,N\}}\max_{k\neq\tilde k}\biggl|\frac{1}{T}\sum_{t=1}^T(\mu_{kt}^0 - \mu_{\tilde kt}^0)'v_{it}\biggr|\gtrsim d\biggr)= o(1) + o(NT^{-\delta}).
\end{align*}
The order $o(NT^{-\delta})$ here is from Lemma \ref{lem-bm}, as was in Theorem \ref{thm-gfe-gfs:gamma-cnst}.

Now, we establish the consistency of $\hat\mu^\ugfe$. This proof amounts to checking if the unit-specific first stage, when considered as the homogeneous model \eqref{eq-gfe-ufs:beta-cnst:homogeneous}, satisfies the required conditions for (ii) of Lemma \ref{lem-sineq:beta-cnst}, which are:
\begin{align}
	\frac{1}{NT}\sum_{i=1}^N\sum_{t=1}^T\|\underbrace{z_{it}'(\Pi_i^0-\hat\Pi_i^\ugfe)}_{\overline z_{it}'(\overline \Pi^0-\hat{\overline \Pi}^\tgfe)}+(\mu_{k_i^0t}^{0\prime}-\hat\mu_{\hat k_i^\ugfe t}^{\ugfe\prime})\|^2\rightarrow_p 0 \label{eq-gfe-ufs:gamma-cnst:fit-fs}
\end{align}
and for all $v\in \mathbb R^{m\times N + T}$,
\begin{align}
	v'\biggl(\frac{1}{N}\sum_{i=1}^N\mathbf 1\{k_i^0=k\}\mathbf 1\{k_i=\tilde k\}\frac{1}{T}	\sum_{t=1}^T\overline z_{it}\overline z_{it}'\biggr)v
	\geq \rho(\gamma,g,\tilde g)\|v_{\{mN+1,\dots,mN+T\}}\|^2. \label{eq-gfe-ufs:gamma-cnst:relev-fs}
\end{align}

Equation \eqref{eq-gfe-ufs:gamma-cnst:fit-fs} holds by Lemma \ref{lem-sineq:fit}, for which due to the first-stage homogeneity, a weaker form of exogeneity $\Exp[\|(NT)^{-1}\sum_{i=1}^N\sum_{t=1}^T\overline z_{it}v_{it}'\|^2]\leq M$ suffices. Recall that we have already derived this condition from the given assumptions in the proof of Theorem \ref{thm-gfe-ufs:beta-cnst}. 

Regarding equation \eqref{eq-gfe-ufs:gamma-cnst:relev-fs}, note that its left-hand side is no smaller than
\begin{align*}
	&v'\begin{pmatrix}
		\begin{matrix}
			\displaystyle\diag_i\frac{1}{N}\mathbf 1\{k_i^0=k\}\mathbf 1\{k_i=\tilde k\}\frac{1}{T}\sum_{t=1}^Tz_{it}z_{it}
		\end{matrix}
		& \mathbf O_{mN\times T}\\
		\mathbf O_{T\times mN} & \frac{1}{N}\sum_{i=1}^N\mathbf 1\{k_i^0=k\}\mathbf 1\{k_i=\tilde k\}I_T
	\end{pmatrix}v\\
	\geq& v_{\{mN+1,\dots,mN+T\}}'\biggl(\frac{1}{N}\sum_{i=1}^N\mathbf 1\{k_i^0=k\}\mathbf 1\{k_i=\tilde k\}I_T\biggr)v_{\{mN+1,\dots,mN+T\}}.
\end{align*}
Thus, by Assumption \ref{ass-gfe-ufs:pi-cnst}\ref{ass-gfe-ufs:relev-fs}, Assumption \ref{ass-sineq}\ref{ass-sineq:relev-ss} is satisfied with $J\supseteq \{mN+1,\dots,mN+T\}$. 
Since Assumption \ref{ass-sineq}\ref{ass-sineq:grpsep-ss} follows from Assumption \ref{ass-gfe-ufs:pi-cnst}\ref{ass-gfe-ufs:grpsep-fs}, by (ii) of Lemma \ref{lem-sineq:beta-cnst}, the desired consistency holds.

The rest of the terms constituting the bound \eqref{eq-ts:gamma-cnst} can be shown $o(NT^{-\delta})(+ o(1))$ as in Theorem \ref{thm-gfe-gfs:gamma-cnst}.

\section{Choice of the number of groups}\label{sec-appdx:num-of-groups}
We provide a set of high-level conditions under which the selection procedure for the number of groups, as introduced in Subsection \ref{sec:select-num-grps}, is valid.
Here, our focus is on the TGFE estimation. We consider a general setting with time-varying fixed effects.
For simplicity, we will omit the ``tgfe'' superscript.

\begin{assumption}\label{ass-ic-fs} As $N$ and $T$ tend to infinity,
\begin{enumerate}[label=(\alph*),ref=(\alph*)]
	\item \label{ass-ic-fs:Kl} for $K<K^0$, there exists $c_{\K}^\tfs>0$ such that 
	\begin{align*}
		\frac{1}{NT}\sum_{i=1}^N\sum_{t=1}^T\|(\Pi_{k_i^0}^0-\hat\Pi_{\hat k_i^{\K}}^{\K\prime})z_{it}+(\mu_{k_i^0}^0-\hat\mu_{\hat k_i^{\K} t}^{\K})\|^2
		\rightarrow_p c_{\K}^\tfs,
	\end{align*}
	while for $K\geq K^0$, 
	\begin{align*}
		\sum_{i=1}^N\sum_{t=1}^T\|(\Pi_{k_i^0}^0-\hat\Pi_{\hat k_i^{\K}}^{\K})'z_{it}+(\mu_{k_i^0 t}^0-\hat\mu_{\hat k_i^{\K} t}^{\K})\|^2 = O_p(T).
	\end{align*}
	\item \label{ass-ic-fs:Kg} for all $K\geq K^{\max}$, $\sum_{i=1}^N\sum_{t=1}^T((\Pi_{k_i^0}^0-\hat\Pi_{\hat k_i^{\K}}^{\K})'z_{it}+(\mu_{k_i^0 t}^0-\hat\mu_{\hat k_i^{\K} t}^{\K}))'v_{it} = O_p(T)$.
	\item \label{ass-ic-fs:c} for all $K\leq K^{\max}$, $c^\tfs(K,N,T)/(NT)\rightarrow 0$. If $K>\tilde K$, 
	\begin{align*}
		(c^\tfs(K,N,T) - c^\tfs(\tilde K, N, T))/T\rightarrow \infty.
	\end{align*}
\end{enumerate}
\end{assumption}
\begin{lemma} \label{thm-ic-fs}
Under Assumption \ref{ass-ic-fs}, as $N$ and $T$ tend to infinity, $\Pr(\hat K = K^0)\rightarrow 1$.
\end{lemma}

\begin{assumption}\label{ass-ic-ss} As $N$ and $T$ tend to infinity,
\begin{enumerate}[label=(\alph*),ref=(\alph*)]
	\item \label{ass-ic-ss:Gl} for $G<G^0$ and $K=K^0$, there exists $c_{\G}^\tss>0$ such that
	\begin{align*}
		\frac{1}{NT}\sum_{i=1}^N\sum_{t=1}^T(y_{it}-(z_{it}'\hat\Pi_{\hat k_i^{\Kp}}^{\Kp}+\hat\mu_{\hat k_i^{\Kp}}^{\Kp})\hat\beta_{\hat g_i^{\GKp}}^{\GKp}-\hat\alpha_{\hat g_i^{\GKp} t}^{\GKp})^2
		\rightarrow_p c_{\G}^\tss,
	\end{align*}
	while for $G\geq G^0$,
	\begin{align*}
		\sum_{i=1}^N\sum_{t=1}^T((z_{it}'\hat\Pi_{\hat k_i^{\Kp}}^{\Kp}+\hat\mu_{\hat k_i^{\Kp}}^{\Kp})(\beta_{g_i^0}^0-\hat\beta_{\hat g_i^{\GpKp}}^{\GpKp})
		+ (\alpha_{g_i^0 t}^0-\hat\alpha_{\hat g_i^{\GpKp} t}^{\GpKp}))^2 = O_p(T)
	\end{align*}
	\item \label{ass-ic-ss:Gg} for $G\geq G^0$ and $K=K^0$, 
	\begin{align*}
		&\sum_{i=1}^N\sum_{t=1}^T((z_{it}'\hat\Pi_{\hat k_i^{\Kp}}^{\Kp}+\hat\mu_{\hat k_i^{\Kp}}^{\Kp})(\beta_{g_i^0}^0-\hat\beta_{\hat g_i^{\GpKp}}^{\GpKp})
		+ (\alpha_{g_i^0 t}^0-\hat\alpha	_{\hat g_i^{\GpKp} t}^{\GpKp}))
		v_{it}' = O_p(T),\\
		&\sum_{i=1}^N\sum_{t=1}^T((z_{it}'\hat\Pi_{\hat k_i^{\Kp}}^{\Kp}+\hat\mu_{\hat k_i^{\Kp}}^{\Kp})(\beta_{g_i^0}^0-\hat\beta_{\hat g_i^{\GpKp}}^{\GpKp})
		+ (\alpha_{g_i^0 t}^0-\hat\alpha_{\hat g_i^{\GpKp} t}^{\GpKp}))
		u_{it} = O_p(T),\\
		&\sum_{i=1}^N\sum_{t=1}^T((\Pi_{k_i^0}^0-\hat\Pi_{\hat k_i^{\Kp}}^{\Kp})'z_{it}+(\mu_{k_i^0 t}^0-\hat\mu_{\hat k_i^{\Kp} t}^{\Kp}))'u_{it} = O_p(T).
	\end{align*}
	\item \label{ass-ic-ss:c} for all $G$, $c^\tss(G,N,T)/(NT)\rightarrow 0$. If $G>\tilde G$, 
	\begin{align*}
		(c^\tss(G,N,T) - c^\tss(\tilde G,N,T))/T\rightarrow 0.
	\end{align*}
\end{enumerate}	
\end{assumption}

\begin{theorem} \label{thm-ic-ss}
Under Assumption \ref{ass-ic-fs}--\ref{ass-ic-ss}, as $N$ and $T$ tend to infinity, $\Pr(\hat G = G^0)\rightarrow 1$. 	
\end{theorem}

If the model does not include any time-varying fixed effects, then the orders in Assumptions \ref{ass-ic-fs}\ref{ass-ic-fs:Kg} and \ref{ass-ic-ss}\ref{ass-ic-ss:Gg} could be weakened into $O_p(NT)$. Accordingly, the second conditions for $c^\tfs$ and $c^\tss$ in Assumptions \ref{ass-ic-fs}\ref{ass-ic-fs:c} and \ref{ass-ic-ss}\ref{ass-ic-ss:c} can be adjusted as: $c^\tfs(K,N,T) - c^\tfs(\tilde K, N, T)\rightarrow\infty$.

\subsection{Proofs of Lemma \ref{thm-ic-fs} and Theorem \ref{thm-ic-fs}}

\subsubsection{Proof of Lemma \ref{thm-ic-fs}}
Define:
\begin{align*}
	A^\tfs(K) \equiv 
	\frac{1}{NT}\sum_{i=1}^N\sum_{t=1}^T\|x_{it}-\hat\Pi_{\hat k_i^{\K}}^{\K\prime}z_{it}-\hat\mu_{\hat k_i^{\K} t}^{\K}\|^2
	- \frac{1}{NT}\sum_{i=1}^N\sum_{t=1}^T\|x_{it} - \Pi_{k_i^0}^{0\prime}z_{it} - \mu_{k_i^0 t}^0\|^2.
\end{align*} 
By Assumption \ref{ass-ic-fs}\ref{ass-ic-fs:Kl}--\ref{ass-ic-fs:Kg}, for $K\geq K^0$,
\begin{align*}
	A^\tfs(K)
	=& \frac{1}{NT}\sum_{i=1}^N\sum_{t=1}^T\|(\Pi_{k_i^0}^0-\hat\Pi_{\hat k_i^{\K}}^{\K})'z_{it}+(\mu_{k_i^0 t}^0-\hat\mu_{\hat k_i^{\K} t}^{\K})\|^2\\
	&\quad + \frac{1}{NT}\sum_{i=1}^N\sum_{t=1}^T((\Pi_{k_i^0}^0-\hat\Pi_{\hat k_i^{\K}}^{\K})'z_{it}+(\mu_{k_i^0 t}^0-\hat\mu_{\hat k_i^{\K} t}^{\K}))'v_{it}
	= O_p(N^{-1}).
\end{align*}

Suppose that $K<K^0$. By Assumption \ref{ass-ic-fs}\ref{ass-ic-fs:Kl} and \ref{ass-ic-fs:c},
\begin{align*}	
	\IC^\tfs(K) - \IC^\tfs(K^0)
	= \underbrace{A^\tfs(K)}_{\rightarrow_p c_{\K}^\tfs} - \underbrace{A^\tfs(K^0)}_{O_p(N^{-1})} + \frac{c^\tfs(K,N,T) - c^\tfs(K^0,N,T)}{NT}
	\rightarrow c_{\K}^\tfs > 0.
\end{align*}
This implies $\Pr(\IC^\tfs(K) - \IC^\tfs(K^0)>0)\rightarrow 1$ for $K< K^0$.

On the other hand, if $K>K^0$, 
\begin{align*}
	\IC^\tfs(K) - \IC^\tfs(K^0) 
	=& A^\tfs(K) - A^\tfs(K^0) \\
	=& \frac{O_p(T) + c^\tfs(K,N,T) - c^\tfs(K^0,N,T)}{NT}.
\end{align*}
It then follows from Assumption \ref{ass-ic-fs}\ref{ass-ic-fs:c} that
\begin{align*}
	\Pr(\IC^\tfs(K) - \IC^\tfs(K^0)>0) = \Pr(c^\tfs(K,N,T) - c^\tfs(K^0,N,T)>-O_p(T))\rightarrow 1.
\end{align*}

Combining results,
\begin{align*}
	&\Pr(\hat K = K^0)\\
	=& \Pr(\IC^\tfs(K) > \IC^\tfs(K^0),\ \forall K\in \{1,\dots,K^{\max}\} \setminus \{K^0\})\\
	\leq& 1 - \sum_{K\in \{1,\dots,K^{\max}\} \setminus \{K^0\}}\Pr(\IC^\tfs(K)\leq \IC^\tfs(K^0))
	\rightarrow 1.
\end{align*}
This completes the proof of Lemma \ref{thm-ic-fs}.

\subsubsection{Proof of Theorem \ref{thm-ic-ss}}
Define:
\begin{align*}
	A^\tss(G,K)
	\equiv& \frac{1}{NT}\sum_{i=1}^N\sum_{t=1}^T(y_{it}-(z_{it}'\hat\Pi_{\hat k_i^{\K}}^{\K}+\hat\mu_{\hat k_i^{\K}}^{\K})\hat\beta_{\hat g_i^{\GK}}^{\GK}-\hat\alpha_{\hat g_i^{\GK} t}^{\GK})^2\\
	&\quad - \frac{1}{NT}\sum_{i=1}^N\sum_{t=1}^T(y_{it}-(z_{it}'\hat\Pi_{\hat k_i^{\K}}^{\K}+\hat\mu_{\hat k_i^{\K}}^{\K})\beta_{g_i^0}^0-\alpha_{g_i^0 t}^0)^2.
\end{align*}
Suppose $K=K^0$. Then,
\begin{align*}
	&A^\tss(G,K^0)\\
	=&\frac{1}{NT}\sum_{i=1}^N\sum_{t=1}^T(y_{it}-(z_{it}'\hat\Pi_{\hat k_i^{\Kp}}^{\Kp}+\hat\mu_{\hat k_i^{\Kp}}^{\Kp})\hat\beta_{\hat g_i^{\GKp}}^{\GKp}-\hat\alpha_{\hat g_i^{\GKp} t}^{\GKp})^2\\
	& - \frac{1}{NT}\sum_{i=1}^N\sum_{t=1}^T(y_{it}-(z_{it}'\hat\Pi_{\hat k_i^{\Kp}}^{\Kp}+\hat\mu_{\hat k_i^{\Kp}t}^{\Kp})\beta_{g_i^0}^0-\alpha_{g_i^0 t}^0)^2\\
	=&\frac{1}{NT}\sum_{i=1}^N\sum_{t=1}^T((z_{it}'\hat\Pi_{\hat k_i^{\Kp}}^{\Kp}+\hat\mu_{\hat k_i^{\Kp}t}^{\Kp})(\beta_{g_i^0}^0-\hat\beta_{\hat g_i^{\GKp}}^{\GKp})
	+ (\alpha_{g_i^0 t}^0-\hat\alpha_{\hat g_i^{\GKp} t}^{\GKp}))^2\\
	&+ \frac{2}{NT}\sum_{i=1}^N\sum_{t=1}^T((z_{it}'\hat\Pi_{\hat k_i^{\Kp}}^{\Kp}+\hat\mu_{\hat k_i^{\Kp}t}^{\Kp})(\beta_{g_i^0}^0-\hat\beta_{\hat g_i^{\GKp}}^{\GKp})
	+ (\alpha_{g_i^0 t}^0-\hat\alpha_{\hat g_i^{\GKp} t}^{\GKp}))\\
	&\hphantom{+ \frac{2}{NT}\sum_{i=1}^N\sum_{t=1}^T}\times (y_{it}-(z_{it}'\hat\Pi_{\hat k_i^{\Kp}}^{\Kp}+\hat\mu_{\hat k_i^{\Kp}t}^{\Kp})\beta_{g_i^0}^0-\alpha_{g_i^0 t}^0)\\
	=&\frac{1}{NT}\sum_{i=1}^N\sum_{t=1}^T((z_{it}'\hat\Pi_{\hat k_i^{\Kp}}^{\Kp}+\hat\mu_{\hat k_i^{\Kp}t}^{\Kp})(\beta_{g_i^0}^0-\hat\beta_{\hat g_i^{\GKp}}^{\GKp})
	+ (\alpha_{g_i^0 t}^0-\hat\alpha_{\hat g_i^{\GKp} t}^{\GKp}))^2\\
	&+ \frac{2}{NT}\sum_{i=1}^N\sum_{t=1}^T((z_{it}'\hat\Pi_{\hat k_i^{\Kp}}^{\Kp}+\hat\mu_{\hat k_i^{\Kp}t}^{\Kp})(\beta_{g_i^0}^0-\hat\beta_{\hat g_i^{\GKp}}^{\GKp})
	+ (\alpha_{g_i^0 t}^0-\hat\alpha_{\hat g_i^{\GKp} t}^{\GKp}))\\
	&\hphantom{+ \frac{2}{NT}\sum_{i=1}^N\sum_{t=1}^T}\times ((z_{it}'(\Pi_{k_i^0}^0-\hat\Pi_{\hat k_i^{\Kp}}^{\Kp})+(\mu_{k_i^0t}^0-\hat\mu_{\hat k_i^{\Kp}t}^{\Kp}))\beta_{g_i^0}^0+(v_{it}'\beta_{g_i^0}^0+u_{it}))
\end{align*}

If $G>G^0$, then by Assumption \ref{ass-ic-ss}\ref{ass-ic-ss:Gl}--\ref{ass-ic-ss:Gg}, and since
\begin{align*}
	&\Biggl|\frac{1}{NT}\sum_{i=1}^N\sum_{t=1}^T((z_{it}'\hat\Pi_{\hat k_i^{\Kp}}^{\Kp}+\hat\mu_{\hat k_i^{\Kp}t}^{\Kp})(\beta_{g_i^0}^0-\hat\beta_{\hat g_i^{\GpKp}}^{\GpKp})
	+ (\alpha_{g_i^0 t}^0-\hat\alpha_{\hat g_i^{\GpKp} t}^{\GpKp}))\\
	&\quad\quad\quad\quad\quad\quad \times (z_{it}'(\Pi_{k_i^0}^0-\hat\Pi_{\hat k_i^{\Kp}}^{\Kp})+(\mu_{k_i^0t}^0 - \hat\mu_{\hat k_i^{\Kp}t}^{\Kp}))\beta_{g_i^0}^0\Biggr|\\
	\leq& \Biggl(\frac{1}{NT}\sum_{i=1}^N\sum_{t=1}^T((z_{it}'\hat\Pi_{\hat k_i^{\Kp}}^{\Kp}+\hat\mu_{\hat k_i^{\Kp}t}^{\Kp})(\beta_{g_i^0}^0-\hat\beta_{\hat g_i^{\GpKp}}^{\GpKp})
	+ (\alpha_{g_i^0 t}^0-\hat\alpha_{\hat g_i^{\GpKp} t}^{\GpKp}))^2\Biggr)^{\frac{1}{2}}\\
	&\quad\quad \times \Biggl(\frac{1}{NT}\sum_{i=1}^N\sum_{t=1}^T(z_{it}'(\Pi_{k_i^0}^0-\hat\Pi_{\hat k_i^{\Kp}}^{\Kp})+(\mu_{k_i^0t}^0 - \hat\mu_{\hat k_i^{\Kp}t}^{\Kp}))^2\Biggr)^{\frac{1}{2}}
	= O_p(N^{-1}),
\end{align*}
where we use Assumption \ref{ass-ic-fs}\ref{ass-ic-fs:Kg} in the equality, $A^\tss(G,K^0)=O_p(N^{-1})$. Thus,
\begin{align*}
	\IC(G,K^0) - \IC(G^0,K^0)
	= \frac{O_p(T) + c^\tss(G,N,T) - c^\tss(G^0,N,T)}{NT},
\end{align*}
and thus by Assumption \ref{ass-ic-ss}\ref{ass-ic-ss:c}, 
\begin{align*}
	\Pr(\IC(G,K^0) - \IC(G^0,K^0)>0)=\Pr(c^\tss(G,N,T) - c^\tss(G^0,N,T)> - O_p(T))\rightarrow 1.	
\end{align*}

On the other hand, if $G<G^0$,
\begin{align*}
	\IC(G,K^0) - \IC(G^0,K^0)
	= \underbrace{A^\tss(G,K^0)}_{\rightarrow_p c_{\G}^\tss} - A^\tss(G^0,K^0) + \frac{c^\tss(G,N,T) - c^\tss(G^0,N,T)}{NT} \rightarrow c_{\G}^\tss>0.
\end{align*}
This implies $\Pr(\IC(G,K^0) - \IC(G^0,K^0)>0)\rightarrow 1$.

Combining results,
\begin{align*}
	&\Pr(\hat G(K^0) = G^0)\\
	=&\Pr(\IC^\tss(G,K^0)>\IC^\tss(G^0,K^0),\ \forall G\in\{1,\dots,G^{\max}\}\setminus\{G^0\})\\
	\leq& 1 - \sum_{G\in\{1,\dots,G^{\max}\}\setminus\{G^0\}}\Pr(\IC^\tss(G,K^0)\leq \IC^\tss(G^0,K^0))
	\rightarrow 1.
\end{align*}
Now, we have
\begin{align*}
	&\Pr(\hat G = G^0)\\
	\geq &\Pr(\hat G(K^0) = G^0 \text{ and } \hat K = K^0)\\
	\geq &\Pr(\hat G(K^0) = G^0) + \Pr(\hat K = K^0) - \Pr(\hat G(K^0) = G^0 \text{ or } \hat K = K^0)\\
	\geq &\Pr(\hat G(K^0) = G^0) + \Pr(\hat K = K^0) - 1 \rightarrow 1,
\end{align*}
where Lemma \ref{thm-ic-fs} is used in the last line. This completes the proof of Theorem \ref{thm-ic-ss}.

\section{More on simulations} \label{sec-appdx:sim}

\subsection{Selection of the number of groups}
We evaluate the performance of the selection procedure for the number of groups by simulation.
The maximum possible numbers of groups in the first- and second-stages are set to $K^{\max} = G^{\max} = 5$.
Table \ref{tab:ic_table} presents the frequency of choosing the correct number when the following penalties are employed:
\begin{align*}
	c^\tfs(K,N,T) &= \hat{\sigma^\tfs}^2 K \log(\min(N,T))\frac{NT}{\min(N,T)} \text{ and}\\
	c^\tss(G,N,T) &= \hat{\sigma^\tss}^2 G \log(\min(N,T))\frac{NT}{\min(N,T)}.
\end{align*}
They are based on the $PC_{p3}$ criterion suggested in \textcite{BaiNg2002}.\footnote{In our settings, the BIC criterion---for example, $c^\tfs(K,N,T)=\hat{\sigma^\tfs}^2(K+N)\log(NT)$---tends to over-select even when $N$ and $T$ are substantially large.}

\begin{table}[htp]
\begin{center}
\resizebox{0.8\columnwidth}{!}{
\begin{threeparttable}
\caption{Frequency of correctly choosing the number of groups}	
\label{tab:ic_table}
\begin{tabular}{lllccccccccccccccc}
  \toprule 
 \multicolumn{3}{c}{} & \multicolumn{15}{c}{$\sigma=0.5$}\\
 \cmidrule(l){4-18}
  & & & \multicolumn{5}{c}{IG} & \multicolumn{5}{c}{TGFE} & \multicolumn{5}{c}{UGFE} \\
 \cmidrule(l){4-8} \cmidrule(l){9-13} \cmidrule(l){14-18}
  & & & 1 & 2 & 3 & 4 & 5 & 1 & 2 & 3 & 4 & 5 & 1 & 2 & 3 & 4 & 5\\
 \midrule \midrule\multirow{6}{*}{\textbf{DGP 1}} & 50 & 5 & 0 & 100 & 0 & 0 & 0 & 55 & 45 & 0 & 0 & 0 & 59 & 41 & 0 & 0 & 0 \\ 
   & 50 & 10 & 0 & 100 & 0 & 0 & 0 & 36 & 64 & 0 & 0 & 0 & 34 & 66 & 0 & 0 & 0 \\ 
   & 50 & 20 & 0 & 100 & 0 & 0 & 0 & 0 & 100 & 0 & 0 & 0 & 0 & 100 & 0 & 0 & 0 \\ 
   & 100 & 5 & 0 & 100 & 0 & 0 & 0 & 47 & 53 & 0 & 0 & 0 & 50 & 50 & 0 & 0 & 0 \\ 
   & 100 & 10 & 0 & 100 & 0 & 0 & 0 & 26 & 74 & 0 & 0 & 0 & 20 & 80 & 0 & 0 & 0 \\ 
   & 100 & 20 & 0 & 100 & 0 & 0 & 0 & 0 & 100 & 0 & 0 & 0 & 0 & 100 & 0 & 0 & 0 \\ 
  \midrule\multirow{6}{*}{\textbf{DGP 2}} & 50 & 5 & 0 & 100 & 0 & 0 & 0 & 48 & 52 & 0 & 0 & 0 & 52 & 48 & 0 & 0 & 0 \\ 
   & 50 & 10 & 0 & 100 & 0 & 0 & 0 & 31 & 69 & 0 & 0 & 0 & 27 & 73 & 0 & 0 & 0 \\ 
   & 50 & 20 & 0 & 100 & 0 & 0 & 0 & 3 & 97 & 0 & 0 & 0 & 2 & 98 & 0 & 0 & 0 \\ 
   & 100 & 5 & 0 & 100 & 0 & 0 & 0 & 50 & 50 & 0 & 0 & 0 & 49 & 51 & 0 & 0 & 0 \\ 
   & 100 & 10 & 0 & 100 & 0 & 0 & 0 & 27 & 73 & 0 & 0 & 0 & 19 & 81 & 0 & 0 & 0 \\ 
   & 100 & 20 & 0 & 100 & 0 & 0 & 0 & 0 & 100 & 0 & 0 & 0 & 0 & 100 & 0 & 0 & 0 \\ 
  \midrule\multirow{6}{*}{\textbf{DGP 3}} & 50 & 5 & 0 & 100 & 0 & 0 & 0 & 44 & 56 & 0 & 0 & 0 & 38 & 62 & 0 & 0 & 0 \\ 
   & 50 & 10 & 0 & 100 & 0 & 0 & 0 & 33 & 67 & 0 & 0 & 0 & 22 & 78 & 0 & 0 & 0 \\ 
   & 50 & 20 & 0 & 100 & 0 & 0 & 0 & 0 & 100 & 0 & 0 & 0 & 0 & 100 & 0 & 0 & 0 \\ 
   & 100 & 5 & 0 & 100 & 0 & 0 & 0 & 32 & 68 & 0 & 0 & 0 & 27 & 73 & 0 & 0 & 0 \\ 
   & 100 & 10 & 0 & 100 & 0 & 0 & 0 & 23 & 77 & 0 & 0 & 0 & 11 & 89 & 0 & 0 & 0 \\ 
   & 100 & 20 & 0 & 100 & 0 & 0 & 0 & 0 & 100 & 0 & 0 & 0 & 0 & 100 & 0 & 0 & 0 \\ 
  \midrule\multirow{6}{*}{\textbf{DGP 4}} & 50 & 5 & 0 & 100 & 0 & 0 & 0 & 38 & 62 & 0 & 0 & 0 & 44 & 56 & 0 & 0 & 0 \\ 
   & 50 & 10 & 0 & 100 & 0 & 0 & 0 & 40 & 60 & 0 & 0 & 0 & 41 & 59 & 0 & 0 & 0 \\ 
   & 50 & 20 & 0 & 100 & 0 & 0 & 0 & 4 & 96 & 0 & 0 & 0 & 3 & 97 & 0 & 0 & 0 \\ 
   & 100 & 5 & 0 & 100 & 0 & 0 & 0 & 37 & 63 & 0 & 0 & 0 & 40 & 60 & 0 & 0 & 0 \\ 
   & 100 & 10 & 0 & 100 & 0 & 0 & 0 & 31 & 69 & 0 & 0 & 0 & 25 & 75 & 0 & 0 & 0 \\ 
   & 100 & 20 & 0 & 100 & 0 & 0 & 0 & 0 & 100 & 0 & 0 & 0 & 0 & 100 & 0 & 0 & 0 \\ 
   \bottomrule 
\end{tabular}

\begin{tablenotes}
	\item Note: The value in each cell denotes the mean of the Rand index across 100 simulations, with 100 starting group memberships randomly generated for each estimate.
\end{tablenotes}
\end{threeparttable}
}
\end{center}
\end{table}

\subsection{Statistical properties of the procedures}
We compare the statistical properties of the estimators introduced in Section \ref{sec:estimators} via simulations.
Specifically, we assess their bias and variance under a different specification for the first stage.
We focus on the case in which the coefficients of the structural equation do not exhibit a grouped pattern. 
That is, now we impose
\begin{align*}
	\beta_i^0 = 1,\ \forall i=1,\dots,N.
\end{align*}
The specification for the first stage is as follows:
\begin{description}
	\item[DGP \thesection1] $\Pi_i^0\sim\mathcal N(\mu,\sigma_\Pi^2)$, where $\mu\in\{0.5,1\}$ and $\sigma_\Pi\in\{1,1.25\}$.
\end{description}
Here, the parameter $\mu$ represents the average strength of the first stage, and $\sigma_\Pi$ captures the degree of first-stage heterogeneity.

As presented in Table \ref{tab:comp1_table}, increasing the number of groups in the first stage, in general, decreases variance, but results in higher bias. Regarding unbiasedness, the 2SLS estimators are ranked in the following order: 2SLS, TGFE$_2$, and UGFE. All variants of 2SLS are consistent. However, the IG estimator suffers from the weak IV problem due to the heterogeneity in the first stage. Regarding variance, the order of the 2SLS estimators is reversed.

\begin{table}[htp]
\begin{center}
\resizebox{0.8\columnwidth}{!}{
\begin{threeparttable}
\caption{Means and standard deviations under DGP \thesection1}	
\label{tab:comp1_table}
\begin{tabular}{lllcccccccc}
  \toprule 
 \multicolumn{3}{c}{} & \multicolumn{8}{c}{$\sigma=0.5$} \\
 \cmidrule(l){4-11} 
  & & & \multicolumn{2}{c}{IG} & \multicolumn{2}{c}{2SLS} & \multicolumn{2}{c}{TGFE$_2$} & \multicolumn{2}{c}{UGFE} \\
 \cmidrule(l){4-5} \cmidrule(l){6-7} \cmidrule(l){8-9} \cmidrule(l){10-11} 
  & & & Mean & SD & Mean & SD & Mean & SD & Mean & SD \\
 \midrule 
 \midrule\multirow{6}{*}{\textbf{$(\mu,\sigma_\Pi)=(0.5,1)$}} & 50 & 5 & 0.562 & 0.101 & 1.057 & 0.438 & 0.896 & 0.121 & 0.875 & 0.102 \\ 
   & 50 & 10 & 0.555 & 0.068 & 1.009 & 0.179 & 0.931 & 0.085 & 0.926 & 0.074 \\ 
   & 50 & 20 & 0.547 & 0.064 & 1.013 & 0.159 & 0.966 & 0.069 & 0.957 & 0.059 \\ 
   & 100 & 5 & 0.562 & 0.071 & 1.025 & 0.202 & 0.884 & 0.082 & 0.870 & 0.070 \\ 
   & 100 & 10 & 0.552 & 0.058 & 1.015 & 0.153 & 0.929 & 0.072 & 0.925 & 0.058 \\ 
   & 100 & 20 & 0.552 & 0.044 & 1.017 & 0.093 & 0.966 & 0.049 & 0.962 & 0.041 \\ 
  \midrule\multirow{6}{*}{\textbf{$(\mu,\sigma_\Pi)=(1,1)$}} & 50 & 5 & 0.672 & 0.087 & 1.000 & 0.112 & 0.930 & 0.092 & 0.917 & 0.088 \\ 
   & 50 & 10 & 0.658 & 0.074 & 1.006 & 0.102 & 0.966 & 0.079 & 0.954 & 0.072 \\ 
   & 50 & 20 & 0.651 & 0.056 & 0.994 & 0.069 & 0.980 & 0.054 & 0.976 & 0.049 \\ 
   & 100 & 5 & 0.661 & 0.060 & 0.998 & 0.086 & 0.926 & 0.067 & 0.904 & 0.060 \\ 
   & 100 & 10 & 0.670 & 0.035 & 0.998 & 0.063 & 0.955 & 0.045 & 0.945 & 0.041 \\ 
   & 100 & 20 & 0.666 & 0.034 & 0.997 & 0.041 & 0.982 & 0.036 & 0.979 & 0.032 \\ 
  \midrule\multirow{6}{*}{\textbf{$(\mu,\sigma_\Pi)=(0.5,1.25)$}} & 50 & 5 & 0.642 & 0.086 & 1.063 & 0.510 & 0.916 & 0.107 & 0.902 & 0.090 \\ 
   & 50 & 10 & 0.644 & 0.068 & 1.012 & 0.191 & 0.960 & 0.074 & 0.952 & 0.066 \\ 
   & 50 & 20 & 0.642 & 0.063 & 1.013 & 0.258 & 0.974 & 0.059 & 0.971 & 0.050 \\ 
   & 100 & 5 & 0.639 & 0.065 & 0.998 & 0.168 & 0.910 & 0.072 & 0.899 & 0.057 \\ 
   & 100 & 10 & 0.637 & 0.053 & 1.000 & 0.147 & 0.946 & 0.056 & 0.944 & 0.045 \\ 
   & 100 & 20 & 0.636 & 0.043 & 1.011 & 0.099 & 0.977 & 0.040 & 0.977 & 0.032 \\ 
  \midrule\multirow{6}{*}{\textbf{$(\mu,\sigma_\Pi)=(1,1.25)$}} & 50 & 5 & 0.701 & 0.077 & 1.005 & 0.144 & 0.934 & 0.095 & 0.917 & 0.073 \\ 
   & 50 & 10 & 0.707 & 0.064 & 0.996 & 0.093 & 0.966 & 0.056 & 0.958 & 0.053 \\ 
   & 50 & 20 & 0.717 & 0.049 & 0.995 & 0.054 & 0.985 & 0.046 & 0.983 & 0.042 \\ 
   & 100 & 5 & 0.729 & 0.058 & 1.009 & 0.094 & 0.950 & 0.061 & 0.938 & 0.056 \\ 
   & 100 & 10 & 0.717 & 0.044 & 0.996 & 0.065 & 0.961 & 0.048 & 0.957 & 0.040 \\ 
   & 100 & 20 & 0.713 & 0.033 & 1.003 & 0.048 & 0.983 & 0.033 & 0.979 & 0.028 \\ 
   \bottomrule 
\end{tabular}

\begin{tablenotes}
	\item Note: The value in each cell denotes the mean of the Rand index across 100 simulations, with 100 starting group memberships randomly generated for each estimate.
\end{tablenotes}
\end{threeparttable}
}
\end{center}
\end{table}

\section{More on empirical results} \label{sec-appdx:emp}

\subsection{Calculation of the standard errors for the pre-estimates}
Let $w_{it}$ be the trade-weighted world income. In our empirical application, we have employed
\begin{align*}
	y_{it-1} =& \pi_{2,k_i}w_{it} + \mu + v_{it}\\
	\intertext{as the first stage for Models 1a--b, and}
	y_{it-1} =& \pi_1 d_{it-1} + \pi_{2,k_i}w_{it} + \mu + v_{it}	
\end{align*}
for Models 2a--b. Only the coefficient of the instrument $w_{it}$ is allowed to vary across groups. We impose it to prevent the first stage from being overly saturated. Henceforth, under this setting, we explain why the second-stage residuals can generate valid standard errors.

We focus on Model 2b. Let $\hat\theta \equiv (\hat\alpha_1, \hat\alpha_2, \hat\beta_1, \hat\beta_{2,1},\hat\beta_{2,2})$ be the estimates for the true coefficient parameter $\theta^0 \equiv (\alpha_1^0, \alpha_2^0, \beta_1^0, \beta_{2,1}^0,\beta_{2,2}^0)$. Denote the regressors of the second stage by $\hat{\mathbf x}_{it}\equiv (\mathbf 1\{\hat g_i=1\}, \mathbf 1\{\hat g_i=2\}, d_{it-1}, \mathbf 1\{\hat g_i=1\}\hat y_{it-1}, \mathbf 1\{\hat g_i=2\}\hat y_{it-1})$. Then, assuming correct classification, i.e., $\hat g_i = g_i^0,\forall i$, we have
\begin{align*}
	d_{it} = \hat{\mathbf x}_{it}\theta^0 + \underbrace{\mathbf 1\{\hat g_i=1\}(y_{it-1}-\hat y_{it-1})'\beta_{2,1}^0 + \mathbf 1\{\hat g_i=2\}(y_{it-1}-\hat y_{it-1})'\beta_{2,2}^0}_{=\hat v_{it}\beta_{2,\hat g_i}^0(=\hat v_{it}\beta_{2,g_i^0}^0)} + u_{it},
\end{align*}
where $\hat v_{it}\equiv y_{it-1} - \hat y_{it-1}$, and therefore
\begin{align*}
	\hat\theta =& \biggl(\frac{1}{NT}\sum_{i=1}^N\sum_{t=1}^T\hat{\mathbf x}_{it}\hat{\mathbf x}_{it}'\biggr)^{-1}
	\biggl(\frac{1}{NT}\sum_{i=1}^N\sum_{t=1}^T\hat{\mathbf x}_{it} d_{it}\biggr)\\
	=& \theta^0 + \biggl(\frac{1}{NT}\sum_{i=1}^N\sum_{t=1}^T\hat{\mathbf x}_{it}\hat{\mathbf x}_{it}'\biggr)^{-1}
	\biggl(\frac{1}{NT}\sum_{i=1}^N\sum_{t=1}^T\hat{\mathbf x}_{it}\hat v_{it}\beta_{2,\hat g_i}^0 + \frac{1}{NT}\sum_{i=1}^N\sum_{t=1}^T\hat{\mathbf x}_{it}u_{it}\biggr).
\end{align*}
In a usual case without group heterogeneity, that is, when $\beta_{2,1}^0=\beta_{2,2}^0$, the first term in the nominator becomes exactly zero, i.e.,
\begin{align*}
	\frac{1}{NT}\sum_{i=1}^N\sum_{t=1}^T\hat{\mathbf x}_{it}\hat v_{it}\beta_{2,\hat g_i}^0	= \mathbf 0_5,
\end{align*}
because $(NT)^{-1}\sum_{i=1}^N\sum_{t=1}^T\hat{\mathbf x}_{it}\hat v_{it}=\mathbf 0_5$. As a result, the second-stage residuals fail to yield valid standard errors. However, in the presence of group heterogeneity, this term generally deviates from zero. Consequently, using the second-stage residuals becomes necessary to righteously capture the variation of $\hat{\mathbf x}_{it}d_{it}$.\footnote{Note also that for each $k\in\mathbb K$, $\sum_{\hat k_i=k}\sum_{t=1}^T\hat{\mathbf x}_{it}\hat v_{it}\neq 0$, since not all parameters in the first stage exhibit heterogeneity across groups. Hence, $\sum_{\hat g_i=g,\hat k_i=k}\sum_{t=1}^T\hat{\mathbf x}_{it}\hat v_{it}\neq 0$ as well.}
 
\subsection{Extension}
An extension of the baseline models, as outlined in Section \ref{sec:empirical}, is proposed. We enlarge them by adding time-varying fixed effects, $\{\alpha_{t}\}_{t=1}^T$, which may also by themselves exhibit group heterogeneity.
\begin{description}
	\item[Model \thesection1a:] $d_{it}=\beta_{2,g_i} y_{it-1} + \alpha_t + u_{it}$
	\item[Model \thesection1b:] $d_{it}=\beta_{2,g_i} y_{it-1} + \alpha_{g_i t} + u_{it}$
	\item[Model \thesection2a:] $d_{it}=\beta_1 d_{it-1} + \beta_{2,g_i} y_{it-1} + \alpha_t + u_{it}$
	\item[Model \thesection2b:] $d_{it}=\beta_1 d_{it-1} + \beta_{2,g_i} y_{it-1} + \alpha_{g_i t} + u_{it}$
\end{description}

Table \ref{tab:m_time-fe_table} presents the estimation results for these models. In Models \thesection1a and \thesection2a, all procedures yield compatible estimates, which are statistically consistent with the post-IV estimates. Groups of larger size demonstrate a more pronounced effect of income---around 0.4 to 0.6 greater in Model \thesection1a and 0.3 in Model \thesection2a. In Model \thesection1b, several pairs of the pre- and post-estimates appear to be statistically different. This is not the case with Model \thesection2b, where the lagged democracy is included as an additional control.

\begin{table}[htp]
\begin{center}
\resizebox{0.65\columnwidth}{!}{
\begin{threeparttable}
\caption{Estimation results for the extended model}
\label{tab:m_time-fe_table}
\begin{tabular}{llccccc}
  \toprule 
  Model & Method & Pre-estimates & SE & Group size & Post-estimates & SE \\
 \midrule 
 \midrule\multirow{1}{*}{\textbf{No groups}} & \multirow{1}{*}{2SLS} &  &  &  & 0.109 & 0.127 \\ 
  \midrule\multirow{10}{*}{\textbf{Model \thesection1a}} & \multirow{2}{*}{IG} &  &  & 47 & 0.178 & 0.036 \\ 
   &  &  &  & 32 & 0.132 & 0.040 \\ 
   & \multirow{2}{*}{2SLS} & 0.154 & 0.072 & 35 & 0.142 & 0.049 \\ 
   &  & 0.220 & 0.072 & 44 & 0.185 & 0.042 \\ 
   & \multirow{2}{*}{TGFE$_2$} & 0.172 & 0.021 & 36 & 0.126 & 0.055 \\ 
   &  & 0.221 & 0.019 & 43 & 0.168 & 0.048 \\ 
   & \multirow{2}{*}{TGFE$_{N/4}$} & 0.214 & 0.010 & 52 & 0.198 & 0.036 \\ 
   &  & 0.174 & 0.013 & 27 & 0.153 & 0.039 \\ 
   & \multirow{2}{*}{UGFE} & 0.212 & 0.012 & 52 & 0.198 & 0.036 \\ 
   &  & 0.171 & 0.015 & 27 & 0.153 & 0.039 \\ 
  \midrule\multirow{10}{*}{\textbf{Model \thesection1b}} & \multirow{2}{*}{IG} &  &  & 34 & 0.079 & 0.026 \\ 
   &  &  &  & 45 & 0.294 & 0.180 \\ 
   & \multirow{2}{*}{2SLS} & 0.135 & 0.069 & 35 & 0.110 & 0.037 \\ 
   &  & 0.622 & 0.148 & 44 & 0.448 & 0.243 \\ 
   & \multirow{2}{*}{TGFE$_2$} & 0.166 & 0.019 & 44 & 0.320 & 0.272 \\ 
   &  & 0.229 & 0.046 & 35 & 0.114 & 0.049 \\ 
   & \multirow{2}{*}{TGFE$_{N/4}$} & 0.168 & 0.028 & 35 & 0.071 & 0.036 \\ 
   &  & 0.180 & 0.016 & 44 & 0.261 & 0.106 \\ 
   & \multirow{2}{*}{UGFE} & 0.140 & 0.028 & 35 & 0.085 & 0.028 \\ 
   &  & 0.171 & 0.017 & 44 & 0.311 & 0.181 \\ 
   \bottomrule 
\end{tabular}

\smallskip
\begin{tabular}{llccccc}
  \toprule 
  Model & Method & Pre-estimates & SE & Group size & Post-estimates & SE \\
 \midrule 
 \midrule\multirow{1}{*}{\textbf{No groups}} & \multirow{1}{*}{2SLS} &  &  &  & -0.013 & 0.052 \\ 
  \midrule\multirow{10}{*}{\textbf{Model \thesection2a}} & \multirow{2}{*}{IG} &  &  & 29 & 0.095 & 0.045 \\ 
   &  &  &  & 50 & 0.124 & 0.046 \\ 
   & \multirow{2}{*}{2SLS} & 0.078 & 0.048 & 34 & 0.079 & 0.041 \\ 
   &  & 0.109 & 0.049 & 45 & 0.108 & 0.040 \\ 
   & \multirow{2}{*}{TGFE$_2$} & 0.058 & 0.017 & 31 & 0.095 & 0.042 \\ 
   &  & 0.088 & 0.017 & 48 & 0.124 & 0.042 \\ 
   & \multirow{2}{*}{TGFE$_{N/4}$} & 0.078 & 0.012 & 29 & 0.095 & 0.045 \\ 
   &  & 0.103 & 0.011 & 50 & 0.124 & 0.046 \\ 
   & \multirow{2}{*}{UGFE} & 0.083 & 0.011 & 27 & 0.095 & 0.047 \\ 
   &  & 0.108 & 0.011 & 52 & 0.123 & 0.048 \\ 
  \midrule\multirow{10}{*}{\textbf{Model \thesection2b}} & \multirow{2}{*}{IG} &  &  & 47 & 0.265 & 0.359 \\ 
   &  &  &  & 32 & 0.008 & 0.044 \\ 
   & \multirow{2}{*}{2SLS} & 0.047 & 0.096 & 51 & 0.058 & 0.128 \\ 
   &  & -0.113 & 0.052 & 28 & -0.125 & 0.063 \\ 
   & \multirow{2}{*}{TGFE$_2$} & 0.035 & 0.022 & 33 & 0.022 & 0.048 \\ 
   &  & 0.052 & 0.018 & 46 & 0.190 & 0.271 \\ 
   & \multirow{2}{*}{TGFE$_{N/4}$} & 0.083 & 0.013 & 49 & 0.118 & 0.106 \\ 
   &  & 0.070 & 0.017 & 30 & 0.029 & 0.042 \\ 
   & \multirow{2}{*}{UGFE} & 0.081 & 0.013 & 49 & 0.118 & 0.106 \\ 
   &  & 0.067 & 0.017 & 30 & 0.029 & 0.042 \\ 
   \bottomrule 
\end{tabular}

\begin{tablenotes}
	\item Note: 10,000 starting group memberships are randomly generated for each estimate. Standard errors clustered at country level.
\end{tablenotes}
\end{threeparttable}
}
\end{center}
\end{table}


\section{Supplement to the proofs} \label{sec-appdx:supp}
\subsection{Proof of Lemma \ref{lem-sineq:gamma-cnst}}
In order to represent the concentration inequality associated with $\min_i\min_{g\neq \tilde g}T^{-1}\sum_{t=1}^T\|x_{it}'(\hat\beta_g-\hat\beta_{\tilde g})+(\hat\alpha_{gt}'-\hat\alpha_{\tilde g}')\|^2$ in terms of $\min_i\min_{g\neq \tilde g}T^{-1}\sum_{t=1}^T\|x_{it}'(\beta_g^0-\beta_{\tilde g}^0)+(\alpha_{gt}^{0\prime}-\alpha_{\tilde g}^{0\prime})\|^2$, we employed the following result:
\begin{align*}
	|h_A(\beta,\alpha)-h_A(\beta^0,\alpha^0)|\lesssim \|A\|_2\max_{g\in\mathbb G}\biggl(\|\beta_g-\beta_g^0\|^2 + \frac{1}{T}\sum_{t=1}^T\|\alpha_{gt}-\alpha_{gt}^0\|^2\biggr)^{\frac{1}{2}}.
\end{align*}
Below is a detailed derivation process of this inequality:
\begin{align*}
	&h(\tilde\beta,\tilde\alpha)-h(\beta,\alpha) \\
	\leq& - \min_{g\neq \tilde g}	\tr\left(\begin{pmatrix}
		(\beta_g-\beta_{\tilde g})-(\tilde \beta_g-\tilde \beta_{\tilde g})\\
		\frac{1}{\sqrt T}(\alpha_{g1}' - \alpha_{\tilde g1}') - \frac{1}{\sqrt T}(\tilde\alpha_{g1}' - \tilde \alpha_{\tilde g1}')\\
		\vdots\\
		\frac{1}{\sqrt T}(\alpha_{gT}' - \alpha_{\tilde gT}') - \frac{1}{\sqrt T}(\tilde\alpha_{gT}' - \tilde\alpha_{\tilde gT}'))
		\end{pmatrix}'A\begin{pmatrix}
		(\beta_g-\beta_{\tilde g})-(\tilde\beta_g-\tilde\beta_{\tilde g})\\
		\frac{1}{\sqrt T}(\alpha_{g1}' - \alpha_{\tilde g1}') - \frac{1}{\sqrt T}(\tilde\alpha_{g1}' - \tilde\alpha_{\tilde g1}')\\
		\vdots\\
		\frac{1}{\sqrt T}(\alpha_{gT}' - \alpha_{\tilde gT}') - \frac{1}{\sqrt T}(\tilde\alpha_{gT}' - \tilde\alpha_{\tilde gT}')
		\end{pmatrix}\right. \\
		&\left.\hphantom{\min_{g\neq \tilde g}	\tr\left( \right.} +  2\begin{pmatrix}
		(\beta_g-\beta_{\tilde g})-(\tilde\beta_g-\tilde\beta_{\tilde g})\\
		\frac{1}{\sqrt T}(\alpha_{g1}' - \alpha_{\tilde g1}') - \frac{1}{\sqrt T}(\tilde\alpha_{g1}' - \tilde\alpha_{\tilde g1}')\\
		\vdots\\
		\frac{1}{\sqrt T}(\alpha_{gT}' - \alpha_{\tilde gT}') - \frac{1}{\sqrt T}(\tilde\alpha_{gT}' - \tilde\alpha_{\tilde gT}')
		\end{pmatrix}'A\begin{pmatrix}
		(\beta_g-\beta_{\tilde g})\\
		\frac{1}{\sqrt T}(\alpha_{g1}' - \alpha_{\tilde g1}')\\
		\vdots\\
		\frac{1}{\sqrt T}(\alpha_{gT}' - \alpha_{\tilde gT}')
		\end{pmatrix}\right)\\
	\leq& \max_{g\neq \tilde g}	\left|\tr\left(\begin{pmatrix}
		(\beta_g-\beta_{\tilde g})-(\tilde \beta_g-\tilde \beta_{\tilde g})\\
		\frac{1}{\sqrt T}(\alpha_{g1}' - \alpha_{\tilde g1}') - \frac{1}{\sqrt T}(\tilde\alpha_{g1}' - \tilde \alpha_{\tilde g1}')\\
		\vdots\\
		\frac{1}{\sqrt T}(\alpha_{gT}' - \alpha_{\tilde gT}') - \frac{1}{\sqrt T}(\tilde\alpha_{gT}' - \tilde\alpha_{\tilde gT}'))
		\end{pmatrix}'A\begin{pmatrix}
		(\beta_g-\beta_{\tilde g})-(\tilde\beta_g-\tilde\beta_{\tilde g})\\
		\frac{1}{\sqrt T}(\alpha_{g1}' - \alpha_{\tilde g1}') - \frac{1}{\sqrt T}(\tilde\alpha_{g1}' - \tilde\alpha_{\tilde g1}')\\
		\vdots\\
		\frac{1}{\sqrt T}(\alpha_{gT}' - \alpha_{\tilde gT}') - \frac{1}{\sqrt T}(\tilde\alpha_{gT}' - \tilde\alpha_{\tilde gT}')
		\end{pmatrix}\right.\right. \\
		&\left.\left.\hphantom{\min_{g\neq \tilde g}	\tr\left( \right.} +  2\begin{pmatrix}
		(\beta_g-\beta_{\tilde g})-(\tilde\beta_g-\tilde\beta_{\tilde g})\\
		\frac{1}{\sqrt T}(\alpha_{g1}' - \alpha_{\tilde g1}') - \frac{1}{\sqrt T}(\tilde\alpha_{g1}' - \tilde\alpha_{\tilde g1}')\\
		\vdots\\
		\frac{1}{\sqrt T}(\alpha_{gT}' - \alpha_{\tilde gT}') - \frac{1}{\sqrt T}(\tilde\alpha_{gT}' - \tilde\alpha_{\tilde gT}')
		\end{pmatrix}'A\begin{pmatrix}
		(\beta_g-\beta_{\tilde g})\\
		\frac{1}{\sqrt T}(\alpha_{g1}' - \alpha_{\tilde g1}')\\
		\vdots\\
		\frac{1}{\sqrt T}(\alpha_{gT}' - \alpha_{\tilde gT}')
		\end{pmatrix}\right)\right|\\
	\leq& \max_{g\neq \tilde g}\left\|\begin{pmatrix}
		(\beta_g-\beta_{\tilde g})-(\tilde \beta_g-\tilde \beta_{\tilde g})\\
		\frac{1}{\sqrt T}(\alpha_{g1}' - \alpha_{\tilde g1}') - \frac{1}{\sqrt T}(\tilde\alpha_{g1}' - \tilde \alpha_{\tilde g1}')\\
		\vdots\\
		\frac{1}{\sqrt T}(\alpha_{gT}' - \alpha_{\tilde gT}') - \frac{1}{\sqrt T}(\tilde\alpha_{gT}' - \tilde\alpha_{\tilde gT}'))
		\end{pmatrix}\right\|
		\left\|A\begin{pmatrix}
		(\beta_g-\beta_{\tilde g})-(\tilde\beta_g-\tilde\beta_{\tilde g})\\
		\frac{1}{\sqrt T}(\alpha_{g1}' - \alpha_{\tilde g1}') - \frac{1}{\sqrt T}(\tilde\alpha_{g1}' - \tilde\alpha_{\tilde g1}')\\
		\vdots\\
		\frac{1}{\sqrt T}(\alpha_{gT}' - \alpha_{\tilde gT}') - \frac{1}{\sqrt T}(\tilde\alpha_{gT}' - \tilde\alpha_{\tilde gT}')
		\end{pmatrix}
		\right\|\\
	& + 2\max_{g\neq \tilde g}\left\|\begin{pmatrix}
		(\beta_g-\beta_{\tilde g})\\
		\frac{1}{\sqrt T}(\alpha_{g1}' - \alpha_{\tilde g1}')\\
		\vdots\\
		\frac{1}{\sqrt T}(\alpha_{gT}' - \alpha_{\tilde gT}')
		\end{pmatrix}\right\|
		\left\|A\begin{pmatrix}
		(\beta_g-\beta_{\tilde g})-(\tilde\beta_g-\tilde\beta_{\tilde g})\\
		\frac{1}{\sqrt T}(\alpha_{g1}' - \alpha_{\tilde g1}') - \frac{1}{\sqrt T}(\tilde\alpha_{g1}' - \tilde\alpha_{\tilde g1}')\\
		\vdots\\
		\frac{1}{\sqrt T}(\alpha_{gT}' - \alpha_{\tilde gT}') - \frac{1}{\sqrt T}(\tilde\alpha_{gT}' - \tilde\alpha_{\tilde gT}')
		\end{pmatrix}
		\right\|\\
	\lesssim& \max_{(\beta_g,\alpha_g)\in\mathcal B\times \mathcal A^T}\left\|\begin{pmatrix}
		\beta_g\\
		\frac{1}{\sqrt T}\alpha_{g1}'\\
		\vdots\\
		\frac{1}{\sqrt T}\alpha_{gT}'
		\end{pmatrix}\right\|
	\|A\|_2\max_{g\neq \tilde g}\left(\left\|\begin{pmatrix}
		(\beta_g-\tilde\beta_g)\\
		\frac{1}{\sqrt T}(\alpha_{g1}' - \tilde\alpha_{g1}')\\
		\vdots\\
		\frac{1}{\sqrt T}(\alpha_{gT}' - \tilde\alpha_{gT}')
		\end{pmatrix}\right\|
		+ \left\|\begin{pmatrix}
		(\beta_{\tilde g}-\tilde\beta_{\tilde g})\\
		\frac{1}{\sqrt T}(\alpha_{\tilde g1}' - \tilde\alpha_{\tilde g1}')\\
		\vdots\\
		\frac{1}{\sqrt T}(\alpha_{\tilde gT}' - \tilde\alpha_{\tilde gT}')
		\end{pmatrix}\right\|\right)\\
	\lesssim& \|A\|_2\max_{g\in\mathbb G}\left\|\begin{pmatrix}
		(\beta_g-\tilde\beta_g)\\
		\frac{1}{\sqrt T}(\alpha_{g1}' - \tilde\alpha_{g1}')\\
		\vdots\\
		\frac{1}{\sqrt T}(\alpha_{gT}' - \tilde\alpha_{gT}')
		\end{pmatrix}\right\|.
\end{align*}
Assumption \ref{ass-sineq}\ref{ass-sineq:cpt-param} is used in the last two inequalities 

\subsection{Proof of Lemma \ref{lem-ts:beta-cnst}}
In the beginning of the proof, we argued that under Assumptions \ref{ass-ts}\ref{ass-ts:cpt-param}--\ref{ass-ts:wkdep-u}, $\Pi_{k_i^0}^{0\prime}z_{it}+\mu_{k_i^0t}^0$, $\alpha_{g_i^0t}^0$, and $v_{it}'\beta_{g_i^0}^0+u_{it}$ satisfy Assumption \ref{ass-sineq}\ref{ass-sineq:wkdep-xe}--\ref{ass-sineq:wkdep-e}, once we consider each of them as $x_{it}$, $\alpha_{g_i^0}$, and $e_{it}$ respectively. Below, we show that this argument is true. 
\begin{itemize}
	\item \textbf{Assumption \ref{ass-sineq}\ref{ass-sineq:wkdep-xe}:} For each $(g,\tilde g)\in\mathbb G^2$ and $\gamma\in\mathbb G^N$, the term
	\begin{align*}
		\biggl\|\frac{1}{NT}\sum_{i=1}^N\mathbf 1\{g_i^0=g\}\mathbf 1\{g_i=\tilde g\}\sum_{t=1}^T\smash[t]{\overbrace{\tilde x_{it}}^{x_{it}}(\overbrace{v_{it}'\beta_g^0+u_{it}}^{e_{it}}})\biggr\|
	\end{align*}
	can be bounded by
	\begin{align*}
		& \biggl\|\frac{1}{NT}\sum_{i=1}^N\mathbf 1\{g_i^0=g\}\mathbf 1\{g_i=\tilde g\}\sum_{t=1}^T(\Pi_{k_i^0}^{0\prime}z_{it}+\mu_{k_i^0t}^0)(v_{it}'\beta_g^0+u_{it})\biggr\| \nonumber\\
		\leq& \max_{\beta_g\in\mathcal B}\|\beta_g\| \frac{1}{N}\sum_{i=1}^N\biggl\|\frac{1}{T}\sum_{t=1}^T(\Pi_{k_i^0}^{0\prime}z_{it})v_{it}'\biggr\| + \frac{1}{N}\sum_{i=1}^N\biggl\|\frac{1}{T}\sum_{t=1}^T(\Pi_{k_i^0}^{0\prime}z_{it})u_{it}\biggr\| \nonumber\\
		& + \max_{\beta_g\in\mathcal B}\|\beta_g\|\sum_{k=1}^K\biggl(\frac{1}{N^2}\sum_{i=1}^N\sum_{j=1}^N\biggl|\frac{1}{T}\sum_{t=1}^Tv_{it}'v_{jt}\biggr|\biggr)^{\frac{1}{2}}
		+ \sum_{k=1}^K\biggl(\frac{1}{N^2}\sum_{i=1}^N\sum_{j=1}^N\biggl|\frac{1}{T}\sum_{t=1}^Tu_{it}u_{jt}\biggr|\biggr)^{\frac{1}{2}},
	\end{align*}
	which neither depends on $g$, $\tilde g$ nor $\gamma$. This equation is derived from the expansion: 
	\begin{align*}
		&\frac{1}{NT}\sum_{i=1}^N\mathbf 1\{g_i^0=g\}\mathbf 1\{g_i=\tilde g\}\sum_{t=1}^T(\Pi_{k_i^0}^{0\prime}z_{it}+\mu_{k_i^0t}^0)(v_{it}'\beta_{g_i^0}^0+u_{it})\\
		=&\biggl(\frac{1}{NT}\sum_{i=1}^N\mathbf 1\{g_i^0=g\}\mathbf 1\{g_i=\tilde g\}\sum_{t=1}^T(\Pi_{k_i^0}^{0\prime}z_{it})v_{it}'\biggr)\beta_g^0
		+ \frac{1}{NT}\sum_{i=1}^N\mathbf 1\{g_i^0=g\}\mathbf 1\{g_i=\tilde g\}\sum_{t=1}^T(\Pi_{k_i^0}^{0\prime}z_{it})u_{it}\\
		& + \biggl(\frac{1}{NT}\sum_{i=1}^N\mathbf 1\{g_i^0=g\}\mathbf 1\{g_i=\tilde g\}\sum_{t=1}^T\mu_{k_i^0t}^0v_{it}'\biggr)\beta_g^0 + \frac{1}{NT}\sum_{i=1}^N\mathbf 1\{g_i^0=g\}\mathbf 1\{g_i=\tilde g\}\sum_{t=1}^T\mu_{k_i^0t}^0u_{it},
	\end{align*}
	and the observation that for any random variable $w_{it}$,
	\begin{align*}
		&\biggl\|\frac{1}{NT}\sum_{i=1}^N\mathbf 1\{g_i^0=g\}\mathbf 1\{g_i=\tilde g\}\sum_{t=1}^T\mu_{k_i^0t}^0w_{it}'\biggr\|\\
		\leq& \sum_{k=1}^K\biggl\|\frac{1}{T}\sum_{t=1}^T\mu_{kt}^0\biggl(\frac{1}{N}\sum_{i=1}^N\mathbf 1\{g_i^0=g\}\mathbf 1\{g_i=\tilde g\}\mathbf 1\{k_i^0=k\}w_{it}'\biggr)\biggr\|\\
		\lesssim& \sum_{k=1}^K\frac{1}{T}\sum_{t=1}^T\biggl\|\frac{1}{N}\sum_{i=1}^N\mathbf 1\{g_i^0=g\}\mathbf 1\{g_i=\tilde g\}\mathbf 1\{k_i^0=k\}w_{it}\biggr\|\\
		\leq& \sum_{k=1}^K\biggl(\frac{1}{T}\sum_{t=1}^T\biggl\|\frac{1}{N}\sum_{i=1}^N\mathbf 1\{g_i^0=g\}\mathbf 1\{g_i=\tilde g\}\mathbf 1\{k_i^0=k\}w_{it}\biggr\|^2\biggr)^{\frac{1}{2}}\\
		\leq& \sum_{k=1}^K\biggl(\frac{1}{N^2}\sum_{i=1}^N\sum_{j=1}^N\biggl|\frac{1}{T}\sum_{t=1}^Tw_{it}'w_{jt}\biggr|\biggr)^{\frac{1}{2}}.
	\end{align*}
	The first two terms are $O_p(T^{-1/2})$ by Assumptions \ref{ass-ts}\ref{ass-ts:cpt-param}, \ref{ass-ts:wkdep-zv}, and \ref{ass-ts:wkdep-zu}, and the remaining two terms are $O_p((1/N+1/T)^{1/2})$ by Assumptions \ref{ass-ts}\ref{ass-ts:cpt-param}, \ref{ass-ts:wkcdep-v}--\ref{ass-ts:wkdep-v} and \ref{ass-ts:wkcdep-u}--\ref{ass-ts:wkdep-u}.
	\item \textbf{Assumption \ref{ass-sineq}\ref{ass-sineq:wkdep-e}:} For each $(g,\tilde g)\in\mathbb G^2$ and $\gamma\in\mathbb G^N$, 
	\begin{align*}
		&\frac{1}{T}\sum_{t=1}^T\biggl\|\frac{1}{N}\sum_{i=1}^N\mathbf 1\{g_i^0=g\}\mathbf 1\{g_i=\tilde g\}(\smash[t]{\overbrace{v_{it}'\beta_{g_i^0}+u_{it}}^{e_{it}}})\biggr\|\\
		\leq& \max_{\beta_g\in\mathcal B}\|\beta_g\|\biggl(\frac{1}{N^2}\sum_{i=1}^N\sum_{j=1}^N\biggl|\frac{1}{T}\sum_{t=1}^Tv_{it}'v_{jt}\biggr|\biggr)^{\frac{1}{2}}
		+\biggl(\frac{1}{N^2}\sum_{i=1}^N\sum_{j=1}^N\biggl|\frac{1}{T}\sum_{t=1}^Tu_{it}u_{jt}\biggr|\biggr)^{\frac{1}{2}},
	\end{align*}
	which neither depends on $g$, $\tilde g$, nor $\gamma$. The same arguments as above can be made.
\end{itemize}

In addition, while establishing the uniform convergence, we made a similar statement for the variables $z_{it}$, $\mu_{k_i^0t}^0$, and $v_{it}$. It can be shown by following virtually the same steps as above.

Now, we conclude by explaining why the uniform convergence was enough for equation \eqref{eq-ts:beta-cnst:min}. Suppose so. Then, 
\begin{align*}
	&\min_{(\beta,\alpha,\gamma)\in\mathcal B^G\times \mathcal A^{TG}\times \mathbb G^N}	\biggl(\frac{1}{NT}\sum_{i=1}^N\sum_{t=1}^T(y_{it}-(\hat\Pi_{\hat k_i}'z_{it}+\hat\mu_{\hat k_it})'\beta_{g_i} - \alpha_{g_it})^2\\
	&\phantom{\min_{(\beta,\alpha,\gamma)\in\mathcal B^G\times \mathcal A^{TG}\times \mathbb G^N}	\biggl(} 
	- \frac{1}{NT}\sum_{i=1}^N\sum_{t=1}^T(y_{it}-(\hat\Pi_{\hat k_i}'z_{it}+\hat\mu_{\hat k_it})'\beta_{g_i} - \alpha_{g_it})^2\\
	&\phantom{\min_{(\beta,\alpha,\gamma)\in\mathcal B^G\times \mathcal A^{TG}\times \mathbb G^N}	\biggl(}
	+ \frac{1}{NT}\sum_{i=1}^N\sum_{t=1}^T(y_{it}-(\Pi_{k_i^0}^{0\prime}z_{it}+\mu_{k_i^0t}^0)'\beta_{g_i} - \alpha_{g_it})^2\biggr)\\
	\geq& \min_{(\beta,\alpha,\gamma)\in\mathcal B^G\times \mathcal A^{TG}\times \mathbb G^N}	\biggl(\frac{1}{NT}\sum_{i=1}^N\sum_{t=1}^T(y_{it}-(\hat\Pi_{\hat k_i}'z_{it}+\hat\mu_{\hat k_it})'\beta_{g_i} - \alpha_{g_it})^2\\
	&\phantom{\min_{(\beta,\alpha,\gamma)\in\mathcal B^G\times \mathcal A^{TG}\times \mathbb G^N}	\biggl(}
	-\max_{(\beta,\alpha,\gamma)\in\mathcal B^G\times \mathcal A^{TG}\times \mathbb G^N}\biggl| \frac{1}{NT}\sum_{i=1}^N\sum_{t=1}^T(y_{it}-(\hat\Pi_{\hat k_i}'z_{it}+\hat\mu_{\hat k_it})'\beta_{g_i} - \alpha_{g_it})^2\\
	&\phantom{\min_{(\beta,\alpha,\gamma)\in\mathcal B^G\times \mathcal A^{TG}\times \mathbb G^N}	\biggl(-\max_{(\beta,\alpha,\gamma)\in\mathcal B^G\times \mathcal A^{TG}\times \mathbb G^N}\biggl|}
	- \frac{1}{NT}\sum_{i=1}^N\sum_{t=1}^T(y_{it}-(\Pi_{k_i^0}^{0\prime}z_{it}+\mu_{k_i^0t}^0)'\beta_{g_i} - \alpha_{g_it})^2\biggr|\biggr)\\
	=&\min_{(\beta,\alpha,\gamma)\in\mathcal B^G\times \mathcal A^{TG}\times \mathbb G^N} \frac{1}{NT}\sum_{i=1}^N\sum_{t=1}^T(y_{it}-(\hat\Pi_{\hat k_i}'z_{it}+\hat\mu_{\hat k_it})'\beta_{g_i} - \alpha_{g_it})^2 + o_p(1)\\
	=&\frac{1}{NT}\sum_{i=1}^N\sum_{t=1}^T(y_{it}-(\hat\Pi_{\hat k_i}'z_{it}+\hat\mu_{\hat k_it})'\hat\beta_{\hat g_i} - \hat\alpha_{\hat g_i t})^2 + o_p(1) \text{ by definition}\\
	\geq& \frac{1}{NT}\sum_{i=1}^N\sum_{t=1}^T(y_{it}-(\Pi_{k_i^0}^{0\prime}z_{it}+\mu_{k_i^0t}^0)'\hat\beta_{\hat g_i} - \hat\alpha_{\hat g_i t})^2\\
	& -\max_{(\beta,\alpha,\gamma)\in\mathcal B^G\times \mathcal A^{TG}\times \mathbb G^N}\biggl|\frac{1}{NT}\sum_{i=1}^N\sum_{t=1}^T(y_{it}-(\hat\Pi_{\hat k_i}'z_{it}+\hat\mu_{\hat k_it})'\hat\beta_{\hat g_i} - \hat\alpha_{\hat g_i t})^2\\
	&\phantom{-\max_{(\beta,\alpha,\gamma)\in\mathcal B^G\times \mathcal A^{TG}\times \mathbb G^N}} - \frac{1}{NT}\sum_{i=1}^N\sum_{t=1}^T(y_{it}-(\Pi_{k_i}^{0\prime}z_{it}+\mu_{k_i^0t}^0)'\hat\beta_{\hat g_i} - \hat\alpha_{\hat g_i t})^2\biggr| + o_p(1)\\
	=& \frac{1}{NT}\sum_{i=1}^N\sum_{t=1}^T(y_{it}-(\Pi_{k_i^0}^{0\prime}z_{it}+\mu_{k_i^0t}^0)'\hat\beta_{\hat g_i} - \hat\alpha_{\hat g_i t})^2 + o_p(1).
\end{align*}

\subsection{Proof of Lemma \ref{lem-ts:gamma-cnst}}

We provide a more detailed explanation for some of the arguments in our proof.

First, we argued that there exists some (large) $\tilde L>0$ such that
\begin{align*}
	\Pr\biggl(\max_{i\in\{1,\dots,N\}}\biggl\|\frac{1}{T}\sum_{t=1}^T(\Pi_{k_i^0}^{0\prime}z_{it}+\mu_{k_i^0t}^0)(\Pi_{k_i^0}^{0\prime}z_{it}+\mu_{k_i^0t}^0)'\biggr\|_2> L\biggr)
	\leq \Pr\biggl(\max_{i\in\{1,\dots,N\}}\biggl\|\frac{1}{T}\sum_{t=1}^Tz_{it}z_{it}'\biggr\|_2\gtrsim \tilde L\biggr).
\end{align*}
This follows from the observation that by Assumption \ref{ass-ts}\ref{ass-ts:cpt-param},
\begin{align*}
	&\biggl\|\frac{1}{T}\sum_{t=1}^T(\Pi_{k_i^0}^{0\prime}z_{it}+\mu_{k_i^0t}^0)(\Pi_{k_i^0}^{0\prime}z_{it}+\mu_{k_i^0t}^0)'\biggr\|_2\\
	\lesssim& \biggl\|\frac{1}{T}\sum_{t=1}^T\mathbf z_{it}\mathbf z_{it}'\biggr\|_2
	\leq \biggl\|\frac{1}{T}\sum_{t=1}^Tz_{it}z_{it}'\biggr\|_2 + 2\underbrace{\left\|\begin{pmatrix}
		\frac{1}{\sqrt T}z_{i1}'\\
		\vdots\\
		\frac{1}{\sqrt T}z_{iT}'
	\end{pmatrix}\right\|_2}_{
	\displaystyle \mathclap{=\left\|\begin{pmatrix}
		\frac{1}{\sqrt T}z_{i1}'\\
		\vdots\\
		\frac{1}{\sqrt T}z_{iT}'
	\end{pmatrix}'\begin{pmatrix}
		\frac{1}{\sqrt T}z_{i1}'\\
		\vdots\\
		\frac{1}{\sqrt T}z_{iT}'
	\end{pmatrix}\right\|_2^{\frac{1}{2}}
	= \biggl\|\frac{1}{T}\sum_{t=1}^Tz_{it}z_{it}'\biggr\|_2^{\frac{1}{2}}
	}} + \|I_T\|_2 
	= \biggl(\biggl\|\frac{1}{T}\sum_{t=1}^Tz_{it}z_{it}'\biggr\|_2^{\frac{1}{2}} + 1\biggr)^2.
\end{align*}

Next, we clarify the origin of the term $L^{\frac{1}{2}}+1$. For illustration, we present the following derivation process of the concentration probability regarding one of the interaction terms:
\begin{align*}
	-\frac{2}{T}\sum_{t=1}^T\bigl((\Pi_{k_i^0}^{0\prime}z_{it}+\mu_{k_i^0t}^0)'(\beta_{g_i^0}^0-\tilde\beta_g) + (\alpha_{g_i^0 t}^0-\tilde\alpha_{g t})\bigr)\bigl((\Pi_{k_i^0}^{0\prime}z_{it} + \mu_{k_i^0t}^0)'(\tilde\beta_g - \hat\beta_g) + (\tilde\alpha_{gt} - \hat\alpha_{gt})\bigr).
\end{align*}
Its absolute value is bounded by
\begin{align*}
	&\left|\begin{pmatrix}
		\beta_{g_i^0}^0 - \tilde\beta_g\\
		\frac{1}{\sqrt T}(\alpha_{g_i^0 1}^0-\tilde\alpha_{g 1})\\
		\vdots\\
		\frac{1}{\sqrt T}(\alpha_{g_i^0 T}^0-\tilde\alpha_{g T})
	\end{pmatrix}
	\biggl(-\frac{2}{T}\sum_{t=1}^T\mathbf{\tilde x}_{it}\mathbf{\tilde x}_{it}'\biggr)
	\begin{pmatrix}
		\tilde\beta_g - \hat\beta_g\\
		\frac{1}{\sqrt T}(\tilde\alpha_{g1}-\hat\alpha_{g1})\\
		\vdots\\
		\frac{1}{\sqrt T}(\tilde\alpha_{gT}-\hat\alpha_{gT})
	\end{pmatrix}
	\right|
	\lesssim \biggl\|\frac{1}{T}\sum_{t=1}^T\mathbf{\tilde x}_{it}\mathbf{\tilde x}_{it}'\biggr\|_2
	\left\|\begin{pmatrix}
		\tilde\beta_g - \hat\beta_g\\
		\frac{1}{\sqrt T}(\tilde\alpha_{g1}-\hat\alpha_{g1})\\
		\vdots\\
		\frac{1}{\sqrt T}(\tilde\alpha_{gT}-\hat\alpha_{gT})
	\end{pmatrix}\right\|,
\end{align*}
again by Assumption \ref{ass-ts}\ref{ass-ts:cpt-param}. Now, note that $\|T^{-1}\sum_{t=1}^T\tilde x_{it}\tilde x_{it}\|_2\leq L$ implies
\begin{align*}
	\biggl\|\frac{1}{T}\sum_{t=1}^T\mathbf{\tilde x}_{it}\mathbf{\tilde x}_{it}'\biggr\|_2^{\frac{1}{2}}
	\leq \left(\biggl\|\frac{1}{T}\sum_{t=1}^T\tilde x_{it}\tilde x_{it}'\biggr\|_2
	+ 2\left\|\begin{pmatrix}
		\frac{1}{\sqrt T}\tilde x_{i1}'\\
		\vdots\\
		\frac{1}{\sqrt T}\tilde x_{iT}'
	\end{pmatrix}\right\|_2 + \|I_T\|_2\right)^{\frac{1}{2}}
	\leq (L + 2L^{\frac{1}{2}} + 1)^{\frac{1}{2}}
	= L^{\frac{1}{2}}+1.
\end{align*}

\printbibliography[category={appx}]

\end{document}